\documentclass{llncs}

\usepackage{amsmath}
\usepackage{pslatex}

\def\finex{{\unskip\nobreak\hfil
\penalty50\hskip1em\null\nobreak\hfil$\diamond$
\parfillskip=0pt\finalhyphendemerits=0\endgraf}}

\newenvironment{myex}{\begin{example}}{\finex\end{example}}

\newcommand{\comment}[1]{}

\usepackage{color,graphics,xcolor}

  \usepackage{ulem}
  \definecolor{shadecolor}{rgb}{1,0.99,0.9}

\newcommand{\qand}{\quad \text{and} \quad}

\newtheorem{DEF}{Definition}

\newcommand{\BNFOR}{\;\; \mid \;\; }
\newcommand{\rulename}[1]{{\scriptsize{\textsc{[#1]}}}}
\newcommand{\st}{\;:\;} %

\newcommand{\defi}{\overset{\text{def}}{=}}
\newcommand{\nada}{\bot}
\newcommand{\goodex}{\text{\ding{52}}}
\newcommand{\badex}{\text{\ding{56}}}
\newcommand{\conndef}[1]{R_{#1}}
\newcommand{\conndefG}{\conndef{\GA}}
\newcommand{\conndefS}{\conndef{\SYSV}}
\newcommand{\bisim}{\approx}
\renewcommand{\sim}{\, \lesssim \,}
\newcommand{\freen}[1]{\mathtt{fv}(#1)}
\newcommand{\boundn}[1]{\mathtt{bv}(#1)}
\newcommand{\msgsort}[1]{\mathtt{#1}}

\newcommand{\subs}[2]{[{#1}/{#2}]}

\newcommand{\coname}[1]{\overline{\chan{#1}}}

\newcommand{\SYSV}{S} %
\newcommand{\PV}{P} %
\newcommand{\GPV}{P} %
\newcommand{\SORTV}{\mathtt{e}} %
\newcommand{\GA}{\mathcal{G}} %
\newcommand{\nosort}{}

\newcommand{\SETPARTS}{\mathbb{P}}
\newcommand{\SETCHANS}{\mathbb{C}}

\newcommand{\GSENDPREF}[3]{\ptp{#1} \rightarrow \ptp{#2} : \chan{#3}}
\newcommand{\GSEND}[4]{\GSENDPREF{#1}{#2}{#3} \langle \msgsort{#4} \rangle}
\newcommand{\ptp}[1]{\mathtt{#1}}
\newcommand{\chan}[1]{#1}
\DeclareMathOperator{\GCH}{+}
\newcommand{\GPAR}{\PPAR}
\newcommand{\GEND}{\PEND}
\newcommand{\GSEP}{.}

\newcommand{\GPROD}[2]{\prod_{#1} #2}
\newcommand{\GWS}{\, ;} %
\newcommand{\GRECN}{\mu \, \chi .}
\newcommand{\GRECP}{\mu \, \chi}
\newcommand{\GREC}[1]{\mu \, #1 .}
\newcommand{\GRECV}{\chi}
\newcommand{\GSENDN}{\GSEND{s}{r}{a}{\SORTV}}
\newcommand{\GSTAR}{*} %

\renewcommand{\GA}{\mathcal{G}}

\newcommand{\intbang}{\text{\ftextinterrobang}}

\newcommand{\PSEP}{.}

\newcommand{\PSEND}[2]{\chan{#1} ! \mathtt{#2} }
\newcommand{\PRECEIVE}[2]{\chan{#1} ? \mathtt{#2}}
\newcommand{\PSERE}[2]{\chan{#1} \intbang \langle \mathtt{#2} \rangle}
\DeclareMathOperator{\PINCH}{\oplus}
\DeclareMathOperator{\POUTCH}{+}
\newcommand{\PSUM}[2]{\sum_{#1}#2}
\newcommand{\PINS}[2]{\bigoplus_{#1}#2}
\DeclareMathOperator{\PPAR}{\; | \;}
\newcommand{\END}{\mathbf{0}}
\newcommand{\PEND}{\mathbf{0}}
\newcommand{\PERROR}{\mathtt{error}}
\newcommand{\PBOX}[2]{\ptp{#1} [ #2 ]}
\newcommand{\PBOXB}[2]{\ptp{#1} \left[ #2 \right]}
\newcommand{\PBOXi}[3]{\ptp{#1_{#2}} [ #3 ]}

\newcommand{\QUEUE}[2]{#1 : #2}
\newcommand{\emptyQ}{[]}

\newcommand{\PPROD}[2]{\prod_{#1} #2}
\newcommand{\PRECN}{\mu \, \PRECVN .}
\newcommand{\PRECE}{\mu \,}

\newcommand{\PRECi}[2]{\mu \, \PRECV{#1}_{#2} .}
\newcommand{\PRECV}[1]{\MakeLowercase{\textbf{#1}}}
\newcommand{\PRECVN}{\PRECV{x}}

\newcommand{\Tbool}{\texttt{bool}}
\newcommand{\Tint}{\texttt{int}}

\newcommand{\labsend}{ \labbox{ \PSEND{a}{\SORTV}}}
\newcommand{\labrec}{ \labbox{\PRECEIVE{a}{\SORTV}}}

\newcommand{\labend}{ \checkmark}
\newcommand{\labpop}{ \mathtt{\SORTV}  \cdot  a}
\newcommand{\labpush}{a  \cdot  \mathtt{\SORTV} }

\newcommand{\labboxn}[2]{ \ptp{ #1 } [ #2 ] }
\newcommand{\labbox}[1]{ \labboxn{n}{#1} }
\newcommand{\labn}[1]{\ptp{n}  :  {#1}}
\newcommand{\labsyns}{\labn{\PSEND{a}{\SORTV}}}
\newcommand{\labsynr}{\labn{\PRECEIVE{a}{\SORTV}}}
\newcommand{\ltsrule}[3]{\inferrule*[left=\rulename{#1}]{#2}{#3}}
\newcommand{\ltsruleS}[2]{\rulename{\textsc{#1}} \; #2}
\newcommand{\ltsarrow}[1]{\overset{{#1}}{\longrightarrow}}

\newcommand{\ltsarrowC}[1]{\overset{{\scriptstyle #1}}{\Longrightarrow}}
\newcommand{\glorule}[3]{\inferrule*[left=\rulename{\ensuremath{#1}}]{#2}{#3}}

\newcommand{\chanT}[3]{\GSENDPREF{#2}{#3}{#1}}
\newcommand{\chanTN}{\chanT{a}{s}{r}}
\newcommand{\chanTA}{\chanT{a}{\_}{\_}}

\newcommand{\addchan}[2]{#1 \! \Yleft #2}
\newcommand{\addchanB}[2]{#1 \! \Yleft #2}

\DeclareMathOperator{\addnodeop}{\star}

\newcommand{\addnode}[2]{#1 \addnodeop   #2}

\newcommand{\chanG}[1]{\mathcal{T}(#1)}
\newcommand{\emptyTree}{\bullet}
\newcommand{\emptyLab}{\circ}
\newcommand{\subtree}[2]{#1(\chan{#2})}
\newcommand{\subtreeC}[1]{C(\chan{#1})}

\newcommand{\nodeV}{c}
\newcommand{\nodeT}[1]{\mathsf{#1}}
\newcommand{\nodeTN}{\nodeT{\nodeV}}
\newcommand{\lab}[1]{\underline{\nodeT{#1}}}
\newcommand{\labN}{\lab{\nodeTN}}

\newcommand{\linprev}{\prec}
\newcommand{\linrel}[1]{\prec_{#1}}
\newcommand{\linII}{\linrel{\mathtt{II}}}
\newcommand{\linIO}{\linrel{\mathtt{IO}}}
\newcommand{\linOO}{\linrel{\mathtt{OO}}}
\newcommand{\prefg}{\nodeTN}

\newcommand{\singleN}[1]{#1}
\newcommand{\onekid}[1]
{
  \begin{array}{c}
    \Treek[-1]{1.5}{
      \K{\ensuremath{\nodeTN}}\B{d} \\
      \K{\ensuremath{#1}}
    }
  \end{array}
}

\newcommand{\onekidR}[2]
{
  \begin{array}{c}
    \Treek[-1.2]{1.5}{
      \K{\ensuremath{#1}}\B{d} \\
      \K{\ensuremath{#2}}
    }
  \end{array}
}

\newcommand{\twokids}[2]
{
  \begin{array}{c}
    \Treek[-1.2]{1.5}{
      &\K{\ensuremath{\nodeTN}}\B{dr}\B{dl} \\
      \K{\ensuremath{#1}}&&\K{\ensuremath{#2}}
    }
  \end{array}
}

\newcommand{\twokidsR}[3]
{
  \begin{array}{c}
    \Treek[-1.2]{1.5}{
      &\K{\ensuremath{#1}}\B{dr}\B{dl} \\
      \K{\ensuremath{#2}}&&\K{\ensuremath{#3}}
    }
  \end{array}
}

\newcommand{\GFOUTE}{\mathtt{F_O}}
\newcommand{\GFOUT}[2]{\GFOUTE(#2  ,  #1)}
\newcommand{\GFOUTS}[1]{\GFOUTE(#1)}

\newcommand{\GFINE}{\mathtt{R}}
\newcommand{\GFIN}[1]{\GFINE(#1)}

\newcommand{\GFINPE}{\mathtt{F_P}}
\newcommand{\GFINP}[1]{\GFINPE(#1)}

\newcommand{\wfjudge}[2]{#1 \vdash \! \! #2}
\newcommand{\wfrule}[3]{\inferrule*[left=\rulename{{#1}}]{#2}{#3}}

\newcommand{\PARTS}[1]{\mathcal{P}(#1)} %
\newcommand{\CHANS}[1]{\mathcal{C}(#1)} %

\newcommand{\Pmerge}[2]{#1 \uplus #2} %
\newcommand{\Pmergec}[2]{#1 \uplus #2} %
\newcommand{\Proj}[2]{#1\! \! \downharpoonright_{\ptp{#2}}}
\newcommand{\Projc}[2]{#1\! \! \downharpoonright_{\chan{#2}}}

\newcommand{\spjudge}[4]{#1 ; \, #2 \; \vdash \, #3 \, \Bumpeq \, #4} %
\newcommand{\prtrack}[2]{\ptp{#1} : #2} %
\newcommand{\prtrackc}[2]{\chan{#1} : \mathtt{#2}} %
\newcommand{\SsplitE}{\mathtt{split}}
\newcommand{\Ssplit}[1]{\SsplitE (#1)} %
\newcommand{\fsplit}[2]{#1 \, \text{\%} \, #2} %
\newcommand{\sppre}[2]{#1 - #2}
\newcommand{\Omegan}[1]{\Omega(\ptp{#1})}
\newcommand{\Omegac}[1]{\Omega(\chan{#1})}
\newcommand{\Omeganp}[1]{\Omega '(\ptp{#1})}

\newcommand{\lnkT}{\stackrel\Theta{\tiny\frown}}
\newcommand{\lnkO}{\stackrel\Omega{\tiny\frown}}

\newcommand{\sprelT}{\leftrightarrow_{\Theta}}
\newcommand{\spR}{\circledast}
\newcommand{\spminus}[1]{\! \setminus \! \{ \ptp{ #1 } \}}

\newcommand{\spcomp}[2]{#1 \asymp #2}
\newcommand{\spmerge}[2]{#1 \sqcup #2}

\newcommand{\splepsilon}{\epsilon}
\newcommand{\splsync}{sync}
\newcommand{\splplusl}{\POUTCH}

\newcommand{\splplus}{\POUTCH}
\newcommand{\splpinch}{\PINCH}
\newcommand{\splax}{ax}
\newcommand{\splend}{\PEND}
\newcommand{\splrem}{rem}
\newcommand{\splqueue}{q}

\newcommand{\spP}{\pi}
\newcommand{\spQ}{\varphi}

\newcommand{\restri}[2]{#1 (\ptp{#2})}
\newcommand{\restric}[2]{#1 (\chan{#2})}
\newcommand{\restriset}[2]{#1 [#2]}

\newcommand{\fire}[1]{\mathtt{R}(#1)}
\newcommand{\firebar}[1]{\mathtt{\overline{R}}(#1)}

\newcommand{\noonestep}[1]{#1 \! \not\updownarrow}
\newcommand{\onestep}[1]{#1 \! \updownarrow}

\newcommand{\chanset}{A}
\newcommand{\emptyctx}{\circ}
\newcommand{\gloturn}{\vdash}
\newcommand{\glotri}{\blacktriangleright}
\newcommand{\judgeC}[5]
{#1 \, ; \, #2  \, ; \, #3 \, \gloturn #4 \, \glotri #5}

\newcommand{\judge}[4]
{\judgeC{\chanset}{#1}{#2}{#3}{#4}}

\newcommand{\mybi}[1]{\ptp{b_{#1}}}
\newcommand{\myba}{\mybi{1}}
\newcommand{\mybb}{\mybi{2}}
\newcommand{\mysi}[1]{\ptp{s_{#1}}}
\newcommand{\myorder}{\msgsort{order}}
\newcommand{\myprice}{\msgsort{price}}
\newcommand{\myaddress}{\msgsort{addr}}

\newcommand{\gipsep}{\PSEP}
\newcommand{\gipinch}{\PINCH}
\newcommand{\gipoutchl}{\GCH}%
\newcommand{\gipoutch}{\GCH}
\newcommand{\gipar}{\PPAR}

\newcommand{\gigws}{\GWS}
\newcommand{\gimua}{\ensuremath{\mu}}
\newcommand{\givar}{\PRECVN}
\newcommand{\giend}{\PEND}
\newcommand{\gieq}{eq}
\newcommand{\giqueue}{\rho}

\newcommand{\wfseq}{wf-$\GWS$}
\newcommand{\wfdot}{wf-$\GSEP$}
\newcommand{\wfpar}{wf-$\GPAR$}
\newcommand{\wfchoice}{wf-$\GCH$}
\newcommand{\wfseqen}{wf-$\GWS $-$ \GEND$}
\newcommand{\wfrec}{wf-$\GRECP$}
\newcommand{\wfvar}{wf-$\GRECV$}
\newcommand{\wfend}{wf-$\GEND$}

\newcommand{\derive}[2]{
\judge{\Gamma}{C}{#1}{#2}} %

\newcommand{\deriveN}{\judge{\Gamma}{C}{\SYSV}{\GA}} %

\newcommand{\proocase}[1]{\noindent \textbf{#1.}}

\newcommand{\thmdecide}{
Typability is decidable.
}

\newcommand{\thmunique}{
If $\deriveN $ and $\derive{\SYSV}{\GA'}$ then $\GA \equiv \GA'$.
}

\newcommand{\thmwf}{
If $\deriveN$ then $\wfjudge{\emptyTree}{\GA}$ and $\GA$ is projectable.
}

\newcommand{\thmsubred}{
If $\judge{\emptyctx}{C}{\SYSV}{\GA}$, $\SYSV \ltsarrow{\lambda} \SYSV'$, and
$\CHANS{\lambda} \not\in C$
then $\judge{\emptyctx}{C}{\SYSV'}{\GA'}$
}

\newcommand{\thmprogress}{
If $\judge{\emptyctx}{C}{\SYSV}{\GA}$
then $\SYSV \ltsarrow{} \SYSV'$, or
$\forall \, \ptp{n} \in \PARTS{\SYSV} \, . \, \restri{\SYSV}{n} = \PEND$,
or $\SYSV \ltsarrow{\labend}$. 
}

\newcommand{\thmsafety}{
If $\judge{\emptyctx}{C}{\SYSV}{\GA}$, then
$\SYSV$ is race free and
$\SYSV \ltsarrowC{} \ltsarrow{} \PERROR$ is not possible.
}

\newcommand{\thmsimu}{
If  $\judge{\emptyctx}{C}{\SYSV}{\GA}$ then $\forall \ptp{n} \in S \, . $
$\Proj{\GA}{n} \sim \restri{S}{n}$.
}

\newcommand{\thmbisim}{ If $\judge{\emptyctx}{C}{\SYSV}{\GA}$ then
  $\PPROD{\ptp{n} \in \PARTS{\SYSV}}{\PBOX{n}{\Proj{\GA}{n}}} \bisim
  \SYSV$.  }

\newcommand{\thmcomplete}{
If $\wfjudge{\emptyTree}{\GA}$ and $\GA$ is projectable,
then there is $\GA' \equiv \GA$ such that
$\derive{\PPROD{\ptp{n} \in \PARTS{\GA}}{\PBOX{n}{\Proj{\GA}{n}}}}{\GA'}$.
}

\usepackage{pslatex}
\usepackage{latexsym}
\usepackage{graphicx}
\usepackage{amssymb}
\usepackage{amsmath}
\usepackage{color}
\usepackage{latexsym}
\usepackage{graphicx}
\usepackage{citesort}
\usepackage{url}
\usepackage{mathpartir}
\usepackage{stmaryrd}
\usepackage{wasysym}
\usepackage{pifont}
\usepackage{xyling} %
\usepackage{amscd} %
\usepackage{array}
\usepackage{multirow}

\usepackage{rotating}

\usepackage{textcomp} %
\newcommand{\ftextinterrobang}{{\fontfamily{txr}\selectfont \textinterrobang}}
\numberwithin{equation}{section}

\begin{document}
\title{
  Synthesising Choreographies from Local Session Types
}
\subtitle{(extended version)}
\author{Julien Lange \and Emilio Tuosto}

\institute{Department of Computer Science, University of Leicester, UK}

\maketitle
\pagestyle{plain}

\begin{abstract}
  Designing and analysing multiparty distributed interactions can be
  achieved either by means of a global view (e.g. in
  choreography-based approaches) or by composing available
  computational entities (e.g. in service orchestration).

  This paper proposes a typing systems which allows, under some
  conditions, to synthesise a choreography (i.e.\ a multiparty global
  type) from a set of local session types which describe end-point
  behaviours (i.e.\ local types).
\end{abstract}

\section{Introduction}\label{sec:intro}
Communication-centred applications are paramount in the design and
implementation of modern distributed systems such as those in
service-oriented or cloud computing.
Session types~\cite{hv98} and their multiparty variants~\cite{hyc08,dy11} offer an
effective formal framework for designing, analysing, and implementing
this class of applications.
Those theories feature rather appealing methodologies that consists of
($i$) designing a global view of the interactions -- aka \emph{global
  type} --, ($ii$) effective analysis of such a global view, ($iii$)
automatic projection of the global view to local end-points -- aka
\emph{local types} --, and ($iv$) type checking end-point code against
local types.
Such theories guarantee that, when the global view enjoys suitable
properties (phase $(ii)$), the end-points typable with local types
enjoy e.g., liveness properties like progress.

A drawback of such approaches is that they cannot be applied when the
local types describing the communication patterns of end-points are
not obtained by an a priori designed global view.
For instance, in service-oriented computing, one typically has
independently developed end-points that have to be combined to form
larger services.
Hence, deciding if the combined service respects its specification
becomes non trivial.
To illustrate this, we introduce a simple example used throughout the
paper.

\newcommand{\sex}{\SYSV_{\text{BS}}}

Consider a system $\sex = \PBOX{\myba}{\PV_1} \PPAR
\PBOX{\mysi{1}}{S_1} \PPAR \PBOX{\mybb}{\PV_2} \PPAR
\PBOX{\mysi{2}}{S_2}$
consisting of two buyers ($\myba$ and $\mybb$) and two servers
($\mysi 1$ and $\mysi 2$) running in parallel, so that
\begin{equation*}\label{eq:ex1local}
\begin{array}{lll}
  \PV_1 & =  &  \PSEND{t_1}{\myorder} \PSEP 
  \PRECEIVE{p_1}{\myprice} \PSEP 
  \PRECEIVE{r}{\myprice} \PSEP 
  (\PSEND{c_1}{\nosort} \PSEP \PSEND{t_1}{\myaddress}
  \PINCH 
  \PSEND{c_2}{\nosort} \PSEP \PSEND{no_1}{\nosort}) \qquad \text{is the behaviour of } \myba
  \\[0.2pc]
  \PV_2 & =  & \PSEND{t_2}{\myorder} \PSEP 
  \PRECEIVE{p_2}{\myprice} \PSEP
  \PSEND{r}{\myprice} \PSEP 
  (\PRECEIVE{c_2}{\nosort} \PSEP \PSEND{t_2}{\myaddress}
  \POUTCH 
  \PRECEIVE{c_1}{\nosort} \PSEP \PSEND{no_2}{\nosort}) \qquad \text{is the behaviour of } \mybb
  \\[0.2pc]
  S_i & =  &   \PRECEIVE{t_i}{\myorder} \PSEP 
  \PSEND{p_i}{\myprice} \PSEP 
  (\PRECEIVE{t_i}{\myaddress} \POUTCH \PRECEIVE{no_i}{\nosort}), \ \ \ \ i \in \{1,2\}
  \qquad\qquad\, \text{is the behaviour of } \mysi i
\end{array}
\end{equation*}
with $\PSEND{a}{\SORTV}$ (resp. $\PRECEIVE{a}{\SORTV}$) representing
the action of sending (resp. receiving) a message of type $\SORTV$ on
a channel $\chan{a}$ (we omit $\SORTV$ when the message is
immaterial), $\PINCH$ representing an internal choice, and $\POUTCH$
a choice made by the environment.

Intuitively, the overall behaviour of $\sex$ should be that either
$\myba$ or $\mybb$ purchase from their corresponding sellers.
A natural question arises: is $\sex$ correct?
Arguably, it is not immediate to decide this by considering the
end-point behaviours.

We propose to construct a global view of distributed end-points like
$\sex$ by a straightforward extension of the multiparty session types
introduced in~\cite{hyc08}.
Such types formalise a global view of the behaviour which, for $\sex$,
resembles the informal diagram below,
where the choreography of the overall protocol becomes much clearer.
\begin{figure}[h!]
\centering
\includegraphics[scale=0.4]{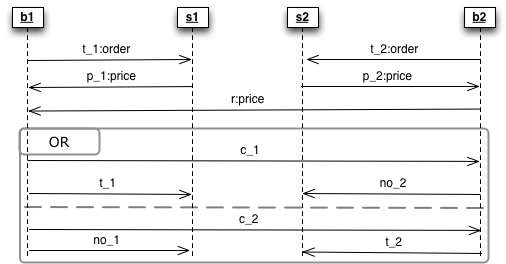}
\end{figure}

An advantage of our approach is that we can reuse the results of
the theory of multiparty session types to prove properties of
end-points e.g.\ safety and progress.
In fact, we show that when the choreography can be constructed,
its projections correspond to the initial end-points.
Therefore, the well-formedness of the synthesised global choreography
guarantees progress and safety properties of end-points.

The extraction of session types from programs has been studied
extensively~\cite{hyc08,cp09,hv98}.
We assume in this work that such session types are readily available
before addressing the construction of a global type.

\paragraph{Contributions.}
We introduce a theory whereby, under some conditions, it is possible
to assign a global type to a set of local types.
If a global type can be constructed from a set of local types, we show
that it is unique (Theorem~\ref{thm:RUNIQUE}) and well-formed
(Theorem~\ref{thm:RWF}).
In addition, we show that typability is preserved by reduction
(Theorem~\ref{thm:RSUBRED}).
Our theory also guarantees progress and safety properties
(Theorems~\ref{thm:RPROGR} and~\ref{thm:RSAFE}).
We also show that the projections of a constructed global type are
equivalent to the original system (Theorem~\ref{thm:RBISIMU}).
Finally, we show that for every well-formed global types, an
equivalent global type can be assigned to its projections
(Theorem~\ref{thm:RCOMPL}).

\paragraph{Synopsis.}
In \S~\ref{sec:local}, we give the syntax and semantics of the local
types from which it is possible to construct a global type.
In \S~\ref{sec:global}, we present an extension of the global
types in~\cite{hyc08}.
In \S~\ref{sec:infer}, we introduce a typing systems for
local types, and we give our main results.
Finally, in \S~\ref{sec:conc} we conclude, and discuss related and future work.

\section{Local Types}\label{sec:local}
We use CCS-like processes (with guarded external and internal choices)
to infer a global type from \emph{local types} that correspond to the
participants in the inferred choreography.
Hereafter, $\SETPARTS$ is a denumerable set of \emph{participant
  names} (ranged over by $\ptp{s}$, $\ptp{r}$, $\ptp{n}$, \ldots) and
$\SETCHANS$ is a denumerable set of \emph{channel names} (ranged over by
$\chan{a}$, $\chan{b}$, \ldots).

\paragraph{Syntax.}
The syntax of local types below is parametrised wrt basic data types
such as $\Tbool, \Tint, \ldots$ (ranged over by $\SORTV$):
\[\begin{array}{l@{\quad}l@{\quad}l}
  \SYSV, T & ::= & \SYSV \PPAR \SYSV'
  \BNFOR \PBOX{n}{\PV}
  \BNFOR \QUEUE{a}{\rho}
  \BNFOR \END
  \\[0.7pc]
  \PV,Q  & ::= &  \PINS{i \in I}{\PSEND{a_i}{\SORTV_i}{\PSEP \PV_i}} \BNFOR
  \PSUM{i \in I}{\PRECEIVE{a_i}{\SORTV_i}{\PSEP \PV_i}} \BNFOR
  \PRECN P \BNFOR
  \PRECVN 
\end{array}\]
A system $\SYSV$ consists of the parallel
composition of \emph{processes} and queues.
A \emph{process} $\PBOX{n}{\PV}$ is a behaviour $\PV$ identified by
$\ptp{n} \in \SETPARTS$; we assume that the participant names are all
different.
A behaviour is either a guarded external choice, a guarded
internal choice, or a recursive process.
An internal choice $\PINS{i \in I}{\PSEND{a_i}{\SORTV_i}{\PSEP
    \PV_i}}$ is guarded by output prefixes $\PSEND{a_i}{\SORTV_i}$
representing the sending of a value of sort $\SORTV_i$ on channel
$\chan{a_i}$.
An external choice $\PSUM{i \in I}{\PRECEIVE{a_i}{\SORTV_i}{\PSEP
    \PV_i}}$ is guarded by input prefixes $\PRECEIVE{a_i}{\SORTV_i}$
representing the reception of a value of type $\SORTV_i$ on channel $a_i$.
We adopt asynchronous (order-preserving) communications and assume
that the channels in the guards of choices are pairwise distinct;
moreover
\[\begin{array}{c}
\PEND
\quad \defi \quad
\PINS{i \in \varnothing}{\PSEND{a_i}{\SORTV_i}{\PSEP\GPV_i}}
\quad = \quad
\PSUM{i \in \varnothing}{\PRECEIVE{a_i}{\SORTV_i} \PSEP \PV_i}
\end{array}\]
Finally, in a recursive behaviour $\PRECN \PV$, all occurrences of
$\PRECVN$ in $\PV$ are bound and prefix-guarded; also, we consider
closed behaviours only that is, behaviours with no free occurrences of
recursion variables.
We assume that bound variables are pairwise distinct.

A \emph{program} is a system with no queues, while a \emph{runtime
  system} is a system having exactly one queue $\QUEUE{a}{\rho}$ per
channel name $\chan{a} \in \SETCHANS$ in $\SYSV$.
In the following, $S, T, \ldots$ denote either a program or runtime system.
\paragraph{Semantics.}
The semantics of local types is given by the labelled transition
system (LTS) whose labels are 
\[
\begin{array}{c}
\lambda ::= 
\alpha \BNFOR
\labend \BNFOR
\labpush \BNFOR
\labpop \BNFOR
\labbox{\alpha} \BNFOR
\labn \alpha
\qquad \text{where} \qquad
\alpha ::= \PSEND{a}{\SORTV} \BNFOR \PRECEIVE{a}{\SORTV}
\end{array}\]
Label $\alpha$ indicates either sending or reception by a process.
Label $\labend$ indicates termination, $\labpush$ and $\labpop$
respectively indicate push and pop operations on queues.
Label $\labbox{\alpha}$ indicates a communication action done by
participant $\ptp n$ while $\labsyns$ and $\labsynr$ indicate a
synchronisation between $\ptp n$ and a queue.

Assume the usual laws for commutative monoids for $\PPAR$ and $\END$
on systems and $\PRECN \PV \equiv \PV \subs{\PRECN \PV}{\PRECVN}$.
The LTS $\ltsarrow \lambda$ is the smallest relation closed under the
following rules:
\newcommand{\ltspush}{push}
\newcommand{\ltspop}{pop}
\newcommand{\ltsend}{end}
\newcommand{\ltseq}{eq-p}
\newcommand{\ltseqs}{eq-s}
\newcommand{\ltsbox}{box}
\newcommand{\ltsext}{ext}
\newcommand{\ltsint}{int}
\newcommand{\ltsout}{out}
\newcommand{\ltsin}{in}
\[
{
\begin{array}{c@{\hspace{0.8cm}}c@{\hspace{0.8cm}}c}
\multicolumn{3}{c}{
  \ltsruleS{\ltsint}
  {\PINS{i \in I}{\PSEND{a_i}{\SORTV_i} \PSEP \PV_i}
    \ltsarrow{\PSEND{a_j}{\SORTV_j}}
    \PV_j}
  \qquad\qquad
  \ltsruleS{\ltsext}
  {\PSUM{i \in I}{\PRECEIVE{a_i}{\SORTV_i} \PSEP \PV_i}
    \ltsarrow{\PRECEIVE{a_j}{\SORTV_j}}
    \PV_{j}}
  \quad j \in I
}
  \\[.7pc]
  \ltsruleS{\ltspush}{\QUEUE{a}{\rho} \ltsarrow{\labpush} \QUEUE{a}{\rho \cdot \SORTV}}
  &
  \ltsruleS{\ltspop}{\QUEUE{a}{\SORTV \cdot \rho} \ltsarrow{\labpop} \QUEUE{a}{\rho}}
  &
  \ltsruleS{\ltsend}{\END \ltsarrow{\labend} \END}
  \\[.7pc]
  \ltsrule{\ltsin}
  {\SYSV \ltsarrow{\labrec} \SYSV' \\ T \ltsarrow{\labpop} T'}
  {\SYSV \PPAR T \ltsarrow{\labsynr} \SYSV' \PPAR T'}
  &
  \ltsrule{\ltsout}
  {\SYSV\ltsarrow{\labsend} \SYSV' \\ T \ltsarrow{\labpush} T' }
  {\SYSV \PPAR T \ltsarrow{\labsyns} \SYSV' \PPAR T'}
  &
  \ltsrule{\ltsbox}
  {\PV \ltsarrow{\alpha} \PV'}
  {\PBOX{n}{\PV} \ltsarrow{ \labbox{ \alpha } } \PBOX{n}{\PV'} }
  \\[.7pc]
  \multicolumn 3 c {
    \ltsrule{\ltseq}
    {\PV \equiv Q \ltsarrow{\alpha} Q' \equiv \PV'}
    {\PV \ltsarrow{\alpha} \PV'}
    \qquad\qquad\qquad
    \ltsrule{\ltseqs}
    {\SYSV \equiv T \ltsarrow{\lambda} T' \equiv \SYSV'}
    {\SYSV \ltsarrow{\lambda} \SYSV'}
  }
\end{array}
}
\]
Rules \rulename{\ltsint} and \rulename{\ltsext} are trivial.
By \rulename{\ltspush} (resp. \rulename\ltspop), a queue receives a
(resp. sends the first) datum (resp. if any).
Processes can synchronise with queues according to rules
\rulename{\ltsin} and \rulename{\ltsout}.
The remaining rules are rather standard.
Let $\SYSV \ltsarrow{}$ iff there are $\SYSV'$ and $\lambda$ s.t.
$\SYSV \ltsarrow{\lambda} \SYSV'$ and $\ltsarrowC{\lambda_1 \ldots
  \lambda_n}$ (resp. $\ltsarrowC{}$) be the reflexive transitive
closure of $\ltsarrow{\lambda}$ (resp. $\ltsarrow{}$).

\section{Global Types}\label{sec:global}
A global type $\GA$ specifies an ordering of the interactions in a
choreography.
The syntax for global types in~\cite{hyc08} is extended with a generalised
sequencing in the following syntax:
\[\begin{array}{c}
\GA \ ::=  \ \GSENDN \GSEP \GA
\BNFOR  \GA \GWS \GA'
\BNFOR \GA \GCH \GA'
\BNFOR \GA \GPAR \GA'
\BNFOR  \GRECN \GA
\BNFOR \GRECV
\BNFOR \GEND
\end{array}\]
The prefix $\GSENDN$ represents an interaction where $\ptp{s} \in
\SETPARTS$ sends a value of sort $\SORTV$ to $\ptp{r} \in \SETPARTS$
on $\chan{a} \in \SETCHANS$ (we let $\iota$ range over interactions
$\GSENDN$ and assume that $\ptp{s} \neq \ptp{r}$).
The production $\GA \GWS \GA'$ indicates a generalised form of
sequencing, where the interactions in $\GA'$ are enabled only after
the ones in $\GA$.
The production $\GA \GCH \GA'$ indicates a (exclusive) choice of interactions.
Concurrent interactions are written $\GA \GPAR \GA'$.
A global type $\GRECN \GA$, indicates a recursive type, where $\GRECV$ is bound in $\GA$.
We assume that global types are closed and that recursion is guarded.
We often omit trailing occurrences of $\GEND$.

\begin{myex}\label{bla}
  The first two interactions between $\mybi i$ and $\mysi i$ in the
  example of \S~\ref{sec:intro} are
  \begin{equation}\label{eq:ex1glo1}
    \GA_i =
    \GSEND{\mybi{i}}{\mysi{i}}{t_i}{\myorder} \GSEP
    \GSEND{\mysi{i}}{\mybi{i}}{p_i}{\myprice}
    \quad i \in \{1,2\}
  \end{equation}
  The type $\GA_i$ says that a participant $\mybi{i}$ sends a message
  of type $\myorder$ to participant $\mysi{i}$ on channel
  $\chan{t_i}$, then $\mysi{i}$ replies with a message of type
  $\myprice$ on channel $\chan{p_i}$.
\end{myex}

The smallest equivalence relation satisfying the laws for commutative
monoids for $\GPAR$, $\GCH$, and $\GEND$ and the axioms below is the
structural congruence for global types:
\[
{\setlength\arraycolsep{1em}
  \begin{array}{c}
    \begin{array}{ccc}
      \GA \GWS \GEND \equiv \GA
      &
      \GEND \GWS \GA \equiv \GA
      &
      (\GA \GWS \GA') \GWS \GA'' \equiv \GA \GWS (\GA' \GWS \GA'')
    \end{array}
    \\[0.3pc]
    \begin{array}{cc}
      \iota \GSEP (\GA \GWS \GA')
      \equiv (\iota \GSEP \GA) \GWS \GA' 
      &
      \GRECN \GA \equiv \GA \subs{\GRECN \GA}{\GRECV}
  \end{array}
\end{array}
}
\]

The syntax of global types may specify behaviours that are not
implementable.
The rest of this section borrows from~\cite{cdp11} and~\cite{hyc08}
and adapts the requirements a global type must fulfil to ensure that
the ordering relation it prescribes is indeed feasible.

\subsection{Channel Usage and Linearity}\label{sub:lin}
It is paramount that no race occurs on the channels of a global type
(i.e.\ a datum sent on a channel is received by its intended recipient).
As in~\cite{hyc08}, we require that a global type is \emph{linear},
that is actions on channels shared by different participants are
temporally ordered.
For this, we use generic environments (ranged over by $C$) which keep
track of channel usage.
Such environments are trees defined as follows:
\[\begin{array}{c@{\qquad\qquad}c@{\qquad\qquad}c}\scriptstyle
\emptyTree
&\scriptstyle
\onekidR{\nodeTN}{C} %
&\scriptstyle
\twokidsR{\nodeTN}{C_1}{C_2}
\\\scriptstyle
\text{root only}
&\scriptstyle
C \text{ is a child of } \nodeTN
&\scriptstyle
C_1 \text{ and } C_2 \text{ are children of } \nodeTN
\end{array}\]
Each node $\nodeTN$ has a label $\lab{\nodeTN}$ of the form
$\emptyLab$, $\chanTN$, or $\mu \GRECV$ respectively representing the root
of choice or concurrent branches, an interaction between $\ptp s$ and
$\ptp r$ on $\chan{a}$, and a recursive behaviour.
We write $\nodeTN \in C$ if $\nodeTN$ is a node in $C$.
We use $\_$ as a wild-card when some of the components of a label are
immaterial, e.g.\ $\chanTA$ matches any label representing an
interaction on $\chan{a}$.
Given a tree $C$, we write $\nodeTN_1 \linprev \nodeTN_2$,
if $\nodeTN_1, \nodeTN_2 \in C$ and
$\nodeTN_2$ is a node in the sub-tree rooted at $\nodeTN_1$.
We adapt the definitions in~\cite{hyc08} to our framework.

\begin{DEF}[Dependency relations~\cite{hyc08}]\label{def:linrelations}
  Fix $C$, we define the following relations:
  \[\begin{array}{ll}\displaystyle
    \prefg_1 \linII \prefg_2
    &\displaystyle \text{if } \prefg_1 \linprev \prefg_2 \text{ and }
    \prefg_i = \chanT{a_i}{s_i}{r} \quad \displaystyle i \in \{1,2\}
    \\\displaystyle
    \prefg_1 \linIO \prefg_2
    &\displaystyle \text{if } \prefg_1 \linprev \prefg_2 \text{ and }
    \prefg_1 = \chanT{a_1}{s_1}{r} \text{ and }
    \prefg_2 = \chanT{a_2}{r}{s_2}
    \\\displaystyle
    \prefg_1 \linOO \prefg_2
    &\displaystyle \text{if } \prefg_1 \linprev \prefg_2 \text{ and }
    \prefg_i = \chanT{a}{s}{r_i} \quad\displaystyle i \in \{1,2\}
  \end{array}
  \]
  An \emph{input dependency} from $\prefg_1$ to $\prefg_2$ is a chain of the form
  $\prefg_1 \linrel{\phi_1} \ldots \linrel{\phi_k} \prefg_2$ ($k \geq 0$)
  such that $\phi_i \in \{\mathtt{II, IO}\}$ for $1 \leq i \leq k-1$
  and $\phi_k = \mathtt{II}$.
  An \emph{output dependency} from $\prefg_1$ to $\prefg_2$ is a chain
  $\prefg_1 \linrel{\phi_1} \ldots \linrel{\phi_k} \prefg_2$ ($k \geq 1$)
  such that $\phi_i \in \{\mathtt{OO, IO}\}$.
\end{DEF}

\begin{DEF}[Linearity~\cite{hyc08}]\label{def:linearity}
$C$ is linear if and only if
whenever  $\nodeTN_1 \linprev \nodeTN_2$
with
$\lab{\nodeTN_1}=\chanTA$ and
$\lab{\nodeTN_2} = \chanT{a}{\_}{\_}$ then
there is both input and output dependencies from
 $\nodeTN_1$ to $\nodeTN_2$.
\end{DEF}
We also define a function $\addnode{\_}{\_}$ to append trees as follows
\[%
\addnode{\onekid{C_0}\hspace{-.3cm}}{C'} \ = \ \onekid{\addnode{C_0}{C'}},
\qquad
\addnode{\emptyTree}{C} = C,
\qquad
\addnode{\twokids{C_1}{C_2}\hspace{-.3cm}}{C'} \ = \  \twokids{\addnode{C_1}{C'}}{\addnode{C_2}{C'}}
\]
and a partial function to append a tree $C'$ to a tree $C$ while
preserving linearity: $\addchan{C}{C'} = \addnode{C}{C'}$ if
$\addnode{C}{C'}$ is linear, otherwise $\addchan{C}{C'} = \nada$.
Also, let $\chanG{\GA}$ be the total function (cf.
Appendix~\ref{app:linearity}) which returns a tree $C$ corresponding
to the use of channels in $\GA$.

\subsection{Well-formed Global Types}
We define the conditions for a global type to be well-formed.
We write $\PARTS{\GA}$ (resp.\ $\CHANS{\GA}$) for the set of
participant (resp.\ channel) names in $\GA$, and $\freen{\GA}$ for the
set of free variables in $\GA$, similarly for a system $\SYSV$.
We give a few accessory functions.
\newcommand{\ppsset}{\mathcal{P}} %
Let
\begin{eqnarray*}
  \GFIN{\GA} & \defi &
  \left\{
    \chanTN \,|\, \GA \equiv (\GSENDN \GSEP \GA_1 \GCH \GA_2 \GPAR \GA_3)
    \GWS \GA_4
  \right\}
  \\[0.3em]
  \GFINP{\GA} & \defi &
  \begin{cases}\displaystyle
    \GFINP{\GA_1} \cup \GFINP{\GA_2}, & \displaystyle \GA = \GA_1 \GPAR \GA_2
    \\\displaystyle
    \{ \PARTS \GA \}, &\displaystyle \text{otherwise} 
  \end{cases}
  \\[0.3em]
  \GFOUT{\GA}{\ppsset} & \defi &
  \begin{cases}\displaystyle
    \GFOUT{\GA_1}{ \{\ptp{s} , \ptp{r}\} \cup \ppsset}
    ,&\displaystyle \GA = \GSENDN \GSEP \GA_1
    \\\displaystyle
    \GFOUT{\GA_1}{\varnothing} \cup \GFOUT{\GA_2}{\varnothing}
    ,&\displaystyle \GA = \GA_1 \GPAR \GA_2
    \\\displaystyle
    \GFOUT{\GA_1}{\ppsset}
    ,&\displaystyle \GA = \GA_1 \GCH \GA_2 
    \text{ and } \GFOUT{\GA_1}{\ppsset} = \GFOUT{\GA_2}{\ppsset}
    \\\displaystyle
    \GFOUT{\GA_1}{\ppsset}
    ,&
    \GA = \GRECN \GA_1
    \\\displaystyle
    \GFOUT{\GA_2}{\varnothing}
    ,&\displaystyle \GA = \GA_1 \GWS \GA_2
    \\\displaystyle
    \{\ppsset\}
    ,&\displaystyle \GA = \GEND \text{ or } \GA = \GRECV
    \\\displaystyle
    \nada
    ,&\displaystyle \text{otherwise}  
  \end{cases}
\end{eqnarray*}
$\GFIN{\GA}$ is the \emph{ready set} of $\GA$.
$\GFINP{\GA}$ is the family of sets of its participants running in different
concurrent branches. That is, $N \in \GFINP{\GA}$ iff all $\ptp n
\in N$ are in a same top-level thread of $\GA$.
$\GFOUT{\ppsset}{\GA}$ is the family of sets of participants of $\GA$,
so that for all $N,M \in \GFOUT \GA \ppsset$, the participants in $N$
and those in $M$ are in different concurrent branches in the last part
of $\GA$; define $\GFOUTS{\GA} \defi \GFOUT{\GA}{\varnothing}$.
Note that $\GFOUT \_ \_ $ is a partial function.
\begin{myex}
  Let $\GA_{i,j} = \GSEND{\myba}{\mybb}{c_i}{\nosort} \GSEP
  (\GSEND{\mybi{i}}{\mysi{i}}{t_i}{\myaddress} \GPAR
  \GSEND{\mybi{j}}{\mysi{j}}{no_j}{\nosort})$ describe each of the
  branches of the \emph{or} box in the example of \S~\ref{sec:intro},
  where $i\neq j \in \{1,2\}$, then
  \[\displaystyle
  \GFIN{\GA_{1,2}} =
  \left\{ \chanT{c_1}{\myba}{\mybb} \right\},
  \;
  \GFINP{\GA_{1,2}} =
  \left\{
    \ptp{\myba},\ptp{\mysi 1},\ptp{\mybb},\ptp{\mysi 2}
    \right\},
  \;
  \GFOUTS{\GA_{1,2}} =
  \left\{
    \{\ptp{\myba},\ptp{\mysi 1}\},
    \{\ptp{\mybb},\ptp{\mysi 2}\}
  \right\}
  \]
  The global type below corresponds to the whole protocol of
  \S~\ref{sec:intro}
  \[\displaystyle
    \GA \ = \
    (\GA_1 \GPAR \GA_2) \GWS 
    \GSEND{\mybb}{\myba}{r}{\myprice} \GSEP 
    (\GA_{1,2} \GCH \GA_{2,1})
    \]
hence
$\GFIN{\GA} = \left\{\chanT{t_i}{\mybi i}{\mysi{i}}\right\}_{ i = 1,2}$,
$\GFINP{\GA} = \GFINP{\GA_{1,2}}$, and
$\GFOUTS{\GA} = \GFOUTS{\GA_{1,2}}$.
\end{myex}
\subsubsection{Well-formedness.}
The well-formedness of a global type $\GA$ depends on how it uses
channels; a judgement of the form $\wfjudge C \GA$ states that $\GA$
is well-formed according to the channel environment $C$ (cf.
\S~\ref{sub:lin});
$\GA$ is \emph{well-formed} if $\wfjudge{\emptyTree}{\GA}$ can be
derived from the rules given in Fig.~\ref{fig:wellformed}.
We assume that each premise of the rules in Fig.~\ref{fig:wellformed}
does not hold if any of the functions used are not defined (e.g., in
\rulename{\wfseq}, if $\GFOUTS{\GA} = \nada$ then $\wfjudge{C}{\GA
  \GWS \GA'}$ is not derivable).
Hereafter, we assume that a node $\nodeTN$ is fresh (i.e.\ $\nodeTN \not\in C$).
The environment $C$ permits to tackle one of the main
requirements for a global type to be well-formed: there should
not be any race on channels.
In the following, we discuss the rules of Fig.~\ref{fig:wellformed},
which are grouped according to three other requirements:
sequentiality, single threaded, and knowledge of choice.

\begin{figure}[t]
  \centering
  \[
  \renewcommand\arraystretch{2}
  \begin{array}{c}
    \wfrule{\wfdot}
    {
      \forall \; \chanT{\_}{s'}{r'} \in \GFIN{\GA} \, : \,
      \{\ptp{s'} , \ptp{r'}\} \cap \{\ptp{s} ,\ptp{r}\} \neq \varnothing
      \\
      \wfjudge{\addchanB{C}{\nodeTN}}{\GA}
      \\
      \labN = \chanTN
    }
    {
      \wfjudge{C}{\GSENDN \GSEP \GA}
    }
    \\
    \wfrule{\wfseq}
    {
      \forall \; \chanT{\_}{s}{r}  \in \GFIN{\GA'} \, . \,
      \exists N_1 \neq N_2 \in \GFOUTS{\GA} \, . \,
      \ptp{s} \in N_1 \land \ptp{r} \in N_2
      \\
      \forall N \in \GFINP{\GA} \, . \,
      \exists N' \in \GFINP{\GA'} \, . \,
      N \cap N' \neq \varnothing
      \\
      \wfjudge{C}{\GA}
      \\
      \wfjudge{\addchan{C}{\chanG{\GA}}}{\GA'}
    }
    {
      \wfjudge{C}{\GA \GWS \GA'}
    }
    \\
    \wfrule{\wfpar}
    {
      \PARTS{\GA} \cap \PARTS{\GA'} = \varnothing
      \\
      \CHANS{\GA} \cap \CHANS{\GA'} = \varnothing
      \\
      \wfjudge{C}{\GA}
      \\
      \wfjudge{C}{\GA'}
    }
    {
      \wfjudge{C}{\GA \GPAR \GA'}
    }
    \\
    \wfrule{\wfrec}
    {
      \GRECV \in \freen{\GA} \Rightarrow
      \# \GFOUTS{\GA} = 1
      \\
      \wfjudge{\addnode{C}{\nodeTN}}{\GA}
      \\
      \labN = \GRECV
    }
    {
      \wfjudge{C}{\GRECN \GA}
    }
    \\
    \wfrule{\wfseqen}
    {\wfjudge{C}{\GA}}
    {\wfjudge{C}{\GA \GWS \GEND}}
    \quad
    \wfrule{\wfvar}
    {\addchan{C}{\subtreeC{\chi}}}
    {
      \wfjudge{C}{\GRECV}
    }
    \quad
    \wfrule{\wfend}
    {}
    {
      \wfjudge{C}{\GEND}
    }
    \\
    \wfrule{\wfchoice}
    {
      \forall \; \chanTN  \in \GFIN{\GA} .
      \forall \; \chanT{b}{s'}{r'} \in \GFIN{\GA'}
      \, . \, \ptp{s} = \ptp{s'}  \land \chan{a} \neq \chan{b}
      \\
      \wfjudge{C}{\GA}
      \\
      \wfjudge{C}{\GA'}
    }
    {
      \wfjudge{C}{\GA \GCH \GA'}
    }
  \end{array}
  \renewcommand\arraystretch{1}
  \]
  \caption{Rules for Well-formedness}
  \label{fig:wellformed}
\end{figure}

\paragraph{Sequentiality~\cite{cdp11}.}
Rules \rulename{\wfdot}, \rulename{\wfseq} and \rulename{\wfseqen}
ensure that sequentiality is preserved.
In \rulename{\wfdot}, there must be an ordering dependency between 
a prefix and its continuation so that it is possible
to implement each participant so that at least one action of the first prefix
always happens before an action of the second prefix.
More concretely, we want to avoid global types of the form, e.g.\
{\small
\[
\GSEND{s_1}{r_1}{a}{\SORTV} \PSEP \GSEND{s_2}{r_2}{b}{\SORTV'}
\quad \badex
\]
}
where, evidently, it is not possible to guarantee that $\ptp{s_2}$ sends
after $\ptp{r_1}$ receives on $\chan{a}$.
Since we are working in an asynchronous setting, we do not want to force both
send and receive actions of the first prefix to happen before both actions of 
the second one.

\noindent
Rule \rulename{\wfseq} requires the following for generalised sequencing.
($i$) For each pair of ``first'' participants in $\GA'$,
there exist two concurrent branches of $\GA$ such that
these two participants appear in different branches.
This is to avoid global types of the form, e.g.\
{%
\[
(\GSEND{s_1}{r_1}{a}{\SORTV} \GPAR \GSEND{s_2}{r_2}{b}{\SORTV})
\GWS
\GSEND{s_1}{r_1}{c}{\SORTV}
\quad \badex
\]
}
since there is no possible sequencing between the prefix on $\chan{b}$
and the one on $\chan{c}$.
($ii$) For all top-level concurrent branches in $\GA$, there is a
participant in that branch which is also in one of the branches of
$\GA'$.  This requirement discards global types of the form, e.g.\
{%
\[
(
\GSEND{s_1}{r_1}{a}{\SORTV}
\GPAR \GSEND{s_2}{r_2}{b}{\SORTV}
\GPAR \GSEND{s_3}{r_3}{c}{\SORTV}
)
\GWS
\GSEND{s_1}{r_2}{d}{\SORTV}
\quad \badex
\]
}
since it is not possible to enforce an order between $\ptp{s_3}$ and
$\ptp{r_3}$ and the others.
($iii$) $\GA$ and $\GA'$ are also well-formed.
Observe that ($i$) implies that
for $\GA \GWS \GA'$ to be well-formed, $\GA$ is of the form
$\GA_1 \GPAR \GA_2$, with $\GA_1 \neq \GEND$ and $\GA_2 \neq \GEND$.
Both  \rulename{\wfdot} and \rulename{\wfseq}
are only applicable when linearity is preserved.
Finally, rule \rulename{\wfseqen} is a special case of $\GA \GWS \GA'$.

\paragraph{Single threaded~\cite{hyc08}.}
A participant should not appear in different concurrent branches of a
global type, so that each participant is single threaded.  This is also
reflected in the calculus of \S~\ref{sec:local}, where parallel
composition is only allowed at the system level.
Therefore, in \rulename{\wfpar}, the participant (resp.\ channel) names
in concurrent branches must be disjoints.
Rule \rulename{\wfrec} adds a new node in $C$ to keep track
of recursive usage of the channels, and requires that $\GA$ is single threaded,
i.e.\ concurrent branches cannot appear under recursion.
If that was the case, a participant would appear in different concurrent
branches of the unfolding of a recursive global type.
Rule \rulename{\wfvar} unfolds $C$ at $\GRECV$ to ensure that the one-time unfolding
of $C$ preserves linearity (see~\cite{hyc08} for details).

\paragraph{Knowledge of choice~\cite{hyc08,cdp11}.}
Whenever a global type specifies a choice of two sets of interactions, the
decision should be made by exactly one participant. For instance,
{%
\[
\GSEND{s_1}{r_1}{a_1}{\SORTV} \GSEP \GA_1
\quad \GCH \quad
\GSEND{s_2}{r_2}{a_2}{\SORTV'} \GSEP \GA_2 \quad \badex
\]
}
specifies a choice made by two participants.
Indeed, $\ptp{s_1}$ is the one making a decision in the first branch,
while $\ptp{s_2}$ makes a decision in the second one; this kind of
choreographies cannot be implemented (without using hidden
interactions).
Also, we want to avoid global types where a participant $\ptp{n}$
behaves differently in two choice branches without being aware of the
choice made by others. For instance, in
{%
\[
\GSENDN \GSEP \GSEND{n}{r}{c}{\SORTV} \GSEP \GA_1
\quad \GCH \quad
\GSEND{s}{r}{b}{\SORTV} \GSEP \GSEND{n}{r}{d}{\SORTV} \GSEP \GA_2
\qquad
\badex
\]
}
where $\ptp{n}$ ignores the choice of $\ptp s$ and behaves differently
in each branch.
On the other hand, we want global types of following form to be accepted.
{%
\[
\begin{array}{l}
  \GSEND{s}{r}{a}{\SORTV}\GSEP \GSEND{n}{s}{b}{\SORTV} \GSEP \GSEND{s}{n}{c}{\SORTV} \GSEP
  \GSEND{n}{r}{d}{\SORTV}
  \\
  \multicolumn{1}{c}{\GCH}
  \\
  \GSEND{s}{r}{a'}{\SORTV} \GSEP \GSEND{n}{s}{b}{\SORTV} \GSEP \GSEND{s}{n}{c'}{\SORTV} \GSEP
  \GSEND{n}{r}{d'}{\SORTV}
\end{array}
\quad \goodex
\]
}
Indeed, in this case $\ptp{n}$ behaves differently in each branch, but
only \emph{after} ``being informed'' by $\ptp s$ about the chosen
branch.

Together with the projection map defined below, rule
\rulename{\wfchoice} guarantees that ``knowledge of choice'' is
respected.
In particular, the rule requires that the participant who makes the
decision is the same in every branch of a choice, while the channels
guarding the choice must be distinct.

\begin{definition}[$\Proj{\_}{\_}$]\label{def:proj}
  The \emph{projection of a global type $\GA$ wrt. $\ptp n \in \PARTS
    \GA$} is defined as
\[
\Proj{\GA}{n} \defi
\begin{cases}
  \PRECEIVE{a}{\SORTV} \PSEP \Proj{\GA'}{n},
  & \text{if } \GA = \GSEND{s}{n}{a}{\SORTV} \GSEP \GA'\\
  \PSEND{a}{\SORTV} \PSEP \Proj{\GA'}{n},
  & \text{if } \GA = \GSEND{n}{r}{a}{\SORTV} \GSEP \GA'\\
  \Proj{\GA'}{n},
  & \text{if } \GA = \GSENDN \GSEP \GA' \text{ and } \ptp{s} \neq \ptp{n} \neq \ptp{r}\\
  \Pmerge{\Proj{\GA_1}{n}}{\Proj{\GA_2}{n}},
  &\text{if } \GA = \GA_1 \GCH \GA_2 \\
  \Proj{\GA_i}{n},
  &\text{if } \GA = \GA_1 \GPAR \GA_2 
  \text{ and }
  \ptp{n} \not\in \PARTS{\GA_j},
  i \neq j \in \{1,2\} \\
  \Proj{\GA_1}{n} \subs{\Proj{\GA_2}{n}}{\PEND},
  &\text{if } \GA = \GA_1 \GWS \GA_2 \\
  \mu \, \GRECV . \Proj{\GA'}{n},
  &\text{if } \GA = \GREC{\GRECV} \GA' \\
  \GA,
  &\text{if } \GA = \GRECV \text{ or }  \GA = \GEND\\
  \nada,
  & \text{otherwise}
\end{cases}
\]
We say that a global type is \emph{projectable} if $\Proj{\GA}{n}$ is
defined for all $\ptp{n} \in \PARTS{\GA}$.
\end{definition}
The projection map is similar to the one given in~\cite{hyc08}, but
for the generalised sequencing case and the use of $\Pmerge{\_}{\_}$
to project choice branches.
Observe that if $\GA = \GA_1 \GWS \GA_2$, we replace $\PEND$ by the
projection of $\GA_2$ in the projection of $\GA_1$.
Function $\Pmerge{\_}{\_}$ basically merges (if possible) the
behaviour of a participant in different choice branches;
$\Pmerge \_ \_$ is defined only when
the behaviour is the same in all branches, or if
it differs after having received enough information about
the branch which was chosen.
The definition of $\Pmerge{\_}{\_}$ is given in Appendix~\ref{app:proj}.
A global type may be projected even if is not well-formed, but in that
case none of the properties given below are guaranteed to hold.

\section{Synthesising Global Types}\label{sec:infer}
We now introduce a typing systems to synthesise a global type $\GA$
from a system $\SYSV$ so that $\SYSV$ satisfies safety and progress
properties (e.g.\ no race on channels and no participant gets stuck).
Also, the set of typable systems corresponds exactly to the set of
systems obtained by projecting well-formed global types.
To synthesise $\GA$ from a system $\SYSV$, a careful analysis of what
actions can occur at each possible state of $\SYSV$ is necessary.

If $\SYSV \equiv \PBOX{n}{\PV} \PPAR \SYSV'$ then $\restri{S}{n}$
denotes $ \PV$ (if $\SYSV \not\equiv \PBOX{n}{\PV} \PPAR \SYSV'$ then
$\restri{S}{n} = \nada$).
We define the \emph{ready set} of a system as follows:
\[
\fire{\SYSV} = \left\{
  \begin{array}{ll}
    \{ a_i| i \in I\} \cup \fire{\SYSV'}
    & \text{if } \SYSV \equiv \PBOX{r}{\PSUM{i \in I}{\PRECEIVE{a_i}{\SORTV_i} \PSEP \PV_i }} \PPAR \SYSV'
    \\[0.5pc]
    \{ \coname{a_i} | i \in I\} \cup \fire{\SYSV'}
    & \text{if } \SYSV \equiv
    \PBOX{s}{\PINS{i \in I}{\PSEND{a_i}{\SORTV_i} \PSEP \PV_i}} \PPAR \SYSV'
    \\[0.5pc]
    \{ \coname{a} \} \cup \fire{\SYSV'}
    & \text{if } \SYSV \equiv \QUEUE{a}{\SORTV \cdot \rho} \PPAR \SYSV'
    \\[0.5pc]
    \varnothing 
    & \text{if } \SYSV \equiv \PEND
  \end{array}
\right.
\]
We overload $\fire{\_}$ on behaviours in the obvious way and define
$\firebar{\SYSV} \defi \{ \chan{a} \in \SETCHANS \, | \, \chan{a} \in
\fire{\SYSV} \text{ or } \coname{a} \in \fire{\SYSV} \}$ and
$\onestep{\SYSV} \iff \exists \, \chan{a} \in \SETCHANS \st \chan{a}
\in \fire{\SYSV} \land \coname{a} \in \fire{\SYSV}$; we write
$\noonestep{\SYSV}$ if $\onestep{\SYSV}$ does not hold.

\subsection{Validation Rules}\label{sub:infer}
A judgement of the form $\judgeC{\chanset}{\Gamma}{C}{\SYSV}{\GA}$
says that the system $\SYSV$ forms a choreography defined by a global
type $\GA$, under the environments $\chanset$, $\Gamma$, and $C$.
The environment $\chanset$ is a superset of the channel names used in
$\SYSV$, and corresponds to the channels $\SYSV$ is entitled to use.
The environment $\Gamma$ maps participant names and local recursion
variables to global recursion variables ($\emptyctx$ is the empty
context $\Gamma$).
The channel environment $C$ records the use of channels.
Hereafter, we use $\cdot$ for the disjoint union of environments.

\paragraph{Programs.}
A global type $\GA$ can be synthesised from the \emph{program} $\SYSV$
if the judgement
\[
\judgeC
{\CHANS{\SYSV}}
{\emptyctx}
{\emptyTree}
{\SYSV}{\GA}
\]
(stating that $\SYSV$ is entitled to use all its channels in empty
environments) is derivable from the rules in Fig.~\ref{fig:gloinfer}
(driven by the ready set of $\SYSV$ and the structure of its
processes).

\begin{figure}[t]
  \centering
  \[
  \renewcommand\arraystretch{2.5}
  \begin{array}{c}
    \glorule
    {\gipsep}
    {
      \judgeC{\{\chan{a}\} \cup \chanset}{\Gamma}{\addchanB{C}{\nodeTN}}
      {\PBOX{s}{P} \PPAR \PBOX{r}{Q} \PPAR \SYSV}
      {\GA}
      \\
      \labN = \chanTN
      \\
      \noonestep{\SYSV}
    }
    {
      \judgeC{\{\chan{a}\} \cup \chanset}{\Gamma}{C}
      {\PBOX{s}{\PSEND{a}{\SORTV} \PSEP P} \PPAR \PBOX{r}{\PRECEIVE{a}{\SORTV} \PSEP Q} \PPAR \SYSV}
      {\GSENDN \GSEP \GA}
    }
    \\
    \glorule
    {\gipar}
    {
      \judgeC{\chanset_1}{\emptyctx}{C}
      {\SYSV}{\GA}\\
      \judgeC{\chanset_2}{\emptyctx}{C}
      {\SYSV'}{\GA'}
      \\
      \chanset_1 \cap \chanset_2 = \varnothing
    }
    {
      \judgeC{\chanset_1 \cup \chanset_2}{\Gamma}{C}
      {\SYSV \PPAR \SYSV'}{\GA \GPAR \GA'}
    }
    \\
     \glorule
    {\gigws}
    {
      \judge{\emptyctx}{C}{\SYSV_1}{\GA_1}
      \\
      \Ssplit{\SYSV} = (\SYSV_1 , \SYSV_2)
      \\
      \judge{\emptyctx}{\addchan{C}{\chanG{\GA_1}}}{\SYSV_2}{\GA_2}
    }
    {
      \judge{\Gamma}{C}{\SYSV}{\GA_1 \GWS \GA_2}
    }

    \\
    \glorule
    {\gipinch}
    {
      \judge{\Gamma}{C}
      {\PBOX{s}{P} \PPAR \SYSV}{\GA}\\
      \judge{\Gamma}{C}
      {\PBOX{s}{Q} \PPAR \SYSV}{\GA'} \\
      \noonestep{\SYSV}
    }
    {
      \judge{\Gamma}{C}
      {\PBOX{s}{P \PINCH Q} \PPAR \SYSV}{\GA \GCH \GA'}\\
    }
    \\

    \glorule
    {\gipoutchl}
    {
      \firebar{Q} \subseteq \chanset
      \\
      \judge{\Gamma}{C}
      {\PBOX{r}{P} \PPAR \SYSV}{\GA}
      \\
      \noonestep{\SYSV}
    }
    {
      \judge{\Gamma}{C}
      {\PBOX{r}{P \POUTCH Q} \PPAR \SYSV}{\GA}
    }

    \\
    
    \glorule
    {\gimua}
    {
      \exists \, 1 \leq i,j \leq k \, . \, \onestep{(\PBOX{n_i}{\PV_i} \PPAR \PBOX{n_j}{\PV_j})}
      \\
      \judge{\Gamma \cdot (\ptp{n}_1,\PRECV{X}_1) : \GRECV, \ldots , (\ptp{n}_k,\PRECV{x}_k) : \GRECV}
      {\addnode{C}{\mu\GRECV}}
      { \PBOX{n_1}{\PV_1 }
        \PPAR \! \ldots \! \PPAR
        \PBOX{n_k}{\PV_k}
      }{\GA}
    }
    {\judge{\Gamma}{C}{
        \PBOX{n_1}{\PRECi{X}{1} \PV_1 }
        \PPAR \! \ldots \! \PPAR
        \PBOX{n_k}{\PRECi{X}{k} \PV_k}
        }{\GRECN \GA }} 

    \\
    \glorule
    {\givar}
    {
      \forall 1 \leq i \leq k \, . \, 
      \Gamma(\ptp{n_i},\PRECV{x}_i) = \GRECV
      \quad
      \addchan{C}{\subtreeC{\mu\GRECV}}
    }
    {
      \judge{\Gamma}{C}{
        \PBOX{n_1}{\PRECV x _1}
        \PPAR \ldots \PPAR
        \PBOX{n_k}{\PRECV x _k}
      }{\GRECV}
    }

    \\
    \glorule
    {\gieq}
    {
      S \equiv S' \\
      \judge{\Gamma}{C}{S'}{\GA}
    }
    {
      \judge{\Gamma}{C}{S}{\GA}
    }
    \qquad
    \glorule
    {\giend}
    {
      \forall \ptp{n} \in \PARTS{\SYSV} \, . \, \restri{\SYSV}{n} =\PEND
      \quad
      \CHANS{\SYSV} = \varnothing
    }
    { \judge{\Gamma}{C}{\SYSV}{\GEND}}
  \end{array}
  \renewcommand\arraystretch{1}
  \]
  \caption{Validation Rules for Programs}
  \label{fig:gloinfer}
\end{figure}

Rule \rulename{$\gipsep$} validates prefixes provided that the system
is entitled to linearly use the channel $\chan{a}$, that the
continuation is typable, and that no other interactions are possible
in $\SYSV$. For instance, \rulename{$\gipsep$} does not apply to
\[
\PBOX{s_1}{\PSEND{a}{\SORTV} \PSEP \PV_1} \PPAR
\PBOX{r_1}{\PRECEIVE{a}{\SORTV} \PSEP Q_1} \PPAR
\PBOX{s_2}{\PSEND{b}{\SORTV} \PSEP \PV_2} \PPAR
\PBOX{r_2}{\PRECEIVE{b}{\SORTV} \PSEP Q_2}
\quad \badex
\]
because there is no ordering relation between the actions on
$\chan{a}$ and $\chan{b}$; in this case either \rulename{$\gipar$} or
\rulename{$\gigws$} should be used.

\noindent
Rule \rulename{$\gipar$} validates concurrent branches when they can
be validated using a partition $\chanset_1$ and $\chanset_2$ of the
channels (recall that $\PARTS{\SYSV} \cap \PARTS{\SYSV'} =
\varnothing$).

\noindent
Rule \rulename{$\gigws$} splits the system into two sequential parts
and it relies on the function $\Ssplit{\_}$ defined in
\S~\ref{sub:split}; for now it suffices to notice that linearity is
checked for in the second part of the split by adding the channel
environment corresponding to $\GA_1$ to $C$ (recall that
$\addchan{C}{C'}$ is undefined if $\addnode{C}{C'}$ is not linear).

\noindent
Rule \rulename{$\gipinch$} introduces the global type choice operator,
it requires that both branches are typable and that no other
interactions are possible in $\SYSV$.

\noindent
Rule \rulename{$\gipoutch$} allows to discharge a branch of an
external choice; together with the premises of \rulename{$\gipar$},
rule \rulename{$\gipoutch$} discards systems such as the one on left
below (due to a race on $\chan{b}$) but permits those like the one on
the right (as only the channels guarding the choice must be in
$\chanset$).
\[
\PBOX{r_1}{\PRECEIVE{a}{\SORTV} \POUTCH \PRECEIVE{b}{\SORTV}} \PPAR
\PBOX{s_2}{\PSEND{b}{\SORTV} } \PPAR
\PBOX{r_2}{\PRECEIVE{b}{\SORTV}}
\; \badex
\;\;
\PBOX{s_1}{\PSEND{a}{\SORTV}} \PPAR
\PBOX{r_1}{\PRECEIVE{a}{\SORTV} \POUTCH \PRECEIVE{c}{\SORTV} \PSEP \PRECEIVE{b}{\SORTV}} \PPAR
\PBOX{s_2}{\PSEND{b}{\SORTV} } \PPAR
\PBOX{r_2}{\PRECEIVE{b}{\SORTV}}
\; \goodex
\]

\noindent
Rules \rulename{$\gimua$} and \rulename{$\givar$} handle recursive
systems.
The former rule ``guesses'' the participants involved in a recursive
behaviour.  If two of them interact, \rulename{$\gimua$} validates the
recursion provided that the system can be typed when such participants
are associated to the global recursion variable $\GRECV$ (assuming
that $\GRECV$ is not in $\Gamma$).
Rule \rulename{$\givar$} checks that all the participants in the
recursion have reached a local recursion variable corresponding to the
global recursion, and that the unfolding of $C$ on $\mu \GRECV$
preserves linearity; for this we define $\subtreeC{\mu \GRECV}$ to be
the subtree of $C$ rooted at the \emph{deepest} node of $C$ labelled
by $\mu \GRECV$ (note that this node is unique since bound variables
are distinct).

\noindent
Rule \rulename{$\giend$} only applies when all the participants in
$\SYSV$ end while \rulename{$\gieq$} validates a system up to
structural congruence.

\begin{theorem}[Decidability]\label{thm:decide}\label{lem:recboth}\label{thm:decidable}
\thmdecide
\end{theorem}
The proofs follows from the fact that the typing is done wrt to the (finite)
partitions of channels in a system, and that the number of required behaviour
unfoldings is finite.

\begin{theorem}[Unique typing]\label{thm:RUNIQUE}
\thmunique
\end{theorem}
\begin{theorem}[Well-formedness]\label{thm:RWF}\label{lem:wellform}
\thmwf
\end{theorem}
The proofs for these two theorems are by induction on the structure of the derivation.

\paragraph{Runtime system.}
In order to have subject reduction for our typing systems, queues have
to be handled effectively; we use a distinguished participant name
$\GSTAR$ to denote an anonymous participant.
Assume $\GSTAR \not\in \SETPARTS$ and
write $ \GSEND{\GSTAR}{r}{a}{\SORTV} \GSEP \GA $ to specify that there
is a message of sort $\SORTV$ on channel $\chan{a}$ for participant $\ptp{r}$.
\begin{myex}
  Let $S = \PBOX{n}{\PSEND{a}{\SORTV}} \PPAR
  \PBOX{s}{\PSEND{b}{\SORTV} \PSEP \PRECEIVE{a}{\SORTV}} \PPAR
  \PBOX{r}{\PRECEIVE{b}{\SORTV}} \PPAR \QUEUE{a}{\emptyQ} \PPAR
  \QUEUE{b}{\emptyQ}$.
  Consider the judgement
  \[
  \judge{\Gamma}{C} S
  {
    \GSEND{s}{r}{b}{\SORTV} \GSEP
    \GSEND{n}{s}{a}{\SORTV}
  }
  \]
  If $S$ evolves to $S' = \PBOX{n}{\PEND} \PPAR
  \PBOX{s}{\PSEND{b}{\SORTV} \PSEP \PRECEIVE{a}{\SORTV}} \PPAR
  \PBOX{r}{\PRECEIVE{b}{\SORTV}} \PPAR \QUEUE{a}{\SORTV} \PPAR
  \QUEUE{b}{\emptyQ}$, the identity of the sender $\ptp n$ is lost.
  However, the judgement
  \[
  \judge{\Gamma}{C}{S'}
  {
    \GSEND{s}{r}{b}{\SORTV} \GSEP
    \GSEND{\GSTAR}{s}{a}{\SORTV}
  }
  \]  
  types $S'$ using $\GSTAR$.
\end{myex}

Runtime systems can be handled by slightly extending
Def.~\ref{def:linrelations} so that we have\footnote{This extension
  makes sense since the order of messages is preserved in the
  calculus.}
\[
\prefg_1 \linOO \prefg_2
 \text{ if } \prefg_1 \linprev \prefg_2 \text{ and }
\prefg_1 = \chanT{a}{\GSTAR}{r} \text{ and }
\prefg_2 = \chanT{a}{s}{r}
\]
and by adding two rules to the validation rules for handling queues:
\[
{\scriptsize
\begin{array}{c}
   \glorule
    {\giqueue}
    {
      \judgeC{\{\chan{a}\} \cup \chanset }{\emptyctx}{\addchan{C}{\nodeTN}}
      {
        \QUEUE{a}{\rho} \PPAR
        \PBOX{r}{\PV} \PPAR
        \SYSV
      }
      {\GA}
      \\
      \labN = \chanT{a}{\GSTAR}{r}
      \\
      \noonestep{\SYSV}
    }
    {
      \judgeC{\{\chan{a}\} \cup \chanset }{\Gamma}{C}
      {
        \QUEUE{a}{\SORTV \cdot \rho} \PPAR
        \PBOX{r}{\PRECEIVE{a}{\SORTV} \PSEP \PV} \PPAR
        \SYSV
      }
      {\GSEND{\GSTAR}{r}{a}{\SORTV} \GSEP \GA}
    }
\qquad%
 \glorule
    {\emptyQ}
    {
     \judge{\Gamma}{C}
      {
        \SYSV
      }
      {\GA}
    }
    {
      \judge{\Gamma}{C}
      {
        \QUEUE{a}{\emptyQ} \PPAR
        \SYSV
      }
      {\GA}
    }
  \end{array}
}
\]
Rule \rulename{$\giqueue$} is similar to rule \rulename{$\gipsep$}, except that a non-empty 
queue replaces the sender, and $\Gamma$ is emptied.
Rule \rulename{$\emptyQ$} simply allows to remove empty queues from the system.
\begin{theorem}\label{thm:RSUBRED}
\thmsubred
\end{theorem}
The proof is by case analysis
on the different types of transitions a system can make. The recursive
case follows from the fact that reduction preserves closeness of behaviours.
\subsection{Splitting Systems}\label{sub:split}
The purpose of systems' splitting is to group participants according
to their interactions.
For this we use judgements of the form
\begin{equation} \label{eq:splitjdg}
  \spjudge{\Psi}{\Theta}{\SYSV}{\Omega}  
\end{equation}
which reads as ``$\SYSV$ splits as $\Omega$ under $\Psi$ and $\Theta$''.
The environment $\Psi$ is a set of (pairwise disjoint)
\emph{ensembles} that is disjoint sets $N \subseteq \PARTS \SYSV$
containing participants that interact with each other for a while; and
then some of them may interact with participants in other ensembles in
$\Psi$.
The environment $\Theta$ is a set of (pairwise disjoint) \emph{duos},
that is two-element sets of participants $\{\ptp s, \ptp r \in \PARTS
\SYSV \st \ptp r \neq \ptp s \}$ representing the first participants able
to interact once the first part of the split is finished.
Under suitable conditions, one could identify when $\ptp n \in N$ has
to interact with a participant of another ensemble.
In other words, one can divide $\restri{\SYSV}{n}$ as $\PBOX n {\PV_1
  \cdot \epsilon \cdot \PV_2}$: the interactions in $\PV_1$ happen
with participants in the ensemble of $\ptp n$, while $\PV_2$ starts
interacting with a participant in another ensemble.
Finally, the environment $\Omega$ assigns behaviours augmented with a
separator $\epsilon$ to participant names, and lists of sorts to queues
$\chan{a}$.

Given a judgement as~\eqref{eq:splitjdg}, we say that $N, M \in \Psi$
are \emph{$\Theta$-linked} ($N \lnkT M$ in symbols) iff $\exists D \in
\Theta \st N \cap D \cap M \neq \emptyset$; also, we say that $\ptp n,
\ptp{m} \in \PARTS \SYSV$ are \emph{$\Omega$-linked} ($\ptp n \lnkO \ptp m$
in symbols) iff $\CHANS{\Omega(\ptp n)} \cap \CHANS{\Omega(\ptp m)}
\neq \emptyset$.
We define
$\restriset{\SYSV}{N} \defi
\PPROD{\ptp{n} \in N}{\PBOX{n}{\restri{\SYSV}{n}}}
\PPAR \PPROD{\chan{a}{ \in \CHANS{\SYSV}}}{\QUEUE{a}{\restric{\SYSV}{a}}}
$.
\begin{definition}\label{def:coherent}
  The judgement $ \spjudge{\Psi}{\Theta}{\SYSV}{\Omega} $ is
  \emph{coherent} if it can be derived from the rules in
  Fig.~\ref{fig:split}, $\Theta \neq \varnothing$, and for all $N \in
  \Psi$, $\onestep{\restriset \SYSV N}$ and the following
  conditions hold
    \begin{equation} 
      \exists ! \ptp n \in N \st \big(
    (\exists ! \ptp m \in N \setminus \{ \ptp n \} \st
    \noonestep{\restriset{\SYSV}{N \setminus \{\ptp{n}\}}}
    \; \wedge \;
    \noonestep{\restriset{\SYSV}{N \setminus \{\ptp{m}\}}})
    \; \text{or} \;
    (\noonestep{\restriset{\SYSV}{N \setminus \{\ptp{n}\}}})
    \big)
    \label{eq:cohecond0}
  \end{equation}
  \begin{equation} 
    \spR \text{ is total on } N
    \; \text{and} \; 
    \sprelT \text{ is total on }\Psi
    \label{eq:cohecond}
  \end{equation}
  where $\sprelT \defi \lnkT^*$ is the reflexive and transitive
  closure of $\lnkT$ and $\spR \defi \lnkO^+$ is the transitive
  closure of $\lnkO$.
\end{definition}

Essentially, Def.~\ref{def:coherent} ensures that rule
\rulename{$\gigws$} is the only rule of Fig.~\ref{fig:gloinfer}
applicable when the system can be split. %
Condition~\eqref{eq:cohecond0} ensures that, in each ensemble $N$,
there is a unique pair of synchronising participants or there is a
unique participant that can synchronise with a queue $a$.
Condition~\eqref{eq:cohecond} is the local counterpart of the
well-formedness rule for global types of the form $\GA \GWS \GA'$.
The totality of $\spR$ on $N$ guarantees that the participants in an
ensemble share channels.
The totality of $\sprelT$ on $\Psi$ guarantees that each ensemble in
$\Psi$ has one ``representative'' which is one of the first
participants to interact in the second part of the split.  Together
with condition $\Theta \neq \varnothing$, the condition on $\sprelT$
ensures that there are (at least) two ensembles of participants in
$\Psi$.
Note that~\eqref{eq:cohecond} also ensures that all the set of
participants in $\Psi$ are interdependent (i.e.\ one cannot divide
them into independent systems, in which case rule \rulename{$\gipar$}
should be used).
\begin{figure}[t]
  \centering
  \[
  \renewcommand\arraystretch{2}
  \begin{array}{c}

\glorule
{\splepsilon}
{
  \ptp{n} \in N, \ptp{m} \in M
  \\
  \onestep{(\PBOX{n}{P} \PPAR \PBOX{m}{Q})}
  \\
  \spjudge{\Psi \cdot N \spminus{n} \cdot M \spminus{m}}
  {\Theta}
  {
    \SYSV
  }
  {
    \Omega
  }
}
{
  \spjudge{\Psi \cdot N \cdot M}
  {\Theta \cdot \{ \ptp{n,m} \} }
  {
    \PBOX{n}{P} \PPAR \PBOX{m}{Q} \PPAR \SYSV
  }
  {
    \Omega \cdot \prtrack{n}{\epsilon}
    \cdot \prtrack{m}{\epsilon}
  }
}

\\ %

\glorule
{\splsync}
{
  \ptp{s},\ptp{r} \in N
  \\
  \spjudge{\Psi \cdot N}
  {\Theta}
  {
    \PBOX{s}{\PV} \PPAR \PBOX{r}{Q} \PPAR \SYSV
  }
  {
    \Omega \cdot \prtrack{s}{\spP} \cdot \prtrack{r}{\spQ}
  } 
}
{
  \spjudge{\Psi \cdot N}
  {\Theta}
  {
    \PBOX{s}{\PSEND{a}{\SORTV} \PSEP \PV} \PPAR 
    \PBOX{r}{\PRECEIVE{a}{\SORTV} \PSEP Q} \PPAR 
    \SYSV
  }
  {
    \Omega \cdot \prtrack{s}{\PSEND{a}{\SORTV} \PSEP \spP}
    \cdot \prtrack{r}{\PRECEIVE{a}{\SORTV} \PSEP\spQ}
  }
}

\\ %

\glorule
{\splplusl}
{
  \ptp{m},\ptp{n} \in N
  \\
  \onestep{(\PBOX{m}{\PV} \PPAR \PBOX{n}{Q})}
  \\
  \spjudge{\Psi \cdot N}
  {\Theta}
  {
    \PBOX{m}{\PV} \PPAR \PBOX{n}{Q} \PPAR \SYSV
  }
  {
    \Omega \cdot \prtrack{m}{\spP}
  } 
}
{
  \spjudge{\Psi \cdot N}
  {\Theta}
  {
    \PBOX{m}{\PV \POUTCH \PV'}  \PPAR \PBOX{n}{Q} \PPAR \SYSV
  }
  {
    {\Omega \cdot \prtrack{m}{\spP}}%
  }
}

\\ %

\glorule
{\splpinch}
{
  \ptp{n},\ptp{m} \in N
  \\
  \onestep{(\PBOX{n}{\PV \PINCH \PV'} \PPAR \PBOX{m}{Q})}
  \\
  \spcomp{\Omega}{\Omega'}
  \\\\
  \spjudge{\Psi \cdot N}
  {\Theta}
  {
    \PBOX{n}{\PV} \PPAR \PBOX{m}{Q} \PPAR \SYSV
  }
  {
    \Omega \cdot \prtrack{n}{\spP}
  } 
  \\
  \spjudge{\Psi \cdot N}
  {\Theta}
  {
    \PBOX{n}{\PV'}  \PPAR \PBOX{m}{Q} \PPAR \SYSV
  }
  {
    \Omega' \cdot \prtrack{n}{\spQ}
  } 
}
{
  \spjudge{\Psi \cdot N}
  {\Theta}
  {
    \PBOX{n}{\PV \PINCH \PV'}  \PPAR \PBOX{m}{Q} \PPAR \SYSV
  }
  {
    \spmerge{\Omega}{\Omega'} \cdot \prtrack{n}{\spP \PINCH \spQ}
  }
}

\\ %

\glorule
{\splax}
{
}
{
\spjudge{\{\varnothing\}}{\varnothing}{\PEND}{\varnothing}
}

\qquad \qquad

\glorule
{\splend}
{
\spjudge{\Psi \setminus \ptp n}{\Theta}{\SYSV}{\Omega}
}
{
\spjudge{\Psi}{\Theta}{\PBOX{n}{\PEND} \PPAR \SYSV}{\Omega \cdot \prtrack{n}{\PEND}}
}

\\ %

\glorule
{\splrem}
{
  \noonestep{(\PBOX{n}{\PV} \PPAR \SYSV)}
  \\
  \PV \not\equiv \PEND
  \\
  \spjudge{\Psi \setminus \ptp n}
  {\Theta}
  {\SYSV}
  {\Omega}
}
{
  \spjudge{\Psi}
  {\Theta}
  {\PBOX{n}{\PV} \PPAR \SYSV}
  {\Omega \cdot \prtrack{n}{\epsilon}}
}
\\
\glorule
{\splqueue}
{
  \ptp{r} \in N
  \\
  \spjudge{\Psi \cdot N}
  {\Theta}
  {
    \PBOX{r}{\PV} \PPAR \SYSV
  }
  {
    \Omega \cdot \prtrack{r}{\spP}
    \cdot \prtrack{a}{\rho}
  } 
}
{
  \spjudge{\Psi \cdot N}
  {\Theta}
  {
    \PBOX{r}{\PRECEIVE{a}{\SORTV} \PSEP \PV} \PPAR 
    \QUEUE{a}{\SORTV \cdot \rho} \PPAR
    \SYSV
  }
  {
    \Omega
    \cdot \prtrack{r}{\PRECEIVE{a}{\SORTV} \PSEP\spP}
    \cdot \prtrack{a}{\SORTV \cdot \rho}
  }
}
\end{array}
\renewcommand\arraystretch{1}
\]
 \caption{Splitting Systems.}
  \label{fig:split}
\end{figure}

A judgement~\eqref{eq:splitjdg} is to be derived with the
rules of Fig.\ref{fig:split} (we omit rules for commutativity and
associativity of systems).
The derivation is driven by the structure of up to two processes in
$\SYSV$, and whether they are in the same ensemble and/or form a duo.

Rule \rulename{$\splepsilon$} marks two processes $\ptp m$ and $\ptp
n$ as ``to be split'' when $\ptp m$ and $\ptp n$ form a duo in
$\Theta$ and are in different ensembles of $\Psi$.
The continuation of the system is to be split as well, with $\ptp m$
and $\ptp n$ removed from the system and from the environments.

\noindent
Rule \rulename{$\splsync$} records in $\Omega$ the interactions of
participants in a same ensemble of $\Psi$.

\noindent
Rule \rulename{$\splplus$} discharges the branch of an external choice
for participants in a same ensemble while \rulename{$\splpinch$} deals
with internal choice.
The premise $\spcomp{\Omega}{\Omega'}$ holds only when $\Omega$ and $\Omega'$ 
have the same domain and differ only up to external choice, i.e.\ for each
$\ptp{n}$ either its split is the same in both branches, or its split is
an external choice (guarded by different channels);
$\spmerge{\Omega}{\Omega'}$ merges $\Omega$ and $\Omega'$ accordingly
(cf.\ Appendix~\ref{app:split}).
The additional premise $\onestep{ \PBOX{s}{\PV \PINCH \PV'} \PPAR
  \PBOX{r}{Q}}$ is required so that the split is done \emph{before} a
branching if a participant cannot interact with one of its peer in $N$
after the branching.

\noindent
Rule \rulename{$\splax$} terminates a derivation (all environments
emptied) while \rulename{$\splend$} completes the split of a process
(abusing notation, $\Psi \setminus \ptp{n}$ denotes the removal of
$\ptp{n}$ from any $N \in \Psi$).

\noindent
Rule \rulename{$\splrem$} marks a process to be split when it cannot
interact with anyone in $\SYSV$. The premise $\PV \not \equiv \PEND$
allows to differentiates a process which terminates after the split,
from others which terminate before.  In the latter case, rule
\rulename{$\splend$} is to be used.

\noindent
Rule \rulename{$\splqueue$} records in $\Omega$ interactions with
non-empty queues.

We now define a (partial) function $\SsplitE$ which splits a system
into two parts.
\begin{definition}[$\Ssplit{\_}$]\label{def:Ssplit}
  Let $\spjudge{\Psi}{\Theta}{\SYSV}{\Omega}$ be a coherent judgement.
  Define $\Ssplit{\SYSV} = (\SYSV_1 , \SYSV_2)$ where
  \begin{itemize}
  \item $\forall \ptp{n} \in \PARTS{\SYSV} . \,
    \restri{\SYSV_1}{n} = \sppre{\restri{\SYSV}{n}}{\Omegan{n}}$
    and
    $\restri{\SYSV_2}{n} = \fsplit{\restri{\SYSV}{n}}{\Omegan{n}}$
  \item $\forall \chan{a} \in \CHANS{\SYSV} . \,
    \restric{\SYSV_1}{a} =  \Omegac{a}$
    and 
    $\restric{\SYSV_2}{a} = \restric{\SYSV}{a} \setminus \Omegac{a}$
  \end{itemize}
  if $\fsplit{\restri{\SYSV}{n}}{\Omegan{n}} \neq \nada$ for all $\ptp
  n \in \PARTS{\SYSV}$, and $\Ssplit{\SYSV} = \nada$ otherwise.
\end{definition}
The auxiliay funtions $\sppre{\_}{\_}$ and $\fsplit{\_}{\_}$ used in
Def.~\ref{def:Ssplit} are defined in Appendix~\ref{app:split}; we give
here their intuitive description.
Let $\ptp{n} \in \PARTS{\SYSV}$,
and $\spjudge{\Psi}{\Theta}{\SYSV}{\Omega}$ be  a coherent judgement.
Function $\sppre{\restri{\SYSV}{n}}{\Omegan{n}}$ returns the ``first
part'' of the split of $\ptp n$, that is the longest common prefix of
$\restri{\SYSV}{n}$ and $\Omegan{n}$, while
$\fsplit{\restri{\SYSV}{n}}{\Omegan{n}}$ is partial and returns the
the remaining part of the behaviour of $\restri{\SYSV}{n}$ after
$\Omegan{n}$.
\begin{myex}
  Taking $\sex$ as in \S~\ref{sec:intro}, we have $\Ssplit{\sex} =
  (\SYSV_1 , \SYSV_2)$ such that
  \[
  \begin{array}{llc@{\qquad}lll}
    \restri{\SYSV_1}{\myba} & = &
    \PSEND{t_1}{\myorder} \PSEP
    \PRECEIVE{p_1}{\myprice} 
    &
    \restri{\SYSV_2}{\myba} & = &
    \PRECEIVE{r}{\myprice} \PSEP(
    \PSEND{c_1}{\nosort} \PSEP \PSEND{t_1}{\myaddress}
    \PINCH
    \PSEND{c_2}{\nosort} \PSEP \PSEND{no_1}{\nosort})
    \\
    \restri{\SYSV_1}{\mysi{i}} & =  &  
    \PRECEIVE{t_i}{\myorder} \PSEP
    \PSEND{p_i}{\myprice}
    & 
    \restri{\SYSV_2}{\mysi{i}} & =  &  
    \PRECEIVE{t_i}{\myaddress} \POUTCH \PRECEIVE{no_i}{\nosort} \\
  \end{array}
  \]
  Note that $\spjudge{\left\{ \{\ptp{\myba},\ptp{\mysi{1}} \},
      \{\ptp{\mybb},\ptp{\mysi{2}} \} \right\}}{\{ \{
    \ptp{\myba},\ptp{\mybb} \} \}}{\sex}{\Omega}$ is coherent.
\end{myex}

\subsection{Properties of Synthesised Global Type}\label{sec:result}

\paragraph{Progress and safety.}
If a system is typable, then it will either terminate
or be able to make further transitions
(e.g.\ if there are recursive processes).
\begin{theorem}\label{thm:RPROGR}
\thmprogress
\end{theorem}
Let us add the rule \rulename{error} below to the semantics given in
\S~\ref{sec:local}.
\[
\begin{array}{c}
\ltsrule{error}
{\SYSV \ltsarrow{\PRECEIVE{a}{e'}} \SYSV' \\ T \ltsarrow{\labpop} T'}
{\SYSV \PPAR T \ltsarrow{} \PERROR}
\end{array}
\qquad
\SORTV \neq \mathtt{e'}
\]
\begin{theorem}\label{thm:RSAFE}
\thmsafety
\end{theorem}
The proofs of Theorems~\ref{thm:RPROGR} and~\ref{thm:RSAFE}
are by contradiction, using Theorem~\ref{thm:RSUBRED}.

\paragraph{Behavioural equivalences.}
We show that there is a correspondence between the original system and the
projections of its global type. First, let us
introduce two relations.
\begin{DEF}[$\sim$ and $\bisim$]\label{def:bisim}
  $\PV \sim Q$ if and only if $Q \ltsarrow{\alpha} Q'$ then there is
  $\PV'$ such that $\PV \ltsarrow{\alpha} \PV'$.
  Also, $\SYSV \bisim T$ iff
  whenever $\SYSV \ltsarrow{\alpha} \SYSV'$
  then $T \ltsarrow{\alpha} T'$ and $\SYSV' \bisim T'$; and
  whenever $T \ltsarrow{\alpha} T'$
  then $\SYSV \ltsarrow{\alpha} \SYSV'$ and $\SYSV' \bisim T'$
  where $\alpha \in \{ \labsyns, \labsynr, \labend \}$.
\end{DEF}
The behaviour of a participant in $\SYSV$ is a simulation
of the projection of a synthesised global type from $\SYSV$ onto this participant.
Intuitively, the other direction
is lost due to rule \rulename{$\gipoutch$}, indeed
external choice branches which are never chosen are not ``recorded''
in the synthesised global type.
\begin{lemma}\label{lem:RSIMU}\label{lem:simu}
\thmsimu
\end{lemma}
The proof is by case analysis on the transitions of $\SYSV$, using
Theorem~\ref{thm:RSUBRED}.

Since the branches that are not recorded in a synthesised global type
are only those which are never chosen, we have the following result.
\begin{theorem}\label{thm:RBISIMU}\label{thm:bissim}
\thmbisim
\end{theorem}
The proof is by case analysis on the transitions of $\SYSV$, using
Theorem~\ref{thm:RSUBRED} and Lemma~\ref{lem:simu}.

Our completeness result shows that every well-formed and projectable
global type is inhabited by the system consisting of the parallel
composition of all its projections.
\begin{theorem}\label{thm:RCOMPL}
\thmcomplete
\end{theorem}
The proof is by induction on the structure of (well-formed) $\GA$.

\section{Concluding Remarks and Related Work}\label{sec:conc}
We presented a typing system that, under some conditions, permits to
synthesise a choreography (represented as global type) from a set of
end-point types (represented as local types).
The synthesised global type is unique and well-formed; moreover, its
projections are equivalent to the original local session types.
We have shown safety and progress properties for the local session
types.
Finally, the derivatives of local types which form a choreography can
also be assigned a global type (subject reduction).

\paragraph{Related work.}
A \emph{bottom-up} approach to build choreographies is studied
in~\cite{myh09}; this work relies on global and local types, but uses
\emph{local and global graphs}.
A local graph is similar to a local type while a global graph is a
disjoint union of family of local graphs.
We contend that global types are more suitable than global graphs to
represent choregraphies; in fact, differently from the approach
in~\cite{myh09}, our work allows us to reuse most of the theories and
techniques based on multiparty global types.

Our work lies on the boundary between theories based on \emph{global
types} (e.g.~\cite{hyc08,dy11,cdp11,bhty10}) and the ones based on
the \emph{conversation types}~\cite{cv09}.
Our work  relies on the formalism of global types, but uses
it the other way around. We start from local types and construct
a global type.
We have discussed the key elements of the global types in \S~\ref{sec:global}.

Conversation types~\cite{cv09} abandon global views of distributed
interactions in favour of a more flexible type structure allowing
participants to dynamically join and leave sessions.
The approach in~\cite{cp09} fills the gap between the theories
based on session types and those based on behavioural
contracts~\cite{cclp06} (where the behaviour of a program is
approximated by some term in a process algebra).
We are also inspired from~\cite{p12}, where session types are viewed
as CCS-like ``projections'' of process behaviours. The approach of
considering local types as processes is similar to ours.  However, the
theory of~\cite{p12} is based on a testing approach.
The \emph{connectedness} conditions for a choreography given
in~\cite{blz08} is similar to our notion of \emph{well-formed} global
type.

\paragraph{Future work.}
We aim to extend the framework so that global types can be constructed
from session types which features name passing and restriction.
We also plan to refine the theory and use it in a methodology
so that if a choreography cannot be synthesised, the designers
are given indications on why this has failed.
Finally, we are considering implementing an algorithm from the rules
of Fig.~\ref{fig:gloinfer} and Fig.~\ref{fig:split}, and integrate it
in an existing tool~\cite{lt10} implementing the framework
from~\cite{bhty10}.

{
  \bibliographystyle{abbrv}
  \bibliography{julien}
}

\appendix
\newpage 
\section{Additional Definitions}
In this section, we give the definitions of the accessory functions
used in the main sections of the paper.

\subsection{Linearity}\label{app:linearity}

\begin{definition}[$\chanG{\_}$]\label{def:chang}
  \[
  {\setlength\arraycolsep{1em}
    \begin{array}{c}
      \begin{array}{ll}
    \chanG{\GSENDN \GSEP \GA} = \onekidR{\chanTN}{\chanG{\GA}}
    &
    \chanG{\GRECN \GA} = \onekidR{\mu \GRECV}{\chanG{\GA}}
    \\
    \chanG{\GA \GCH \GA'} = \twokidsR{\emptyLab}{\chanG{\GA}}{\chanG{\GA'}}
    &
    \chanG{\GA \GPAR \GA'} = \twokidsR{\emptyLab}{\chanG{\GA}}{\chanG{\GA'}}
    \\[3pc]
    \chanG{\GA \GWS \GA'} = \addnode{\chanG{\GA}}{\chanG{\GA'}}
    &
    \chanG{\GEND} = \chanG{\GRECV} = \singleN{\emptyTree}
  \end{array}
\end{array}
}
\]
\end{definition}

The function $\chanG{\_}$ returns a channel environment corresponding to a global type.

\subsection{Projections}\label{app:proj}

\begin{definition}[$\Pmerge{\_}{\_}$]\label{def:merge}
\[
\Pmerge{\PV}{Q} =
\begin{cases}
  \PV \POUTCH Q,
  & \text{if } \PV = \PSUM{i \in I}{\PRECEIVE{a_i}{\SORTV_i} \PSEP \PV'_i}
  \text{ and } Q = \PSUM{j \in J}{\PRECEIVE{a_j}{\SORTV_i} \PSEP Q'_j} \\
  & \text{ and } \forall
  i \in I . \, \forall j \in J . \, \chan{a_i} \neq \chan{a_j}
  \text{ and } I, J \neq \varnothing
  \\
  \PV \PINCH Q,
  & \text{if } \PV = \PINS{i \in I}{\PSEND{a_i}{\SORTV_i} \PSEP \PV'_i}
  \text{ and } Q = \PINS{j \in J}{\PSEND{a_j}{\SORTV_i} \PSEP Q'_j}\\
  & \text{ and } \forall
  i \in I . \, \forall j \in J . \, \chan{a_i} \neq \chan{a_j}
  \text{ and } I, J \neq \varnothing
  \\
  \PSERE{a}{\SORTV} \PSEP (\Pmerge{\PV'}{Q'}),
  & \text{if } \PV = \PSERE{a}{\SORTV} \PSEP \PV' \text{ and }
  Q = \PSERE{a}{\SORTV} \PSEP Q'
  \quad \intbang \in \{!, ?\}
  \\
  \PV,
  & \text{if } \PV \equiv Q
  \\
  \nada,
  & \text{otherwise}
\end{cases}
\]
\end{definition}
The function merges the behaviour of a participant in different choice branches.
In the first two cases, it merges two guarded internal (resp.\ external)
choices, if their sets of guard channels are disjoint.
In the third case, the function merges
the continuation of both processes, if both are guarded by the same prefix.
Note that it is a partial function, e.g.\ it might be the case that a participant behaves
differently in two branches without being aware of which branch was chosen, in which
case the projection of that participant is undefined.

\begin{definition}[Substitution]
  The substitution $\PV \subs{Q}{R}$,
  where $R$ is $\PEND$.%
\[
\PV \subs{Q}{R} =
\begin{cases}
  \PSEND{a}{\SORTV} \PSEP (\PV' \subs{Q}{R})
  & \text{if } \PV = \PSEND{a}{\SORTV} \PSEP \PV'
  \\
  \PRECEIVE{a}{\SORTV} \PSEP (\PV' \subs{Q}{R})
  & \text{if } \PV = \PRECEIVE{a}{\SORTV} \PSEP \PV'
  \\
  \PV_1 \subs{Q}{R} \PINCH \PV_2 \subs{Q}{R}
  & \text{if } \PV = \PV_1 \PINCH \PV_2 
  \\
  \PV_1 \subs{Q}{R} \POUTCH \PV_2 \subs{Q}{R}
  & \text{if } \PV = \PV_1 \POUTCH \PV_2 
  \\
  \PRECN (\PV' \subs{Q}{R})
  & \text{if } \PV = \PRECN  \PV'
  \\
  Q
  & \text{if } \PV = R = \PEND 
  \\
  \PV
  &\text{otherwise}
\end{cases}
\]
\end{definition}

Substitution is used in the projection map.

\subsection{Splitting Systems}\label{app:split}

Omitted rules in Fig.~\ref{fig:split}:
\[
\glorule
{com}
{
\spjudge{\Psi}{\Theta}{\SYSV' \PPAR \SYSV}{\Omega}
}
{
\spjudge{\Psi}{\Theta}{\SYSV\PPAR \SYSV'}{\Omega}
}
\qquad
\glorule
{com}
{
\spjudge{\Psi}{\Theta}{(\SYSV \PPAR T) \PPAR T'}{\Omega}
}
{
\spjudge{\Psi}{\Theta}{\SYSV \PPAR (T \PPAR T')}{\Omega}
}
\]

\begin{definition}[$\spcomp{\_}{\_}$]
  $\spcomp{\Omega}{\Omega'}$
  holds if and only if $\forall \ptp{n} \in \PARTS{\Omega} \cup \PARTS{\Omega'}$
either
  \[
  \Omegan{n} \equiv \Omeganp{n}
  \quad\text{or}\quad
  (
  \Omegan{n} \equiv \PSUM{i \in I}{\PRECEIVE{a_i}{\SORTV_i} \PSEP \PV_j}
  \; \text{ and } \;
  \Omeganp{n} \equiv \PSUM{j \in j}{\PRECEIVE{a_j}{\SORTV_j} \PSEP \PV_j}
  )
  \]
where  $\forall i \in I . \, \forall j \in J . \, \chan{a_i} \neq \chan{a_j}$.
\end{definition}

The boolean function $\spcomp{\_}{\_}$ holds only if two maps $\Omega$ differ wrt 
external choice.

\newcommand{\spmergen}[2]{\mathtt{merge}(#1,#2)}
\begin{definition}[$\spmerge{\_}{\_}$]

  $\Omega = \spmerge{\Omega_0}{\Omega_1}$ is defined only if
  $\spcomp{\Omega_0}{\Omega_1}$ holds,
  in which case
  \[\forall \ptp{n} \in \PARTS{\Omega_0} \cup \PARTS{\Omega_1} \; . \;
  \Omegan{n} = \spmergen{\Omega_0(\ptp{n})}{\Omega_1(\ptp{n})}
  \]
  where
  \[
  \spmergen{P}{Q}
  \begin{cases}
    P & \text{if } P \equiv Q
    \\
    P \POUTCH Q & \text{if } P \equiv  \PSUM{i \in I}{\PRECEIVE{a_i}{\SORTV_i} \PSEP \PV_j}
    \text{ and } Q \equiv \PSUM{j \in j}{\PRECEIVE{a_j}{\SORTV_j} \PSEP \PV_j}
  \end{cases}
  \]
\end{definition}

The function $\spmerge{\_}{\_}$ merges two $\Omega$ maps, if $\spcomp{\_}{\_}$ holds.

\begin{definition}[$\sppre{\_}{\_}$]\label{def:sppre}
\[
\sppre{\PV}{Q} = 
\begin{cases}
  \PINS{(k,k') \in K}{\PSEND{a_k}{\SORTV_k} \PSEP (\sppre{\PV_k}{Q_{k'}})}
  & \text{if }
  P \equiv \PINS{i \in I}{\PSEND{a_i}{\SORTV_i} \PSEP \PV_i}
  \text{ and }
  Q \equiv \PINS{j \in J}{\PSEND{a_j}{\SORTV_j} \PSEP Q_j}
  \\[1pc]
  \PSUM{(k,k') \in K}{\PRECEIVE{a_k}{\SORTV_k} \PSEP (\sppre{\PV_k}{Q_{k'}})}
  \POUTCH 
  \PSUM{n \in N}{\PRECEIVE{a_n}{\SORTV_n} \PSEP \PV_n}
  & \text{if } 
  P \equiv \PSUM{i \in I}{\PRECEIVE{a_i}{\SORTV_i} \PSEP \PV_i}
  \text{ and }
  Q \equiv \PSUM{j \in J}{\PRECEIVE{a_j}{\SORTV_j} \PSEP Q_j}
  \\[1pc]
  \PEND,
  & \text{if } \PV \equiv \PEND \text{ and } Q \equiv \PEND, \text{ or } Q = \epsilon,
\end{cases}
\]
where
$K = \{ (i,j) \in I \times J \; | \;  \chan{a_i} = \chan{a_j} \}
\qand
N = \{ n \in I \; | \; \forall j \in J . \, \chan{a_n} \neq \chan{a_j} \}$.
\end{definition}

The function $\sppre{\_}{\_}$ computes the first part of a split behaviour,
given the original behaviour, e.g.\ $\restri{\SYSV}{n}$ and its prefix
in $\Omegan{n}$. The case for external choices keeps the branches from the original
behaviour which do not appear in $\Omegan{n}$.
The rationale is that even if some branches are never ``chosen'' in the
system, they might still induce e.g.\ races and therefore they need to be
taken into account in the main system.

\begin{definition}[$\fsplit{\_}{\_}$]\label{def:fsplit}
\[
\fsplit{\PV}{Q} = 
\begin{cases}
  \PV_0
  & \text{if }
  \PV \equiv \PINS{i \in I}{\PSEND{a_i}{\SORTV_i} \PSEP \PV_i}
  \text{ and } 
  Q \equiv  \PINS{j \in J}{\PSEND{a_j}{\SORTV_j} \PSEP Q_j}
  \\
  \PV_0
  & \text{if }
  \PV \equiv \PSUM{i \in I}{\PRECEIVE{a_i}{\SORTV_i} \PSEP \PV_i}
  \text{ and } 
  Q \equiv  \PSUM{j \in J}{\PRECEIVE{a_j}{\SORTV_j} \PSEP Q_j}  
  \\
  \PEND,
  & \text{if } \PV \equiv \PEND \text{ and } Q \equiv \PEND,%
  \\
  \PV,
  & \text{if } Q = \epsilon,
  \\
  \nada, & \text{otherwise}
\end{cases}
\]
with $\PV_0$ defined as follows
\[
\PV_0 = 
\begin{cases}
  \fsplit{\PV_i}{Q_j} \text{ with } (i,j) \in K
  & \text{if }
  \forall (i,j)(k,l) \in K \, . \,
  \fsplit{\PV_i}{Q_j} \equiv \fsplit{\PV_k}{Q_l}
  \\
  \nada
  & \text{otherwise}
\end{cases}
\]
where
$K = \{ (i,j) \in I \times J \; | \;  \chan{a_i} = \chan{a_j} \}$

\end{definition}

The function $\fsplit{\_}{\_}$ computes the second part of a split behaviour.
Essentially, it returns the ``rest'' of a behaviour after $\Omegan{n}$.
Note that if $\Omegan{n}$ is a branching behaviour, then the rest must
be the same in all branches (since only \emph{one} behaviour can be
returned), e.g. $\fsplit{\restri{\SYSV}{n}}{\Omegan{n}} = \nada$, if
\[
\restri{\SYSV}{n} =
\PSEND{a}{\SORTV} \PSEP \PSEND{b}{\SORTV} \PINCH 
\PSEND{c}{\SORTV} \PSEP \PSEND{d}{\SORTV}
\qand
\Omegan{n} = \PSEND{a}{\SORTV} \PSEP \epsilon \PINCH 
\PSEND{c}{\SORTV} \PSEP \epsilon
\]

\subsection{Results}\label{app:result}
\begin{definition}
  There is a race in $\SYSV$ if and only if
  there is $\SYSV \ltsarrowC{} \SYSV'$ such that
  $
  \exists \; \chan{a} \in \firebar{\SYSV'}
  $ such that
  \[
  \exists \; \{\ptp{n,m}\} \in \PARTS{\SYSV} \st
  \quad
  \chan{a} \in \fire{\restri{\SYSV}{n}} \text{ and }
  \chan{a} \in \fire{\restri{\SYSV}{m}}
  \quad\text{or}\quad
  \coname{a} \in \fire{\restri{\SYSV}{n}} \text{ and }
  \coname{a} \in \fire{\restri{\SYSV}{m}}
  \]
\end{definition}

\newpage
\section{Definitions Used in the Proofs}

\begin{definition}[Connected - $\conndef{}$]\label{def:connected}
  \begin{itemize}
  \item Two participants are connected in a system $\SYSV$ if $(\ptp{n,m}) \in \conndef{\SYSV}$
    \[
    (\ptp{n,m}) \in \conndef{\SYSV}
    \iff
    \CHANS{\restri{\SYSV}{n}} \cap \CHANS{\restri{\SYSV}{m}} \neq \varnothing
    \text{ or }
    \exists n' . \, (\ptp{n,n'}) \in \conndef{\SYSV}  \land (\ptp{n',m}) \in \conndef{\SYSV}
    \]
  \item Two participants are connected in a global type
    $\GA$ if $(\ptp{n,m}) \in \conndef{\GA}$
    \[
    (\ptp{n,m}) \in \conndef{\GA}
    \iff
    \CHANS{\Proj{\GA}{n}} \cap \CHANS{\Proj{\GA}{m}} \neq \varnothing
    \text{ or }
    \exists n' . \, (\ptp{n,n'}) \in \conndef{\GA}  \land (\ptp{n',m}) \in \conndef{\GA}
    \]
  \end{itemize}
 
\end{definition}

\begin{definition}[Projections with queues]\label{def:projchan}
  \[
  \Projc{\GA}{a} = 
  \begin{cases}
    \SORTV \cdot \Projc{\GA'}{a},
    & \text{if } \GA = \GSEND{\GSTAR}{a}{r}{\SORTV} \GSEP \GA'\\
    \Projc{\GA'}{a},
    & \text{if } \GA = \GSENDN \GSEP \GA'\\
    \Pmerge{\Projc{\GA_1}{a}}{\Projc{\GA_2}{a}},
    &\text{if } \GA = \GA_1 \GCH \GA_2 \\
    \Projc{\GA_i}{a},
    &\text{if } \GA = \GA_1 \GPAR \GA_2 
    \text{ and }
    \chan{a} \not\in \CHANS{\GA_j},
    i \neq j \in \{1,2\} \\
    \Projc{\GA_1}{a} \cdot \, \Projc{\GA_2}{a},
    &\text{if } \GA = \GA_1 \GWS \GA_2 \\
    \emptyQ,
    &\text{if } \GSEND{\GSTAR}{r}{a}{\SORTV} \not\in \GA \\
    \nada,
    & \text{otherwise}
  \end{cases}
  \]
\end{definition}
\begin{definition}[$\Pmergec{\_}{\_}$]\label{def:mergechan}
\[
\Pmergec{\rho_1}{\rho_2} =
\begin{cases}
  \rho_1,
  & \text{if } \rho_1 = \rho_2
  \\
  \nada,
  & \text{otherwise}
\end{cases}
\]
\end{definition}

\section{Proofs for Theorem~\ref{thm:decidable} (Decidability)}
 \thmdecide

\begin{proof}
  The typing systems is decidable from the fact that the ready set of a system,
  the number of participants, and their behaviours are finite.
  Here, we show that the number of behaviour unfoldings needed to type a system
  is also finite.

  \newcommand{\nctx}{C}
  \newcommand{\nhole}[2]{\nctx_{#1}\left[#2\right]}
  \newcommand{\nholep}[2]{\nctx'_{#1}\left[#2\right]}
  \newcommand{\unfold}[2]{\mathtt{unfold}_{#1} ( #2 )}
  \newcommand{\distx}[1]{\lvert #1 \rvert_{\PRECVN}}
  \newcommand{\Sstar}{\SYSV^{*}}
  
  Let (non-recursive) behaviour context $\nhole{}{\_}$ defined as follows
  \[
  \nhole{}{\_} ::=
  \PINS{i \in I}{\PSEND{a_i}{\SORTV_i}{\PSEP \nhole{i}{\_}}} \BNFOR
  \PSUM{i \in I}{\PRECEIVE{a_i}{\SORTV_i}{\PSEP \nhole{i}{\_}}} \BNFOR
  [\_] \BNFOR
  \PRECN P \BNFOR
  \PRECVN 
  \]
  and (possibly recursive) behaviour context $\nholep{}{\_}$ defined as follows
  \[
  \nholep{}{\_} ::=
  \PINS{i \in I}{\PSEND{a_i}{\SORTV_i}{\PSEP \nholep{i}{\_}}} \BNFOR
  \PSUM{i \in I}{\PRECEIVE{a_i}{\SORTV_i}{\PSEP \nholep{i}{\_}}} \BNFOR
  [\_] \BNFOR
  \PRECN \nholep{}{\_} \BNFOR
  \PRECVN 
  \]

  Let $\unfold{i}{\PV}$ be the i$^{th}$ unfolding of $\PV$, defined as follows
  \[
  \begin{array}{lll}
    \unfold{i}{\PV} &=& \unfold{1}{\unfold{i-1}{\PV}}
    \quad i > 1
    \\
    \unfold{1}{\PV}&=&
    \begin{cases}
      \PINS{i \in I}{\PSEND{a_i}{\SORTV_i}{\PSEP \unfold{1}{\PV_i}}}
      & \text{if }
      \PV \equiv \PINS{i \in I}{\PSEND{a_i}{\SORTV_i}{\PSEP \PV_i}}
      \\
      \PSUM{i \in I}{\PRECEIVE{a_i}{\SORTV_i}{\PSEP \unfold{1}{\PV_i}}}
      & \text{if }
      \PV \equiv \PSUM{i \in I}{\PRECEIVE{a_i}{\SORTV_i}{\PSEP \PV_i}}
      \\
      \PV \subs{\PRECN \PV}{\PRECVN}
      & \text{if }
      \PV \equiv \PRECN \PV
    \end{cases}
  \end{array}
  \]

  The need for unfolding occurs whenever a recursive participants interact with another participant,
  while not all the participants feature directly a recursive behaviour.
  In this case, we need to unfold some participants (rule \rulename{$\gieq$}), then use rules 
  \rulename{$\gipinch$},
  \rulename{$\gipoutch$},
  \rulename{$\gipsep$}, and/or
  \rulename{$\giend$}
  until rule \rulename{$\gimua$} is applicable.
  Note that rules \rulename{$\gigws$}, \rulename{$\gipar$} and \rulename{$\giqueue$}
  cannot be used under recursion.
  
  Consider the following system
  \[
  \SYSV = \SYSV_0 \PPAR \SYSV_1
  \]
  where
  \begin{eqnarray}
    \SYSV_0& = &
    \PBOXB{n_1}{\nhole{1}{\PRECN \nholep{1}{\PRECVN}}}
    \PPAR \ldots \PPAR
    \PBOXB{n_j}{\nhole{j}{\PRECN \nholep{j}{\PRECVN}}}
    \\
    \SYSV_1& =&] \PBOX{n_{j+1}}{\PRECN \nholep{j+1}{\PRECVN}} \PPAR \ldots \PPAR \PBOXi{n}{k}{\PRECN \nholep{j+k}{\PRECVN}} 
  \end{eqnarray}
  $\onestep{\SYSV}$, $\noonestep{\SYSV_0}$,
  and there is exactly one $\ptp{n} \in \PARTS{\SYSV}$ such that 
  \[
  \SYSV \equiv \PBOX{n}{\restri{\SYSV}{n}} \PPAR T 
  \qand
  \noonestep{T}
  \]

  Let $\nctx _i$ for $j \leq i \leq j+k$ be the empty context, we can rewrite $\SYSV$ such that
  \[
  \SYSV \equiv
  \PPROD{i \in I}{\PBOXB{n_i}{\nhole{i}{\PRECN \nholep{i}{\PRECVN}}}}
  \qquad I = \{i \; | \; 1\leq i \leq j+k \}
  \]

  Given $\SYSV$ as above, we define $\distx{i}$ to be the smallest $k$ such that
  $\nhole{i}{\PRECVN}$ is a sub-tree of $\unfold{k}{\nholep{i}{\PRECVN}}$,  $\distx{i} = \nada$
  if there is no such $k$.
  Note that  $\distx{i}$ must be smaller than the length of $\nhole{i}{\PEND}$
  (since recursion is guarded).
  If one  $\distx{i}$ is not defined, then $\SYSV$ is not typable.
  
  We also defined
  $M = \mathtt{max} \{ \distx{i} \; | \; i \in I \}$, and $K(i) = M - \distx{i}$.
  We can unfold each behaviour so that all of them are unfolded to the same extent,
  let 
  \[
  \Sstar \equiv \PPROD{i \in I}{\PBOXB{n_i}{
      \unfold{K(i)}{\nhole{i}{\PRECN  \nholep{i}{\PRECVN}}}}
      }
  \]
  
  We show that
  \[
  \derive{\Sstar}{\GA} \Rightarrow \derive{\SYSV}{\GA}
  \]
  
  By definition of $\unfold{}{\_}$, and since $\nhole{}{\_}$ does not contain
  recursive definition, we have
  \begin{eqnarray}
    \Sstar
    &\equiv &\PPROD{i \in I}{\PBOXB{n_i}{
        \nhole{i}{\unfold{K(i)}{\PRECN  \nholep{i}{\PRECVN}}}}
    } \label{eq:proorec1}
    \\
    &\equiv&
    \PPROD{i \in I}{\PBOXB{n_i}{
        \nhole{i}{
          \nholep{i}{
            \nholep{i}{
              \ldots
              \nholep{i}{\PRECN \nholep{i}{\PRECVN}}
            }
            \ldots
          }
        }
      }
    }\label{eq:proorec2}
  \end{eqnarray}

  Where in~\eqref{eq:proorec2}, $\nholep{i}{\_}$ has been unfolded $K(i)$ times.
  It is easy to see that 
  $\Sstar$ is typable if 
  \[
   \PPROD{i \in I}{\PBOXB{n_i}{
        \nhole{i}{
          \nholep{i}{
            \nholep{i}{
              \ldots
              \nholep{i}{\PEND}
            }
            \ldots
          }
        }
      }
    }
    \qand
     \PPROD{i \in I}{\PBOXB{n_i}{
              \nholep{i}{\PEND}
         }
    }
    \]
    are typable themselves, note that rule \rulename{$\gieq$} does not need to be used
    to unfold the left-hand side system, since it is recursion free;
    and there is exactly one recursion less in the right hand side.

    In fact, if we would unfold~\eqref{eq:proorec1} once more, we would not get
    more chances to type $\Sstar$ since it would amount to add the sub-derivation
    of the right-hand side to one of the left-hand side.

\end{proof}

\section{Proofs for Lemma~\ref{thm:RUNIQUE} (Uniqueness)}

\thmunique

\begin{proof}
 The proof is by case analysis. We show that every time one rule
 from Fig.\ref{fig:gloinfer} is applicable, either no other rule is applicable,
 or the derivation produces an equivalent global type.
 
 Due to their syntactic restriction and the condition $\noonestep{\SYSV}$,
 the cases for rules
 \rulename{$\gipsep$},
 \rulename{$\gipinch$}, and
 \rulename{$\gimua$}
 are straightforward.
 In addition, the cases for rule
 \rulename{$\gipoutch$} is easy since it does
 not affect $\GA$.
 The cases for rules 
 \rulename{$\givar$} and 
 \rulename{$\giend$} are trivial.

 The case for rule  \rulename{$\gieq$} follows from the fact that associativity and
 commutativity in $\SYSV$ do not affect $\GA$.
 In addition, if one unfold behaviour once more, we have the result since
 $\GRECN \GA \equiv \GA \subs{\GRECN \GA}{\GRECV}$.

 The interesting part of the proof is to show that
 \rulename{$\gipar$} and \rulename{$\gigws$} are mutually exclusive.
 In fact, if \rulename{$\gipar$} is applicable, \rulename{$\gigws$} cannot
 be used because $\spR$ and $\sprelT$ must be total on $\Psi$ and
 $\Theta \neq \varnothing$. If a system $\SYSV$ could be separated in
 two sub-system by \rulename{$\gipar$}, these two conditions could not hold.
 If \rulename{$\gigws$} is applicable, it means that it is not possible to 
 split the participant in two totally independent sub-systems, and therefore
 \rulename{$\gipar$} is not applicable.
 Finally, observe that by Lemma~\ref{lem:splitunique} the split is unique.

\end{proof}

\section{Proofs for Theorem~\ref{thm:RWF} (Well-formedness)}
\thmwf

\begin{proof}
  The proof is by induction on the derivation $\derive{\SYSV}{\GA}$.
  We make a case analysis on the last rule used.

    \textbf{Case \rulename{$\gipsep$}.}
    We have 
    \[\GA = \GSENDN \GSEP \GA'
    \quad
    \text{ and }
    \quad
    \SYSV = \PBOX{s}{\PSEND{a}{\SORTV} \PSEP P} \PPAR
    \PBOX{r}{\PRECEIVE{a}{\SORTV} \PSEP Q} \PPAR \SYSV'
    \]
    \begin{itemize}
    \item \textit{WF.}
      We show that we have
      \[
      \forall \; \chanT{\_}{n_1}{n_2} \in \GFIN{\GA'} \, . \,
      \{\ptp{s,r}\} \cap \{\ptp{n_1,n_2}\} \neq \varnothing
      \]
      by contradiction. By IH, we know that
      \[
      \judgeC{\{a\} \cup \chanset}{\Gamma}{\addchan{C}{\chanTN}}{
          \PBOX{s}{\PV} \PPAR \PBOX{r}{Q} \PPAR \SYSV'
        }{\GA'}
      \]
      If we had $\GA' \equiv \GSEND{n_1}{n_2}{b}{\SORTV'} \GSEP \GA_0 \GCH \GA_1$,
      with $\ptp{n_i} \neq \ptp{s}$ and $\ptp{n_i} \neq \ptp{r}$ with $i \in \{1,2\}$,
      then we would have
      \[
      \SYSV'  \equiv \PBOX{n_1}{\PSEND{b}{\SORTV'} \PSEP \PV'_0 \PINCH \PV'_1}
      \PBOX{n_2}{\PRECEIVE{b}{\SORTV'} \PSEP Q'_0 \POUTCH Q'_1} \PPAR \SYSV''
      \]
      which is in contradiction with the premise $\noonestep{\SYSV'}$.
      
      By Lemma~\ref{lem:wfimplies}, the result above and since
      $\addchan{C}{\chanTN}$ is defined,
      we have $\wfjudge{\emptyTree}{\GA}$.
    \item \textit{Projection.}
      By Def.~\ref{def:proj},
      we have that $\Proj{\GA}{s} = \PSEND{a}{\SORTV} \PSEP \Proj{\GA'}{s}$,
      $\Proj{\GA}{r} = \PRECEIVE{a}{\SORTV} \PSEP \Proj{\GA'}{r}$, and
      $\Proj{\GA}{n} =\Proj{\GA'}{n}$, for $\ptp{s} \neq \ptp{n} \neq \ptp{r}$.
    \end{itemize}

    \textbf{Case \rulename{$\gipinch$}.}
    We have
    \[
    \GA = \GA_0 \GCH \GA_1 \quad \text{and} \quad
    \SYSV = \PBOX{s}{\PV \PINCH Q} \PPAR \SYSV'
    \]
    \begin{itemize}
    \item \textit{WF.}
      Observe that we have that all the guard channels are disjoint by definition
      of processes.
      We have to show that
      \[
      \forall (\ptp{n}_1,\ptp{n}_2) \in \GFIN{\GA_0} .
      \forall (\ptp{n}'_1,\ptp{n}'_2) \in \GFIN{\GA_1}
      \, . \, \ptp{n}_1 = \ptp{n}'_1
      \]
      i.e.\ that for all prefixes in $\GA_0$ and $\GA_1$, $\ptp{s}$ is the sender.
      In other words, the participant who makes the internal choice must be the same in all
      the branches.
      If that was not the case, we would have
      \[
      \SYSV' \equiv \PBOX{n_1}{\PSEND{b}{\SORTV} \PSEP Q_0 \PINCH Q_1}
      \PPAR \PBOX{n_2}{\PRECEIVE{b}{\SORTV} \PSEP Q_2 \POUTCH Q_3}
      \PPAR \SYSV''
      \]
      which is in contradiction with the premise $\noonestep{\SYSV'}$.

      By Lemma~\ref{lem:wfimplies}, the result above and 
      $\wfjudge{C}{\GA_0}$ and $\wfjudge{C}{\GA_1}$ by IH, we
      have $\wfjudge{\emptyTree}{\GA}$.

    \item \textit{Projection.}
      We have to prove that
      \[
      \forall \ptp{n} \in \PARTS{\GA_0 \GCH \GA_1} . \,
      \Pmerge{\Proj{\GA_0}{n}}{\Proj{\GA_1}{n}} \neq \nada
      \]
      By IH, we have that both $\Proj{\GA_0}{n}$ and $\Proj{\GA_1}{n}$ exist.
      There are two cases where the projection does not exist
      ($i \neq j \in \{0,1\}$ in the following):
      \begin{itemize}
      \item $\Proj{\GA_i}{n} = \PSEND{a}{\SORTV} \PSEP \PV$
        and
        $\Proj{\GA_j}{n} = \PRECEIVE{b}{\SORTV'} \PSEP Q$. This cannot
        happen by definition of behaviours, i.e.\
        such projections could only come from behaviours of the form
        $(\PSEND{a}{\SORTV} \PSEP \PV) \PINCH (\PRECEIVE{b}{\SORTV'} \PSEP Q)$,
        which is not syntactically correct.
      \item $\Proj{\GA_i}{n} = \PSUM{k \in K}{\PRECEIVE{a_k}{\SORTV_k} \PSEP \PV_k}$
        and $\Proj{\GA_j}{n} = \PSUM{k' \in K'}{\PRECEIVE{a_{k'}}{\SORTV_{k'}} \PSEP \PV_{k'}}$,
        with $\chan{a_k} = \chan{a_{k'}}$ for some $k$ and $k'$.
        This cannot happen by definition of behaviours, i.e.\
        such projections could only come from behaviours of the form
        $\PRECEIVE{a}{\SORTV} \PSEP \PV_0 \POUTCH \PRECEIVE{a}{\SORTV} \PSEP \PV_1$
        which is not a syntactically correct behaviour.
        The same reasoning applies for internal choice.
      \end{itemize}
      Note that we have
      $\Proj{\GA}{s} = \Proj{\GA_0}{s} \PINCH \Proj{\GA_1}{s}$.
    \end{itemize}

   \textbf{Case \rulename{$\gipoutchl$}}
    \begin{itemize}
    \item \textit{WF.} By induction hypothesis, we have $\wfjudge{C}{\GA}$
      and by Lemma~\ref{lem:wfimplies}, we have $\wfjudge{\emptyTree}{\GA}$.
    \item \textit{Projection.} By induction hypothesis.
    \end{itemize}

    \textbf{Case \rulename{$\gipar$}}
    We have 
    \[
    \GA = \GA_0 \GPAR \GA_1
    \quad \text{and} \quad
    \SYSV = \SYSV_0 \PPAR \SYSV_1
    \]
    \begin{itemize}
    \item \textit{WF.} We have to show that
      \[
      \PARTS{\GA_0} \cap \PARTS{\GA_1} = \varnothing
      \quad \text{and} \quad
      \CHANS{\GA_0} \cap \CHANS{\GA_1} = \varnothing
      \]
      By definition of systems, we know that there cannot be two participants
      of the same name in a same system, since $\SYSV \PPAR \SYSV'$ is a system
      we have that 
      \[
      \PARTS{\SYSV_0} \cap \PARTS{\SYSV_1} = \varnothing
      \]
      and by Lemma~\ref{lem:partis}, we have $\PARTS{\GA_0} \cap \PARTS{\GA_1} = \varnothing$.
      By Lemma~\ref{lem:chansA}, the premise $\chanset_0 \cap \chanset_1$, we
      have $\CHANS{\GA_0} \cap \CHANS{\GA_1} = \varnothing$.
      We have $\wfjudge{C}{\GA_0}$ and $\wfjudge{C}{\GA_1}$ by assumption,
      with the result above and Lemma~\ref{lem:wfimplies}, we
      have  $\wfjudge{\emptyTree}{\GA}$.
    \item \textit{Projection.}
      For all $\ptp{n} \in \PARTS{\GA}$, $\Proj{\GA}{n}$ is defined by IH
      and since  $\PARTS{\GA_0} \cap \PARTS{\GA_1} = \varnothing$.
    \end{itemize}

   \textbf{Case \rulename{$\gigws$}.}
   We have
   \[
   \GA = \GA_0 \GWS \GA_1
   \quad \text{and} \quad
   \addchan{C}{\chanG{\GA_0}}
   \quad \text{and} \quad
   \Ssplit{\SYSV} = (\SYSV_0 , \SYSV_1)
   \]
    \begin{itemize}
    \item \textit{WF.}
      We have to show that  
      \begin{equation}\label{eq:spcase1}
        \forall \; \chanT{\_}{n_1}{n_2} \in \GFIN{\GA_1} \, . \,
        \exists N_1 \neq N_2 \in \GFOUTS{\GA_0} \, . \,
        \ptp{n}_1 \in N_1 \land \ptp{n}_2 \in N_2 
      \end{equation}
      and
      \begin{equation}\label{eq:spcase2}
        \forall N \in \GFINP{\GA_0} \, . \,
        \exists N' \in \GFINP{\GA_1} \, . \,
        N \cap N' \neq \varnothing 
      \end{equation}
      We know that 
      $
      \Ssplit{\SYSV} = (\SYSV_0 , \SYSV_1)
      $
      therefore there is $\spjudge{\Psi}{\Theta}{\SYSV}{\Omega}$ coherent.
      We show~\eqref{eq:spcase1} first.
      We first show that
      \[
      \chanTN \in \GFIN{\GA_1}
      \iff
      \{\ptp{s,r}\} \in \Theta
      \]
      We start with 
      \[
      \chanT{\_}{n_i}{n_j} \in \GFIN{\GA_1}
      \iff
      \onestep{\restri{\SYSV_1}{n_i} \PPAR \restri{\SYSV_1}{n_j}}
      \]
      ($\Rightarrow$)
      If $ \chanT{\_}{n_i}{n_j} \in \GFIN{\GA_1}$ then we must have
      $\GA_1 = ((\GSEND{n_i}{n_j}{a}{\SORTV} \GSEP \GA_2 \GCH \GA_3) \GPAR \GA_4) \GWS \GA_5$
      (by definition of $\GFINE$). And by Lemma~\ref{lem:svsg},
      this implies that $\restri{\SYSV_1}{n_i} = \PSEND{a}{\SORTV} \PSEP \PV \PINCH \PV'$
      and
      $ \restri{\SYSV_1}{n_j} = \PRECEIVE{a}{\SORTV} \PSEP Q  \POUTCH Q'$.
      Thus we have  $\onestep{\restri{\SYSV_1}{n_i} \PPAR \restri{\SYSV_1}{n_j}}$.
      
      ($\Leftarrow$)
      If $\onestep{\restri{\SYSV_1}{n_i} \PPAR \restri{\SYSV_1}{n_j}}$,
      we must have 
      $\restri{\SYSV_1}{n_i}  = \PSEND{a}{\SORTV} \PSEP \PV \PINCH \PV'$
      and
      $ \restri{\SYSV_1}{n_j} = \PRECEIVE{a}{\SORTV} \PSEP Q  \POUTCH Q'$
      and since $\GA_1$ is well-formed by IH,
      we have the required result by Lemma~\ref{lem:svsg} and
      the definition of $\GFINE$.
      Now, let us show that
      \begin{equation}\label{eq:proctheta}
         \onestep{\restri{\SYSV_1}{n_i} \PPAR \restri{\SYSV_1}{n_j}}
         \iff
         \{\ptp{n_i,n_j}\} \in \Theta
      \end{equation}
    ($\Rightarrow$)
      Since the processes interact, we know that they are $\neq \PEND$.
      Moreover, since they appear in the second part of the split,
      the following rule must be applied so that
      $\restri{\SYSV_1}{n_i}  = \fsplit{\restri{\SYSV}{n_i}}{\Omegan{n_i}}$
      and 
      $ \restri{\SYSV_1}{n_j} = \fsplit{\restri{\SYSV}{n_j}}{\Omegan{n_j}}$
      ( $\neq \nada$ by assumption)
      with $\Omegan{n_i}$ and $\Omegan{n_j}$ ending with $\epsilon$.
      \[
      \glorule
      {\splepsilon}
      {
        \ptp{n_i} \in N, \ptp{n_j} \in M
        \\
        \onestep{\PBOXi{n}{i}{\restri{\SYSV_1}{n_i}}
          \PPAR \PBOXi{n}{j}{\restri{\SYSV_1}{n_j}}}
        \\
        \spjudge{\Psi \cdot N \spminus{n} \cdot M \spminus{m}}
        {\Theta}
        {
          \SYSV
        }
        {
          \Omega
        }
      }
      {
        \spjudge{\Psi \cdot N \cdot M}
        {\Theta \cdot \{\ptp{n_i}, \ptp{n_j}\}}
        {
          \PBOXi{n}{i}{\restri{\SYSV_1}{n_i}} \PPAR
          \PBOXi{n}{j}{\restri{\SYSV_1}{n_j}} \PPAR \SYSV
        }
        {
          \Omega \cdot \prtrack{n_i}{\epsilon}
          \cdot \prtrack{n_j}{\epsilon}
        }
      }
      \]      
      which gives us the expected result.
      Note that the rule \rulename{$\splrem$} cannot be applied
      because the processes interact with each other we do not have
      $\noonestep{\PBOX{n}{\PV} \PPAR \SYSV}$.
      
      ($\Leftarrow$)
      If $\{\ptp{n_i,n_j}\} \in \Theta$ then, the rule \rulename{$\splepsilon$}
      must also be applied
      (the axiom cannot be reach with $\Theta \neq \varnothing$),
      since the rule requires that
      $\onestep{\PBOXi{n}{i}{\restri{\SYSV_1}{n_i}}
        \PPAR \PBOXi{n}{j}{\restri{\SYSV_1}{n_j}}}$, we have
      the expected result.
      
      Let us now show that
      \begin{equation}\label{eq:sptheta}
        \forall \{\ptp{n}_i,\ptp{n}_j\} \in \Theta \, . \,
        \exists N_i \neq N_j \in \GFOUTS{\GA_0} \, . \,
        \ptp{n}_i \in N_j \land \ptp{n}_j \in N_j 
      \end{equation}
      First we show that
      \begin{equation}\label{eq:spgf}
        N \in \GFOUTS{\GA_0} \Rightarrow
        \exists ! N' \in \Psi \st N \cap N' \neq \varnothing
      \end{equation}
      By Lemma~\ref{lem:conntotal}, we have that
      $\forall N \in \GFOUTS{\GA_0} \, . \, \conndef{\GA_0}$ is total on $N$.
      By Lemma~\ref{lem:connGtoSpl}, we have that
      $(\ptp{n,m}) \in \conndef{\GA_0} \Rightarrow (\ptp{n,m}) \in \spR $.
      By Lemma~\ref{lem:spiff}, we have that
      $(\ptp{n,m}) \in \spR \Rightarrow \{\ptp{n,m}\} \in N \in \Psi$.
      Therefore we have
      \[
      N \in \GFOUTS{\GA_0} \Rightarrow
      \exists  N' \in \Psi \st N \cap N' \neq \varnothing
      \]
      and $N'$ is unique since $\Psi$ is a set of pairwise disjoint sets.
      
      We finalise the proof of~\eqref{eq:spcase1}, by showing~\eqref{eq:sptheta}
      by contradiction.
      If we had
      \[
      \exists \{\ptp{n_1,n_2}\} \in \Theta \, .
      \exists N \in \GFOUTS{\GA_0} \, . \, \{\ptp{n_1,n_2}\} \in N
      \]
      by Lemma~\ref{lem:sppsitheta}, we have that
      $\exists N_1 \neq N_2 \in \Psi \st \ptp{n_1} \in N_1 \in \Psi$ and $\ptp{n_2} \in N_2 \in \Psi$
      which is contradiction with~\eqref{eq:spgf}.
    
    \medskip

    Let us now show \eqref{eq:spcase2}, i.e.
    \[
    \forall N \in \GFINP{\GA_0} \, . \, \exists  N' \in \GFINP{\GA_1}
    \st N \cap N' \neq \varnothing
    \]
    We first show that
    \[
    \GFINP{\GA} = \Psi
    \]

    By definition of $ \GFINPE $, we have that
    \[
    \GA = \GPROD{N \in \GFINP{\GA}}{\GA_N}
    \quad
    \text{with }\GA_N \not\equiv \GA'_N \GPAR \GA''_N.
    \]
    Therefore, by Lemma~\ref{lem:conntotal},
    $\conndef{\GA_N}$ is total on $\PARTS{\GA_N}$ for each $N$.
    By Lemma~\ref{lem:connGtoSpl}, we
    have that $\spR$ is total on each $\PARTS{\GA_N}$ as well,
    and by Lemma~\ref{lem:spiff},
    we have 
    \[
    N \in \Psi \iff \PARTS{\GA_N} = N
    \]
    
    Now, we show that 
    \[
    \forall N \in \Psi . \, \exists N' \in \GFINP{\GA_1} . \, N \cap N' \neq \varnothing
    \]
    by contradiction, if we had
    \[
    \exists N \in \Psi . N \cap \PARTS{\GA_1} = \varnothing
    \]
    by Lemma~\ref{lem:sppsitheta}, we must have
    $N \neq M \in \Psi$ such that $\{\ptp{n,m}\} \in \Theta$
    and $\ptp{n} \in N$ and $\ptp{m} \in M$.
    Since $\ptp{n} \in N$ is also in $\Theta$, we must have
    $\onestep{\restri{\SYSV_1}{n} \PPAR \restri{\SYSV_1}{m}}$ which implies 
    that $\restri{\SYSV_1}{n} \equiv \PEND$ and therefore $\ptp{n} \in \PARTS{\GA_1}$,
    by Lemma~\ref{lem:partis}. Thus  $N \cap \PARTS{\GA_1} \neq \varnothing$
    
    From~\eqref{eq:spcase1},~\eqref{eq:spcase2}, the fact that by assumption
    $\addchan{C}{\chanG{\GA_0}}$ is defined, and Lemma~\ref{lem:wfimplies}
    we have $\wfjudge{\emptyTree}{\GA}$.
  \item \textit{Projection.}
    By IH, we have that for all $\ptp{n}$ $\Proj{\GA_0}{n}$ and $\Proj{\GA_1}{n}$ exist,
    thus we have that $\Proj{\GA_0}{n}\subs{\Proj{\GA_1}{n}}{\PEND}$ is defined
    for all $\ptp{n}$.
    Note that since $\wfjudge{\emptyTree}{\GA_0}$, we have
    $\#(\GFOUTS{\GA_0}) > 1$ (cf.~\ref{eq:spcase1}).
    Therefore, by Lemma~\ref{lem:recF}, $\freen{\GA_1} = \varnothing$.
    In addition, by Lemma~\ref{lem:recB}, we have $\boundn{\SYSV} = \varnothing$
    (since the $\Ssplit{\SYSV}$ by assumption),
    and by Lemma~\ref{lem:recBB}, $\boundn{\GA_0} = \varnothing$.
    Thus, every branch in $\GA_0$ ends with $\GEND$, and all the projections of $\GA_0$
    end with $\PEND$.
  \end{itemize}
  
   \textbf{Case \rulename{$\gimua$}}
   We have
   \[
   \GA = \GRECN \GA_0
   \quad \text{and} \quad
   \SYSV =
   \PBOX{n_1}{\PRECi{X}{1} \PV_1 }
   \PPAR \! \ldots \! \PPAR
   \PBOX{n_k}{\PRECi{X}{k} \PV_k}
   \quad \text{and} \quad
   \]
    \begin{itemize}
    \item \textit{WF.}
      We have to show that
      \[
      \GRECV \in \freen{\GA'} \Rightarrow
      \# \GFOUTS{\GA'} = 1
      \]
      we can apply Lemma~\ref{lem:recF} and we have the result directly.
      Observe that the recursion is prefix guarded since
      $\exists \, i,j \st \onestep{\PBOX{n_i}{\PV_i} \PPAR \PBOX{m_j}{\PV_j}}$ holds.
      
    \item \textit{Projection.} By induction hypothesis.
    \end{itemize}

  \textbf{Case \rulename{$\givar$}}
  We have
  \[
  \GA = \GRECV
  \quad \text{and} \quad
  \forall \ptp{n} \in \PARTS{\SYSV} . \restri{\SYSV}{n} = \PRECVN_{\ptp{n}}
  \]
  We have to show that $\addchan{C}{\subtree{C}{\GRECV}}$,
  which follows directly from the premises of rule \rulename{$\givar$}.
  
  \textbf{Case \rulename{$\giend$}}
   Trivial.

 \textbf{Case \rulename{$\giqueue$}}
 We have 
    \[\GA = \GSEND{\GSTAR}{r}{a}{\SORTV} \GSEP \GA'
    \quad
    \text{ and }
    \quad
    \SYSV = \PBOX{r}{\PRECEIVE{a}{\SORTV} \PSEP Q} \PPAR
    \QUEUE{a}{\SORTV \cdot \rho} \PPAR \SYSV'
    \]
    \begin{itemize}
    \item \textit{WF.}
      We show that we have
      \[
      \forall \; \chanT{\_}{n_1}{n_2} \in \GFIN{\GA'} \, . \,
      \ptp{n_i} = \ptp{r}
      \quad \text{with } i \in \{1,2\}
      \]
      by contradiction. By IH, we know that
      \[
      \judgeC{\{a\} \cup \chanset}{\Gamma}{\addchan{C}{\chanTN}}{
          \PBOX{r}{Q} \PPAR \QUEUE{a}{\rho} \PPAR \SYSV'
        }{\GA'}
      \]
      If we had $\GA' \equiv \GSEND{n_1}{n_2}{b}{\SORTV'} \GSEP \GA_0 \GCH \GA_1$,
      with $\ptp{n_i} \neq \ptp{r}$ with $i \in \{1,2\}$,
      then we would have
      \[
      \SYSV'  \equiv \PBOX{n_1}{\PSEND{b}{\SORTV'} \PSEP \PV'_0 \PINCH \PV'_1}
      \PBOX{n_2}{\PRECEIVE{b}{\SORTV'} \PSEP Q'_0 \POUTCH Q'_1} \PPAR \SYSV''
      \]
      which is in contradiction with the premise $\noonestep{\SYSV'}$.
      
      By Lemma~\ref{lem:wfimplies}, the result above, and since
      $\addchan{C}{\chanT{a}{\GSTAR}{r}}$ is defined,
      we have $\wfjudge{\emptyTree}{\GA}$.
    \item \textit{Projection.}
      By Def.~\ref{def:proj},
      we have that
      $\Proj{\GA}{r} = \PRECEIVE{a}{\SORTV} \PSEP \Proj{\GA'}{r}$, and
      $\Proj{\GA}{n} =\Proj{\GA'}{n}$, for $\ptp{n} \neq \ptp{r}$.
    \end{itemize}
\end{proof}

\section{Proofs for Theorem~\ref{thm:RSUBRED} (Subject Reduction)}

\thmsubred 

\begin{proof}
  There are three cases to consider.
  
  \proocase{Case 1}
  $\SYSV \equiv \SYSV_1 \PPAR \SYSV_2$
  with $\SYSV_1 \ltsarrow{\labsend} \SYSV_1'$
  and  $\SYSV_2 \ltsarrow{\labpush} \SYSV_2'$
  
  In this case, $\SYSV$ must have the following form
  \[
  \SYSV \equiv \PBOX{n}{\PSEND{a}{\SORTV} \PSEP \PV \PINCH \PV'}
  \PPAR \QUEUE{a}{\rho} 
  \PPAR T
  \quad
  \text{and}
  \quad
  \SYSV' \equiv \PBOX{n}{\PV}
  \PPAR \QUEUE{a}{\rho \cdot \SORTV} 
  \PPAR T
  \]
  and by Lemma~\ref{lem:subred} we have that $\judge{\emptyctx}{C}{\SYSV'}{\GA'}$.

  \medskip
  \proocase{Case 2}
  $\SYSV \equiv \SYSV_1 \PPAR \SYSV_2$
  with $\SYSV_1 \ltsarrow{\labrec} \SYSV_1'$
  and  $\SYSV_2 \ltsarrow{\labpop} \SYSV_2'$

  In this case, $\SYSV$ must have the following form
  \[
  \SYSV = \PBOX{n}{\PRECEIVE{a}{\SORTV} \PSEP \PV \POUTCH \PV'}
  \PPAR \QUEUE{a}{\SORTV \cdot \rho} 
  \PPAR T
  \quad
  \text{and}
  \quad
  \SYSV' = \PBOX{n}{\PV}
  \PPAR \QUEUE{a}{\rho} 
  \PPAR T
  \]

  and by Lemma~\ref{lem:subred} we have that $\judge{\emptyctx}{C}{\SYSV'}{\GA'}$

  \medskip
  \proocase{Case 3}
  $\SYSV \ltsarrow{\labend} \SYSV$
  in this case, we have $\judge{\emptyctx}{C}{\SYSV'}{\GA'}$ trivially since $\SYSV' \equiv \SYSV$.

  \medskip
  \proocase{Case 4 (unfolding)}
  If $\SYSV \equiv \PBOX{n}{\PRECN \PV} \PPAR \SYSV_0$ and
  $\SYSV' \equiv  \PBOX{n}{\PV\subs{\PRECN \PV}{\PRECVN}} \PPAR \SYSV_0$
  then $\GA = \GRECN \GA_0$, and $\forall \ptp{m} \in \PARTS{\SYSV_0} \, . \, \restri{\SYSV_0}{m} \equiv \PRECN Q$.
  We have the result by unfolding all the participants in $\SYSV_0$ wiht rule \rulename{$\gieq$},
  so to have $\GA' = \GA_0 \subs{\GRECN \GA_0}{\GRECV}$.
  
  In the other direction, we have the result by folding all the participants in $\PARTS{\SYSV_0}$.

  \medskip
  \proocase{Case 5 (commutativity and associativity)}
  If $\SYSV \equiv \SYSV'$, then we have the result with rule \rulename{$\gieq$}.
  
\end{proof}

\begin{lemma}\label{lem:subred}
  The following holds:
  \begin{enumerate}
  \item \label{lem:subred1}
    $
    \judge{\emptyctx}{C}{
      \PBOX{s}{\PSEND{a}{\SORTV} \PSEP \PV \PINCH \PV'}
      \PPAR \QUEUE{a}{\rho} \PPAR T
    }
    {\GA}
    \Rightarrow
    \judge{\emptyctx}{C}{
      \PBOX{s}{\PV}
      \PPAR \QUEUE{a}{\rho \cdot \SORTV } \PPAR T
    }{\GA'}
    $
    
  \item \label{lem:subred2}
    $
    \judge{\emptyctx}{C}{
      \PBOX{r}{\PRECEIVE{a}{\SORTV} \PSEP \PV \POUTCH \PV'}
      \PPAR \QUEUE{a}{\SORTV \cdot \rho} \PPAR T
    }{\GA}
    \Rightarrow
    \judge{\emptyctx}{C}{
      \PBOX{r}{\PV}
      \PPAR \QUEUE{a}{\rho} \PPAR T
    }{\GA'}
    $
  \end{enumerate}
  with $\chanTA \not\in C$ in~\ref{lem:subred1} and~\ref{lem:subred2}.
\end{lemma}

\begin{corollary}\label{cor:subred}
  \[
  \Proj{\GA}{n} = (\PSEND{a}{\SORTV} \PSEP \Proj{\GA_0}{n}) \PINCH \Proj{\GA_1}{n}
  \Rightarrow \Proj{\GA'}{n} = \Proj{\GA_0}{n}
  \]
  and
  \[
  \Proj{\GA}{n} = \PRECEIVE{a}{\SORTV} \PSEP \Proj{\GA_0}{n}
  \Rightarrow \Proj{\GA'}{n} = \Proj{\GA_0}{n}
  \]

\end{corollary}

\begin{proof}

  We show that~\ref{lem:subred1} and~\ref{lem:subred2} hold by contradiction.

  \proocase{Case~\ref{lem:subred1}}
  Assume 
  \[
  \judge{\emptyctx}{C}{
      \PBOX{s}{\PSEND{a}{\SORTV} \PSEP \PV \PINCH \PV'}
      \PPAR \QUEUE{a}{\rho} \PPAR \SYSV
    }
    {\GA}
   \text{ is derivable.}
   \]
   \[
    \judge{\emptyctx}{C}{
      \PBOX{s}{\PV}
      \PPAR \QUEUE{a}{\rho \cdot \SORTV } \PPAR \SYSV
    }{\GA'}
    \text{ is not.}
  \]
  Take $\SYSV \equiv \PBOX{r}{\PRECEIVE{a}{\SORTV} \PSEP Q} \PPAR \SYSV'$,
  $\rho = \emptyQ$, and $\noonestep{\SYSV'}$.
  We must have the following sub-derivation for $\GA$
  \begin{equation}\label{eq:neqsubred1}
  \inferrule* [Left=\rulename{\gipsep}]
  {
    \inferrule*
    {\vdots}
    {
      \judge{\emptyctx}{\addchan{C}{\chanT{a}{s}{r}}}{
        \PBOX{s}{\PV} \PPAR \QUEUE{a}{\rho} \PPAR
        \PBOX{r}{\PRECEIVE{a}{\SORTV} \PSEP Q} \PPAR \SYSV'
      }
      {\GA_0}
    }
  }
  {
    \judge{\emptyctx}{C}{
      \PBOX{s}{\PSEND{a}{\SORTV} \PSEP \PV} \PPAR \QUEUE{a}{\rho}
      \PPAR
      \PBOX{r}{\PRECEIVE{a}{\SORTV} \PSEP Q} \PPAR \SYSV'
    }
    {\GSEND{s}{r}{a}{\SORTV} \GSEP \GA_0}
  }
\end{equation}
Consider the non-derivable judgement for $\GA'$
\begin{equation}\label{eq:neqsubred2}
  \inferrule* [Left=\rulename{$\giqueue$}]
  {
    \inferrule*
    {\nada}
    {
      \judge{\emptyctx}{\addchan{C}{\chanT{a}{\GSTAR}{r}}}{
        \PBOX{s}{\PV}
        \PPAR \QUEUE{a}{\rho} \PPAR
         \PBOX{r}{\PRECEIVE{a}{\SORTV} \PSEP Q} \PPAR \SYSV'
      }
      {\GA'_0}
    }
  }
  {
    \judge{\emptyctx}{C}{
      \PBOX{s}{\PV}
      \PPAR \QUEUE{a}{\SORTV} \PPAR
      \PBOX{r}{\PRECEIVE{a}{\SORTV} \PSEP Q} \PPAR \SYSV'
    }
    {\GSEND{\GSTAR}{r}{a}{\SORTV} \GSEP \GA'_0}
  }
\end{equation}

  Where the rule \rulename{$\giqueue$} must be applicable here, 
  since the only difference with the above system is 
  $\addchan{C}{\chanT{a}{\GSTAR}{r}}$ which is defined since
  $\chanTA \not\in C$. 
  
  We have a contradiction here since 
  \[
  \PBOX{s}{\PV} \PPAR \QUEUE{a}{\rho} \PPAR
  \PBOX{r}{\PRECEIVE{a}{\SORTV} \PSEP Q} \PPAR \SYSV'
  \text{ is derivable in~\eqref{eq:neqsubred1}.}
  \]
  while
  \[
  \PBOX{s}{\PV}
  \PPAR \QUEUE{a}{\rho} \PPAR
  \PBOX{r}{\PRECEIVE{a}{\SORTV} \PSEP Q} \PPAR \SYSV'
  \text{ is not derivable in~\eqref{eq:neqsubred2}.}
  \]
  
  \medskip

  \proocase{Case~\ref{lem:subred2}}
  Assume 
  \[
  \judge{\emptyctx}{C}{
    \PBOX{r}{\PRECEIVE{a}{\SORTV} \PSEP \PV \POUTCH \PV'}
    \PPAR \QUEUE{a}{\SORTV \cdot \rho} \PPAR \SYSV
  }
  {\GA}
  \text{ is derivable.}
  \]
  \[
  \judge{\emptyctx}{C}{
    \PBOX{r}{\PV}
    \PPAR \QUEUE{a}{\rho } \PPAR \SYSV
  }{\GA'}
  \text{ is not.}
  \]
  Take $\SYSV$ such that $\noonestep{\SYSV}$.
  We must have the following sub-derivation for $\GA$
  \begin{equation}\label{eq:newsubred3}
    \inferrule* [Left=\rulename{$\giqueue$}]
    {
      \inferrule*
      {\vdots}
      {
      \judge{\emptyctx}{\addchan{C}{\chanT{a}{\GSTAR}{r}}}{
        \PBOX{r}{\PV}
          \PPAR \QUEUE{a}{\rho} \PPAR \SYSV
        }
        {
          \GA_0
        }
      }
    }
    {
      \judge{\emptyctx}{C}{
        \PBOX{r}{\PRECEIVE{a}{\SORTV} \PSEP \PV}
        \PPAR \QUEUE{a}{\SORTV \cdot \rho} \PPAR \SYSV
        }
        {\GSEND{\GSTAR}{r}{a}{\SORTV} \GSEP \GA_0}
    }
  \end{equation}

  This induces a contradiction since we would have the following for $\GA'$
   \begin{equation}\label{eq:newsubred4}
    \inferrule* [Left=\rulename{$\giqueue$}]
    {
      \nada
    }
    {
      \judge{\emptyctx}{C}{
        \PBOX{r}{\PV}
        \PPAR \QUEUE{a}{\rho} \PPAR \SYSV
        }
        {\GA'_0}
    }
  \end{equation}
  Note that the lack of $\chanT{a}{\GSTAR}{r}$ in $C$ does not 
  affect since  $\chanTA \not\in C$.

\end{proof}

\section{Proofs for Theorem~\ref{thm:RPROGR} (Progress)}
\thmprogress

\begin{proof}
  The proof is by contradiction.
  If we had $\deriveN$ and $\SYSV \ltsarrowC{} \SYSV'$,
  with $\noonestep{\SYSV'}$, $\SYSV' \neq \PEND$ and 
  $\exists \ptp{n} \in \PARTS{\SYSV'} \st \restri{\SYSV'}{n} \neq \PEND$.
  By Theorem~\ref{thm:RSUBRED}, we should have $\derive{\SYSV'}{\GA'}$.
  Let us take $\SYSV' \equiv \PBOX{n}{\PSEND{a}{\SORTV} \PSEP \PV} \PPAR \SYSV''$
  (with $\noonestep{\SYSV'}$).
  No rule from Fig.~\ref{fig:gloinfer} is applicable for this process,
  and therefore $\SYSV'$ is not typable.  
\end{proof}

\section{Proofs for Theorem~\ref{thm:RSAFE} (Safety)}
\thmsafety

\begin{proof}
  
  \proocase{No error}
  The proof is by contradiction.
  Assume we have $\deriveN$ and $\SYSV \ltsarrowC{} \SYSV'$
  with
  \[
  \SYSV' \equiv \QUEUE{a}{\SORTV \cdot \rho}
  \PPAR \PBOX{r}{\PRECEIVE{a}{\SORTV'} \PSEP Q}
  \PPAR \SYSV''
  \qand
  \SORTV \neq \SORTV'
  \]
  so that $\SYSV \ltsarrow{} \PERROR$.
  By Theorem~\ref{thm:RSUBRED}, we should have $\derive{\SYSV'}{\GA'}$,
  however, no rule is applicable for $\SYSV'$.
  Indeed we have $\onestep{\SYSV}$, but $\SORTV \neq \SORTV'$.

  \proocase{No Race}
  Straightforward by contradiction with Theorem~\ref{thm:RSUBRED},
  the following is not typable
  \[
  \SYSV' \equiv
  \PBOX{n}{\PRECEIVE{a}{\SORTV} \PSEP \PV}
  \PPAR \PBOX{m}{\PRECEIVE{a}{\SORTV} \PSEP Q}
  \PPAR \SYSV''
  \]
  due to the condition on $C$, the premise $\noonestep{\SYSV}$ in \rulename{$\gipsep$},
  and the fact that the set of channels must be disjoint in concurrent branches.
  The other case (i.e.\ two sends) is similar.
\end{proof}

\begin{lemma}\label{lem:norace}
  If $\derive{\SYSV}{\GA}$, then,
  $  \forall \; \chanTN \in \GFINP{\GA} \st$, either
  \[
  \exists ! \{\ptp{s,r}\} \in \PARTS{\SYSV} \st
  \chan{a} \in \fire{\restri{\SYSV}{s}} \text{ and } \coname{a} \in \fire{\restri{\SYSV}{r}}
  \quad \text{or} \quad
  \exists ! (\ptp{r} , \chan{a}) \in \PARTS{\SYSV} \times \CHANS{\SYSV} \st
  \chan{a} \in \fire{\restri{\SYSV}{s}}
  \]
\end{lemma}
\begin{proof}
  By straightforward induction on the derivation. Each case follows by definition of 
  $\GFINP{\GA}$, and the premise $\noonestep{\SYSV}$.
\end{proof}

\section{Proofs for Lemma~\ref{lem:RSIMU} and Theorem~\ref{thm:RBISIMU} (Equivalences)}

\thmsimu

\begin{proof}
  Let $B$ be a binary relation on processes defined as follows
  \[
  (\PV , Q) \in B
  \iff
  \derive{\SYSV}{\GA} \text{ and }
  \exists \ptp{n} \in \PARTS{\SYSV} . \, \Proj{\GA}{n} = \PV \text{ and }
  \restri{\SYSV}{n} = Q
  \]
  Let us show that $B$ is a simulation.
  \begin{itemize}
  \item If $\Proj{\GA}{n} \ltsarrow{\PSEND{a}{\SORTV}} \PV_1$
    then $\Proj{\GA}{n} \equiv \PSEND{a}{\SORTV} \PSEP \PV_1 \PINCH \PV_2 $
    and by Lemmas~\ref{lem:gvsproj} and~\ref{lem:gvssys}
    we have that $\restri{\SYSV}{n} = \PSEND{a}{\SORTV} \PSEP Q_1 \PINCH Q_2$
    and thus $\restri{\SYSV}{n} \ltsarrow{\PSEND{a}{\SORTV}} Q_1$

    Now we have to show that $(\PV_1 , Q_1) \in B$, i.e.
    \[
    \derive{\SYSV'}{\GA'} \text{ with } 
    \Proj{\GA'}{n} = \PV_1 \text{ and }
    \restri{\SYSV'}{n} = Q_1
    \]
    Pose
    \[\SYSV' \equiv
    \PBOX{n}{Q_1} \PPAR \QUEUE{a}{\SORTV} \PPAR
    \PPROD{\ptp{m}\neq \ptp{n} \in \PARTS{\SYSV}}{\restri{\SYSV}{m}}
    \]
    by Lemma~\ref{lem:subred}, $\derive{\SYSV'}{\GA'}$,
    and by Corollary~\ref{cor:subred}, we have
    $\Proj{\GA'}{n} = \PV_1$, as required.

  \item If $\Proj{\GA}{n} \ltsarrow{\PRECEIVE{a}{\SORTV}} \PV_1$
    then $\Proj{\GA}{n} \equiv \PRECEIVE{a}{\SORTV} \PSEP \PV_1 \POUTCH \PV_2 $
    and by Lemmas~\ref{lem:gvsproj} and~\ref{lem:gvssys}
    we have that $\restri{\SYSV}{n} = \PRECEIVE{a}{\SORTV} \PSEP Q_1 \POUTCH Q_2$
    and thus $\restri{\SYSV}{n} \ltsarrow{\PRECEIVE{a}{\SORTV}} Q_1$
    
    Now we have to show that $(\PV_1 , Q_1) \in B$, i.e.
    \[
    \derive{\SYSV'}{\GA'} \text{ with } 
    \Proj{\GA'}{n} = \PV_1 \text{ and }
    \restri{\SYSV'}{n} = Q_1
    \]
    Pose
    \[
    \SYSV \equiv 
    \PBOX{n}{\PRECEIVE{a}{\SORTV} \PSEP Q_1 \POUTCH Q_2} \PPAR
    \QUEUE{a}{\SORTV \cdot \rho} \PPAR
    \PPROD{\ptp{m}\neq \ptp{n} \in \PARTS{\SYSV}}{\restri{\SYSV}{m}}
    \]
    and 
    \[
    \SYSV' \equiv 
    \PBOX{n}{Q_1} \PPAR 
    \QUEUE{a}{\rho} \PPAR
    \PPROD{\ptp{m}\neq\ptp{n} \in \PARTS{\SYSV}}{\restri{\SYSV}{m}}
    \]
    
    by Lemma~\ref{lem:subred}, $\derive{\SYSV'}{\GA'}$,
    and by Corollary~\ref{cor:subred}, we have
    $\Proj{\GA'}{n} = \PV_1$, as required.
    
  \item If  $\Proj{\GA}{n} \ltsarrow{\labend} \PV_1$
    then $\Proj{\GA}{n} \equiv \PEND$ and $\PV_1 = \PEND$,
    thus $\ptp{n} \not\in \PARTS{\GA}$
    and by Lemma~\ref{lem:partis} this means that $\restri{\SYSV}{n} = \PEND$,
    we then have $\restri{\SYSV}{n} \ltsarrow{\labend} \PEND$.
  \end{itemize}
\end{proof}

\begin{lemma}\label{lem:sysproj}
  The following holds:
  \begin{enumerate}
  \item \label{en:lem:sysproj1}
    If $\derive{\PBOX{n}{\PSEND{a}{\SORTV} \PSEP \PV \PINCH \PV'} \PPAR \SYSV}{\GA}$
    then $\Proj{\GA}{n} \equiv \PSEND{a}{\SORTV} \PSEP Q \PINCH Q' $
   
  \item \label{en:lem:sysproj2}
    If $\derive{\PBOX{n}{\PRECEIVE{a}{\SORTV} \PSEP \PV \POUTCH \PV'} \PPAR \SYSV}{\GA}$
    then either $\Proj{\GA}{n} \equiv \PRECEIVE{a}{\SORTV} \PSEP Q \POUTCH Q'$,
    or  $\derive{\PBOX{n}{\PV'} \PPAR \SYSV}{\GA}$
  \end{enumerate}
\end{lemma}
\begin{proof}
  The proof of~\ref{en:lem:sysproj1} is by Lemma~\ref{lem:svsg},
  and the proof of~\ref{en:lem:sysproj2}
  is by Lemma~\ref{lem:svsg} and rule \rulename{$\gipoutch$}.
\end{proof}
\medskip

\thmbisim

\begin{proof}
For this proof we pose
$T = \PPROD{\ptp{m} \in \PARTS{\GA}}{\PBOX{m}{\Proj{\GA}{m}}} \PPAR
\PPROD{\ptp{b} \in \CHANS{\GA}}{\QUEUE{b}{\Projc{\GA}{b}}}$.

\proocase{$\SYSV$ sends}
Assume $\SYSV \equiv \SYSV_0 \PPAR \SYSV_1$,
$\SYSV_0 \ltsarrow{\labsend}$ and
$\SYSV_1 \ltsarrow{\labpush}$.
We have
\begin{equation}\label{eq:corr1}
  \SYSV \equiv \PBOX{n}{\PSEND{a}{\SORTV} \PSEP \PV \PINCH \PV'}
  \PPAR \QUEUE{a}{\rho}
  \PPAR \SYSV''
\end{equation}
and 
\begin{equation}\label{eq:corr2}
  \SYSV' \equiv \PBOX{n}{\PV}
  \PPAR \QUEUE{a}{\rho \cdot \SORTV}
  \PPAR \SYSV''
\end{equation}
By Lemma~\ref{lem:sysproj}, $\Proj{\GA}{n} \equiv \PSEND{a}{\SORTV} \PSEP Q \PINCH Q'$,
thus $\Proj{\GA}{n} \ltsarrow{\labsend}$,
and by Lemma~\ref{lem:qproj} $\Projc{\GA}{a} = \rho$.

We then have (note that $\chan{a} \in \CHANS{\GA}$)
\begin{equation}\label{eq:corr3}
  T \equiv
  \PBOX{n}{\PSEND{a}{\SORTV} \PSEP Q \PINCH Q'} \PPAR
  \QUEUE{a}{\rho} \PPAR
  \PPROD{\ptp{m} \neq \ptp{n} \in \PARTS{\GA}}{\PBOX{m}{\Proj{\GA}{m}}} \PPAR
  \PPROD{\ptp{b} \neq \chan{a} \in \CHANS{\GA} }{\QUEUE{b}{\Projc{\GA}{b}}}
\end{equation}
and by definition of $\ltsarrow{}$,
\begin{equation}\label{eq:corr4}
  T' \equiv
  \PBOX{n}{Q} \PPAR
  \QUEUE{a}{\rho \cdot \SORTV} \PPAR
  \PPROD{\ptp{m} \neq \ptp{n} \in \PARTS{\GA}}{\PBOX{m}{\Proj{\GA}{m}}} \PPAR
  \PPROD{\ptp{b} \neq \chan{a} \in \CHANS{\GA} }{\QUEUE{b}{\Projc{\GA}{b}}}
\end{equation}

Let us now show that
\begin{equation}\label{eq:corr5}
  \derive{\SYSV'}{\GA'} \text{ with }
  T' \equiv  \PPROD{\ptp{m} \in \PARTS{\GA'}}{\PBOX{m}{\Proj{\GA'}{m}}} \PPAR
  \PPROD{\ptp{b} \in \CHANS{\GA'}}{\QUEUE{b}{\Projc{\GA'}{b}}}
\end{equation}
By Lemma~\ref{lem:subred} we know that $\derive{\SYSV'}{\GA'}$,
we have that
$\Proj{\GA'}{n} \equiv Q$, by Corollary~\ref{cor:subred}, and
$\Projc{\GA'}{a} \equiv \rho \cdot \SORTV$ by Lemma~\ref{lem:qproj}.

\proocase{$T$ sends}
Assume $T \equiv T_0 \PPAR T_1$, $T_0 \ltsarrow{\labsend}$ and 
$T_1 \ltsarrow{\labpush}$. We have
\begin{equation}\label{eq:corrA}
  T \equiv
  \PBOX{n}{\PSEND{a}{\SORTV} \PSEP Q \PINCH Q'} \PPAR
  \QUEUE{a}{\rho} \PPAR
  \PPROD{\ptp{m} \neq \ptp{n} \in \PARTS{\GA}}{\PBOX{m}{\Proj{\GA}{m}}} \PPAR
  \PPROD{\ptp{b} \neq \chan{a} \in \CHANS{\GA}}{\QUEUE{b}{\Projc{\GA}{b}}}
\end{equation}
and
\begin{equation}\label{eq:corra}
  T' \equiv
  \PBOX{n}{Q} \PPAR
  \QUEUE{a}{\rho \cdot \SORTV} \PPAR
  \PPROD{\ptp{m} \neq \ptp{n} \in \PARTS{\GA}}{\PBOX{m}{\Proj{\GA}{m}}} \PPAR
  \PPROD{\ptp{b} \neq \chan{a} \in \CHANS{\GA}}{\QUEUE{b}{\Projc{\GA}{b}}}
\end{equation}

By Lemma~\ref{lem:simu}, we have $\restri{\SYSV}{n} \ltsarrow{\labsend}$,
and since, by Lemma~\ref{lem:chans}, $\chan{a} \in \CHANS{\GA} \Rightarrow \chan{a} \in \CHANS{\SYSV}$, there is a queue $\chan{a}$ in $\SYSV$. Note that a queue $\chan{a}$ can always make
a transition $\QUEUE{a}{\rho'} \ltsarrow{\labpush}$ (regardless of $\rho'$).
By Lemma~\ref{lem:qproj}, $\restric{\SYSV}{a} = \rho$.

Therefore, we must have
\begin{equation}
  \label{eq:corrb}
  \SYSV \equiv
  \PBOX{n}{\PSEND{a}{\SORTV} \PSEP \PV \PINCH \PV'} \PPAR
  \QUEUE{a}{\rho} \PPAR \SYSV''
\end{equation}

And by definition of $\ltsarrow{}$, we have
\begin{equation}
  \label{eq:corrc}
  \SYSV' \equiv
  \PBOX{n}{\PV} \PPAR
   \QUEUE{a}{\rho \cdot \SORTV } \PPAR \SYSV''
\end{equation}

Finally, we have
\begin{equation}
  \label{eq:corrd}
  \derive{\SYSV'}{\GA'}
  \text{ with }
  T' \equiv
  \PPROD{\ptp{m} \in \PARTS{\GA'}}{\PBOX{m}{\Proj{\GA'}{m}}} \PPAR
  \PPROD{\ptp{b} \in \CHANS{\GA'}}{\QUEUE{b}{\Projc{\GA'}{b}}}
\end{equation}
by Lemma~\ref{lem:subred}.

\proocase{$\SYSV$ receives}
Assume $\SYSV \equiv \SYSV_0 \PPAR \SYSV_1$,
$\SYSV_0 \ltsarrow{\labrec}$ and
$\SYSV_1 \ltsarrow{\labpop}$.
We have
\begin{equation}\label{eq:corr11}
  \SYSV \equiv \PBOX{n}{\PRECEIVE{a}{\SORTV} \PSEP \PV \POUTCH \PV'}
  \PPAR \QUEUE{a}{\SORTV \cdot \rho}
  \PPAR \SYSV''
\end{equation}
and 
\begin{equation}\label{eq:corr12}
  \SYSV' \equiv \PBOX{n}{\PV}
  \PPAR \QUEUE{a}{\rho}
  \PPAR \SYSV''
\end{equation}
By Lemma~\ref{lem:sysproj} and~\eqref{eq:corr11}, we have either
\begin{equation}
  \label{eq:corr13}
  \Proj{\GA}{n} \equiv \PRECEIVE{a}{\SORTV} \PSEP Q \POUTCH Q'
\end{equation}
or
\begin{equation}
  \label{eq:corr14}
  \derive{
    \PBOX{n}{\PV'}
    \PPAR \QUEUE{a}{\SORTV \cdot \rho}
    \PPAR \SYSV''
  }
  {\GA}
\end{equation}
However, by assumption we have $\derive{\SYSV}{\GA}$, with $\SYSV$ as in~\eqref{eq:corr11},
therefore~\eqref{eq:corr14} cannot hold by Lemma~\ref{lem:qbranch}.

By Lemma~\ref{lem:qproj}, we have that $\Projc{\GA}{a} \ltsarrow{\labpop}$
and by~\eqref{eq:corr13}, $\Proj{\GA}{n} \ltsarrow{\PRECEIVE{a}{\SORTV}}$.
By definition of $\ltsarrow{}$, we have
\begin{equation}
  \label{eq:corr15}
  T \equiv
  \PBOX{n}{\PRECEIVE{a}{\SORTV} \PSEP Q \PINCH Q'} \PPAR
  \QUEUE{a}{\SORTV \cdot \rho} \PPAR
  \PPROD{\ptp{m} \neq \ptp{n} \in \PARTS{\GA}}{\PBOX{m}{\Proj{\GA}{m}}} \PPAR
  \PPROD{\ptp{b} \neq \chan{a} \in \CHANS{\GA}}{\QUEUE{b}{\Projc{\GA}{b}}}
\end{equation}

Let us now show that
\begin{equation}
  \label{eq:corr16}
  \derive{\SYSV'}{\GA'}
  \text{ with }
  T' \equiv
  \PPROD{\ptp{m} \in \PARTS{\GA'}}{\PBOX{m}{\Proj{\GA'}{m}}} \PPAR
  \PPROD{\ptp{b} \in \CHANS{\GA'}}{\QUEUE{b}{\Projc{\GA'}{b}}}
\end{equation}
By Lemma~\ref{lem:subred} we know that $\derive{\SYSV'}{\GA'}$,
we have that
$\Proj{\GA'}{n} \equiv Q$, by Corollary~\ref{cor:subred}, and
$\Projc{\GA'}{a} \equiv \SORTV \cdot \rho$ by Lemma~\ref{lem:qproj}.

\proocase{$T$ receives}
Assume $T \equiv T_0 \PPAR T_1$, $T_0 \ltsarrow{\labrec}$ and 
$T_1 \ltsarrow{\labpop}$. We have
\begin{equation}\label{eq:corrAR}
  T \equiv
  \PBOX{n}{\PRECEIVE{a}{\SORTV} \PSEP Q \PINCH Q'} \PPAR
  \QUEUE{a}{\SORTV \cdot \rho} \PPAR
  \PPROD{\ptp{m} \neq \ptp{n} \in \PARTS{\GA}}{\PBOX{m}{\Proj{\GA}{m}}} \PPAR
  \PPROD{\ptp{b} \neq \chan{a} \in \CHANS{\GA}}{\QUEUE{b}{\Projc{\GA}{b}}}
\end{equation}
and
\begin{equation}\label{eq:corraR}
  T' \equiv
  \PBOX{n}{Q} \PPAR
  \QUEUE{a}{\rho} \PPAR
  \PPROD{\ptp{m} \neq \ptp{n} \in \PARTS{\GA}}{\PBOX{m}{\Proj{\GA}{m}}} \PPAR
  \PPROD{\ptp{b} \neq \chan{a} \in \CHANS{\GA}}{\QUEUE{b}{\Projc{\GA}{b}}}
\end{equation}
\end{proof}
By Lemma~\ref{lem:simu}, we have $\restri{\SYSV}{n} \ltsarrow{\PRECEIVE{a}{\SORTV}}$,
and by Lemma~\ref{lem:qproj}, we have $\restric{\SYSV}{a} = \rho$,
therefore
\begin{equation}
  \label{eq:corr21}
  \SYSV \equiv \PBOX{n}{\PRECEIVE{a}{\SORTV} \PSEP \PV \POUTCH \PV'}
  \PPAR \QUEUE{a}{\SORTV \cdot \rho}
  \PPAR \SYSV''
\end{equation}
and 
\begin{equation}
  \label{eq:corr22}
  \SYSV \equiv \PBOX{n}{\PV}
  \PPAR \QUEUE{a}{\rho}
  \PPAR \SYSV''
\end{equation}

We now have to show that
\begin{equation}
  \label{eq:corr23}
  \derive{\SYSV'}{\GA'}
  \text{ with }
  T' \equiv
  \PPROD{\ptp{m} \in \PARTS{\GA'}}{\PBOX{m}{\Proj{\GA'}{m}}} \PPAR
  \PPROD{\ptp{b} \in \CHANS{\GA'}}{\QUEUE{b}{\Projc{\GA'}{b}}}
\end{equation}
as before, we have $\Proj{\GA'}{n} = \PV$ by Corollary~\ref{cor:subred} and
$\Projc{\GA'}{a} = \rho$ by Lemma~\ref{lem:qproj}.

\proocase{End}
If $\SYSV \ltsarrow{\labend} \SYSV$, then $\SYSV \equiv \PEND$ and $\GA \equiv \GEND$,
and vice versa if  $T \ltsarrow{\labend} T$.

\begin{lemma}\label{lem:qbranch}

  If 
  \[
  \derive{
    \PBOX{r}{\PRECEIVE{a}{\SORTV} \PSEP \PV \POUTCH \PV'} \PPAR 
    \QUEUE{a}{\SORTV \cdot \rho} \PPAR
    \SYSV}{\GA}
  \]
  is derivable then
  \[
  \derive{\PBOX{r}{\PV'} \PPAR 
    \QUEUE{a}{\SORTV \cdot \rho}\PPAR
    \SYSV}{\GA'}
  \] is \emph{not} derivable.
\end{lemma}

\begin{proof}
  Assume $\PV' \equiv \PRECEIVE{b}{\SORTV} \PSEP \PV''$.
  We show that we must have
  \[
  \GA \equiv (\GSEND{\GSTAR}{r}{a}{\SORTV} \GSEP \GA_0 \GPAR \GA_1) \GWS \GA_2
  \]
  and the derivation of $\GA$ must have the following form
  \[
  \inferrule* [Left=\rulename{$\gigws$}]
  {
    \inferrule* [Left=\rulename{$\gipar$}]
    {
      \inferrule* [Left=\rulename{$\gipoutch$}]
      {
        \inferrule* [Left=\rulename{$\gipsep$}]
        {
          \inferrule*
          {\vdots}
          {
            \judgeC{\chanset_0}{\emptyctx}{C}{
              \PBOX{r}{\PV_0} \PPAR 
              \QUEUE{a}{\rho_0} \PPAR
              \SYSV_{00}}{\GA_0} 
          }
        }
        {
          \judgeC{\chanset_0}{\emptyctx}{C}{
            \PBOX{r}{\PRECEIVE{a}{\SORTV} \PSEP \PV_0} \PPAR 
            \QUEUE{a}{\SORTV \cdot \rho_0} \PPAR
            \SYSV_{00}}{\GSEND{\GSTAR}{r}{a}{\SORTV} \GSEP \GA_0}
        }
      }
      {
        \judgeC{\chanset_0}{\emptyctx}{C}{
          \PBOX{r}{\PRECEIVE{a}{\SORTV} \PSEP \PV_0 \POUTCH \PV'_0} \PPAR 
          \QUEUE{a}{\SORTV \cdot \rho_0} \PPAR
          \SYSV_{00}}{\GSEND{\GSTAR}{r}{a}{\SORTV} \GSEP \GA_0}
      }
      \\
      \inferrule*
      {\vdots}
      {
        \judgeC{\chanset_1}{\emptyctx}{C}{\SYSV_{01}}{\GA_1}
      }
    }
    {
      \judge{\emptyctx}{C}{
        \PBOX{r}{\PRECEIVE{a}{\SORTV} \PSEP \PV_0 \POUTCH \PV'_0} \PPAR 
        \QUEUE{a}{\SORTV \cdot \rho_0} \PPAR
        \SYSV_0}{\GSEND{\GSTAR}{r}{a}{\SORTV} \GSEP \GA_0 \GPAR \GA_1}
    }
    \\
    \inferrule*
    {\vdots}
    {\GA_2}
  }
  {
    \derive{
      \PBOX{r}{\PRECEIVE{a}{\SORTV} \PSEP \PV \POUTCH \PV'} \PPAR 
      \QUEUE{a}{\SORTV \cdot \rho} \PPAR
      \SYSV}{\GA}
  }
  \]
where we must have
\begin{itemize}
\item $\onestep{\SYSV}$ or $\GA_1 = \GA_2 = \GEND$ and
  $\onestep{\SYSV_0}$ or $\SYSV_{01} \equiv \PEND$.
\item $\Ssplit{\SYSV} = (\PBOX{r}{\PRECEIVE{a}{\SORTV} \PSEP \PV_0 \POUTCH \PV'_0} \PPAR 
  \QUEUE{a}{\SORTV \cdot \rho_0} \PPAR
  \SYSV_0 , \_)$ (the second part of the split does not matter)
\item $\coname{b} \not\in \fire{\SYSV}$ (by Lemma~\ref{lem:twooutsQ})
\item $\chanset_0 \cap \chanset_1 = \varnothing$ and $\chan{b} \not\in \chanset_2$
  because of rules \rulename{$\gipar$} and \rulename{$\gipoutch$}
\end{itemize}

Now let us discuss a derivation for $\GA'$.
Since we have  $\coname{b} \not\in \fire{\SYSV}$, we must have
$\Ssplit{\PBOX{r}{\PV'} \PPAR 
\QUEUE{a}{\SORTV \cdot \rho} \PPAR
\SYSV} = (\SYSV_1 , \SYSV_2)$ such that $\restri{\SYSV_1}{r} = \PV'$,
and $\restric{\SYSV_1}{a} = \SORTV \cdot \rho$,
if the split does exist.
If it does, we have
$\spjudge{\Psi}{\Theta}{\PBOX{r}{\PV'} \PPAR 
  \QUEUE{a}{\SORTV \cdot \rho}\PPAR
  \SYSV}{\Omega}$ such that $\forall N \in \Psi \, . \, \ptp{r} \not\in N$. Therefore, the
split for the rest of the system is the same as in the other derivation.

Again, we can divide the system using \rulename{$\gipar$} if need be such that we get
\[
\judgeC{\chanset_0}{\emptyctx}{C}{
  \PBOX{r}{\PV'_0} \PPAR 
  \QUEUE{a}{\SORTV \cdot \rho_0} \PPAR
  \SYSV_{00}}{\GSEND{\GSTAR}{r}{a}{\SORTV} \GSEP \GA_0}
\]
with $\noonestep{\SYSV_{00}}$ therefore no rule is applicable for
this judgement, and the derivation does not exist.

\end{proof}

\begin{lemma}\label{lem:twoouts}
  Let $\SYSV$ a system such $\noonestep{\SYSV}$, and
  $ \chan{a,b} \not\in \fire{\SYSV}$
  the following is \emph{not} derivable
  \[
  T \equiv
    \PBOX{s_1}{\PSEND{a}{\SORTV} \PSEP \PV_1} \PPAR
    \PBOX{s_2}{\PSEND{b}{\SORTV'} \PSEP \PV_2} \PPAR
    \PBOX{r}{\PRECEIVE{a}{\SORTV} \PSEP Q \POUTCH \PRECEIVE{b}{\SORTV'} \PSEP Q '} \PPAR
    \SYSV
  \]
\end{lemma}
\begin{proof}
  We show this by contradiction.
  Given $T$ as above, the only rule applicable is \rulename{$\gipoutch$} on $\ptp{r}$
  either selecting the branch on $\chan{a}$ or on $\chan{b}$.
  Therefore, the following should be derivable
  \[
  \inferrule* [Left=\rulename{$\gipoutch$}]
  {
    \inferrule* [Left=\rulename{$\gipsep$}]
    {
      \inferrule
      {
        \vdots
      }
      {
        \derive{ \PBOX{s_1}{\PV_1} \PPAR
          \PBOX{s_2}{\PSEND{b}{\SORTV'} \PSEP \PV_2} \PPAR
          \PBOX{r}{Q} \PPAR
          \SYSV}
        {\GA}
      }
    }
    {
      \derive{ \PBOX{s_1}{\PSEND{a}{\SORTV} \PSEP \PV_1} \PPAR
        \PBOX{s_2}{\PSEND{b}{\SORTV'} \PSEP \PV_2} \PPAR
        \PBOX{r}{\PRECEIVE{a}{\SORTV} \PSEP Q} \PPAR
        \SYSV}
      {\GSEND{s_1}{r}{a}{\SORTV} \GSEP \GA}
    }
  }
{
  \derive{T}{\GSEND{s_1}{r}{a}{\SORTV} \GSEP \GA}
}
\]
and, we must have $\chan{a, b} \in \chanset$ and
$\onestep{\PBOX{s_1}{\PV_1} \PPAR
          \PBOX{r}{Q} \PPAR
          \SYSV}$.
And the other derivation as the form:
\[
\inferrule* [Left=\rulename{$\gipoutch$}]
{
  \inferrule* [Left=\rulename{$\gipsep$}]
  {
    \inferrule
    {
      \vdots
    }
    {
      \derive{
        \PBOX{s_1}{\PSEND{a}{\SORTV} \PSEP\PV_1} \PPAR
        \PBOX{s_2}{\PV_2} \PPAR
        \PBOX{r}{Q} \PPAR
        \SYSV}
      {\GA}
    }
  }
  {
    \derive{
      \PBOX{s_1}{\PSEND{a}{\SORTV} \PSEP \PV_1} \PPAR
      \PBOX{s_2}{\PSEND{b}{\SORTV'} \PSEP \PV_2} \PPAR
      \PBOX{r}{\PRECEIVE{b}{\SORTV'} \PSEP Q'} \PPAR
      \SYSV}
    {\GSEND{s_2}{r}{b}{\SORTV'} \GSEP \GA'}
  }
}
{
  \derive{T}{\GSEND{s_2}{r}{b}{\SORTV'} \GSEP \GA'}
}
\]
where we have $\chan{a, b} \in \chanset$ and
$\onestep{\PBOX{s_2}{\PV_2} \PPAR
  \PBOX{r}{Q} \PPAR
  \SYSV}$.
However, this is clearly in contradiction with Theorem~\ref{thm:RUNIQUE}, i.e.\
\[
\GSEND{s_1}{r}{a}{\SORTV} \GSEP \GA
\not\equiv
\GSEND{s_2}{r}{b}{\SORTV'} \GSEP \GA'
\]
\end{proof}

\begin{lemma}\label{lem:twooutsQ}
  Let $\SYSV$ a system such $\noonestep{\SYSV}$, and
  $ \chan{a,b} \not\in \fire{\SYSV}$
  the following is \emph{not} derivable
  \[
  T \equiv
  \PBOX{s_1}{\PSEND{a}{\SORTV} \PSEP \PV_1} \PPAR
  \QUEUE{b}{\SORTV' \cdot \rho} \PPAR
  \PBOX{r}{\PRECEIVE{a}{\SORTV} \PSEP Q \POUTCH \PRECEIVE{b}{\SORTV'} \PSEP Q '} \PPAR
  \SYSV
  \]
\end{lemma}
\begin{proof}
  The proof is similar to the one of Lemma~\ref{lem:twooutsQ}
  where $\QUEUE{b}{\SORTV' \cdot \rho}$ replaces  
  $\PBOX{s_2}{\PSEND{b}{\SORTV'} \PSEP \PV_2}$.
\end{proof}

\section{Proofs for Theorem~\ref{thm:RCOMPL} (Completeness wrt $\GA$)}
\thmcomplete

\begin{proof}
By Lemma~\ref{lem:complete}, with $P = \PARTS{\GA}$ and $\Gamma = \emptyctx$
since $\GA$ is closed by assumption.
\end{proof}

\newcommand{\allproj}[2]{\mathtt{Proj}(#1,#2)}
\begin{lemma}\label{lem:complete}

  Let 
  $
  \allproj{\GA}{P} = \GPROD{\ptp{n} \in P}{\PBOX{n}{\Proj{\GA}{n}}}
  $ with $\PARTS{\GA} \subseteq P$, and if $\Proj{\GA}{n} = \PEND$, then
  $\ptp{n}$ is not in $P$.
  
  If $\wfjudge{C}{\GA}$, $\GA$ is projectable,
  and $\forall \, \GRECV \in \freen{\GA} \, . \,
  \forall \, \ptp{n} \in P \, . \, 
  \exists \, (\ptp{n},\GRECV) : \GRECV \in \Gamma$
  then
  \[
  \judgeC{\CHANS{\GA}}{\Gamma}{C}{\allproj{\GA}{P}}{\GA'}
  \quad  \text{with } \GA \equiv \GA'
  \]

\end{lemma}
\newcommand{\myset}[1]{\SYSV_{\{\ptp{#1}\}}}
\begin{proof}
  We show this by induction on the structure of $\GA$. Let
  \[
  \SYSV \equiv
  \GPROD{\ptp{n} \in P}{\Proj{\GA}{n}}
  \qquad
  \SYSV_Q \equiv
  \GPROD{\ptp{n} \in P \setminus Q}{\Proj{\GA}{n}}
  \qquad
  \chanset = \CHANS{\GA}
  \]

  \proocase{$\GA \equiv \GSENDN \GSEP \GA_0 \GCH \GA_1$}
  
  By definition of projection, we have
  \[
  \SYSV \equiv
  \PBOX{s}{\PSEND{a}{\SORTV} \PSEP \Proj{\GA_0}{s} \PINCH \Proj{\GA_1}{s}}
  \PPAR
  \PBOX{r}{\PRECEIVE{a}{\SORTV} \PSEP  \Proj{\GA_0}{s} \POUTCH \Proj{\GA_1}{s}}
  \PPAR \myset{s,r}
  \]
  We can apply rules \rulename{$\gipinch$},
 \rulename{$\gipoutch$} (twice) and
 \rulename{$\gipsep$} in order to have the result, i.e.
 
{\small
 \[
 \inferrule*  %
 {
   \inferrule* [Left=\rulename{$\gipsep$}]
   {
     \inferrule* 
     {\text{by IH}}
     {
       \judge{\Gamma}{C_a}{
         \PBOX{s}{\Proj{\GA_0}{s}}
         \PPAR
         \PBOX{r}{\Proj{\GA_0}{s}}
         \PPAR \myset{s,r}
       }{\GA_0}
     }
   }
   {
     \judge{\Gamma}{C}{
       \PBOX{s}{\PSEND{a}{\SORTV} \PSEP \Proj{\GA_0}{s}}
       \PPAR
       \PBOX{r}{\PRECEIVE{a}{\SORTV} \PSEP  \Proj{\GA_0}{s}}
       \PPAR \myset{s,r}
   }{\GA'}
   }
   \\
   \inferrule*
   {\text{by IH}}
   {
     \judge{\Gamma}{C}{
       \allproj{\GA_1}{P}
     }{\GA_1}
   }
 }
 {
   \judge{\Gamma}{C}{\SYSV}{\GA}
 }
 \]
 }
 with $\GA' = \GSENDN \GSEP \GA_0$, and $C_a = \addchan{C}{\chanTN}$ note that the later
 is defined by $\wfjudge{C}{\GA}$.
 Observe that $\noonestep{\myset{s,r}}$ otherwise that would mean that
 $\exists \, \chanT{b}{s'}{r'} \in \GFIN{\GA_0}$ such that
 $\{\ptp{s,r}\} \cap \{\ptp{s',r'}\} = \varnothing$ which is in contradiction with
 $\wfjudge{C}{\GA}$.
 Finally, it is obvious that
 \[
 \chan{a} \in \chanset
 \qand
 \firebar{\Proj{\GA_i}{r}} \subseteq \chanset
 \quad i \in \{0,1\}
 \]
 since $\chanset = \CHANS{\GA}$.

 \medskip
 \proocase{$\GA \equiv \GA_0 \GPAR \GA_1$}

 We have $\SYSV$ of the form, by definition of projections
 (and well-formedness)
 \[
 \SYSV \equiv \allproj{\GA_0}{P_0} \PPAR \allproj{\GA_1}{P_1}
 \quad \text{with } P_0 \cap P_1 = \varnothing
 \]
 Note that since $\#(\GFOUTS{\GA}) > 1$, we have $\freen{\GA} = \varnothing$,
 therefore, by IH, we have
 \[
 \judgeC{\chanset_i}{\emptyctx}{C}{
   \allproj{\GA_i}{P}
 }{\GA'_i}
 \qand \GA_i \equiv \GA'_i
 \qand \chanset_i = \CHANS{\GA_i}
 \]

 We have the result by applying rule \rulename{$\gipar$}:

 \[
 \inferrule
 {
   \inferrule
   {\text{by IH}}
   {
     \judgeC{\chanset_0}{\emptyctx}{C}{
       \allproj{\GA_0}{P_0}
     }{\GA_1}
   }
   \\
   \inferrule
   {
     \text{by IH}
   }
   {
     \judgeC{\chanset_1}{\emptyctx}{C}{
       \allproj{\GA_1}{P_1}
     }{\GA_1}
   }
 }
 {
   \judgeC{\chanset_1 \cup \chanset_2}{\Gamma}{C}{S}{\GA}
 }
 \]
 By well-formedness we have, $\chanset_0 \cap \chanset_1 = \varnothing$,
 and Lemmas~\ref{lem:chans} and~\ref{lem:chans} guarantee that each $\chanset_i$
 is large enough.

 \medskip
 \proocase{$\GA \equiv \GRECN \GA'$}
 
 By definition of projections, we have
 \[
 \SYSV \equiv \PPROD{\ptp{n} \in P}{\PBOX{n}{\PRECE \GRECV . \Proj{\GA'}{n}}}
 \]
 
 Since $\GA'$ is prefix-guarded, there must be $\ptp{s,r} \in \PARTS{\GA'}$ such 
 that 
 \[
 \onestep{\PBOX{s}{\PRECE \GRECV . \Proj{\GA'}{s}} \PPAR 
   \PBOX{r}{\PRECE \GRECV . \Proj{\GA'}{r}}}
 \]

 Therefore, rule \rulename{$\gimua$} is applicable here
 \[
 \inferrule
 {
   \judge
   {\Gamma'
   }
   {C}{
     \PBOX{s}{\Proj{\GA'}{s}} \PPAR 
     \PBOX{r}{\Proj{\GA'}{r}} \PPAR
     \PPROD{\ptp{n} \in P \setminus \{\ptp{s,r}\}}{\PBOX{n}{\Proj{\GA'}{n}}}
   }
   {\GA'}
 }
 {
   \judge{\Gamma}{C}{
     \PBOX{s}{\PRECE \GRECV . \Proj{\GA'}{s}} \PPAR 
     \PBOX{r}{\PRECE \GRECV . \Proj{\GA'}{r}} \PPAR
     \myset{s,r}
   }
   {\GA}
 }
 \]
 where
 \[
 \Gamma' = \Gamma \cdot (\ptp{s},\GRECV) : \GRECV), (\ptp{r},\GRECV) : \GRECV \cdot \Gamma_S
 \qand
 \Gamma_S \text{ such that } \forall \, \ptp{n} \in \PARTS{\myset{s,r}} . \Gamma(\ptp{n},\GRECV) = \GRECV
 \]
 The rest follows by induction hypothesis.

 \medskip
 \proocase{$\GA = \GRECV$}
 Then, we have
 \[
 \SYSV \equiv \PPROD{\ptp{n} \in P}{\PBOX{n}{\GRECV}}
 \]
 By assumption, we have 
 $\forall \, \ptp{n} \in P \, . \, 
 \exists \, (\ptp{n},\GRECV) : \GRECV \in \Gamma$, hence we can apply rule \rulename{$\givar$}
 and we are done.

 \medskip
 \proocase{$\GA \equiv \GA_0 \GWS \GA_1$}
 Then we have
 \[
 \SYSV \equiv \PPROD{\ptp{n} \in P}{\Proj{\GA_0}{n}\subs{\Proj{\GA_1}{n}}{\PEND}}
 \]
 Let us show that $\Ssplit{\SYSV} = (\SYSV_0, \SYSV_1)$
 and
 \[
 \SYSV_0 \equiv \PPROD{\ptp{n} \in P}{\Proj{\GA_0}{n}} = \allproj{\GA_0}{P}
 \qand
 \SYSV_1 \equiv \PPROD{\ptp{n} \in P}{\Proj{\GA_1}{n}} = \allproj{\GA_1}{P}
 \]
 Since $\GA$ is well-formed, we have $\spjudge{\Psi}{\Theta}{\SYSV}{\Omega}$ coherent
 if we pose
 \[
 \Psi = \{ N | N \in \GFINP{\GA_0}\} 
 \qand
 \Theta = \{ \{\ptp{s,r}\} | \chanTN \, \in \GFIN{\GA_1} \}
 \]
 For each $\GA_i$ a top level concurrent branch of $\GA_0$, 
 we have that
\[
\spjudge{\{\PARTS{\GA_i}\}}{\varnothing}{\allproj{\GA_i}{\PARTS{\GA_i}}}{\Omega_i}
\]
 is derivable  by Lemma~\ref{lem:allinpsi}, since $ \allproj{\GA_i}{\PARTS{\GA_i}}$
 is typable by IH. In addition, we have
 $\Proj{\GA_i}{n} = \Omega_i(\ptp{n}) \subs{\PEND}{\epsilon}$.
 
 By construction, we have
 \[
 \spjudge{\Psi'}{\Theta}{\allproj{\GA_0}{P}}{\Omega'}
 \quad \text{with } \forall \ptp{n} \in P \, . \,\Omeganp{n} = \epsilon
 \]
 with $\Psi ' = \{ N' | \exists N \in \Psi \st N' \subseteq N\}$.

 Finally, we have $\forall \ptp{n} \in P \, . \, \fsplit{\restric{\SYSV}{n}}{\Omegan{n}} \neq \nada$
 since the same suffix $\Proj{\GA_1}{n}$ is added to each branch of a behaviour,
 and  $\fsplit{\restric{\SYSV}{n}}{\Omegan{n}} = \Proj{\GA_1}{n}$.
 We have the required result by IH and rule \rulename{$\gigws$}.

 \medskip
 \proocase{$\GA \equiv \GEND$}
 
 We have 
 \[
 \SYSV \equiv \PPROD{\ptp{n} \in P}{\PBOX{n}{\PEND}}
 \]
 and the results holds by rule \rulename{$\giend$}.
 
\end{proof}

\section{Accessory Results}
\subsection{Linearity}

\begin{lemma}\label{lem:linpref}
  If $\addchan{C}{\chanTN}$ is defined, then
  either
  \begin{enumerate}
  \item\label{en:lem:linpref1} $\chanTA \not\in C$,
  \item\label{en:lem:linpref2} $\subtreeC{\chanTA} = \onekidR{\chanTN}{C'}$,
  \item\label{en:lem:linpref3}  $\subtreeC{\chanTA} = \onekidR{\chanT{a}{\GSTAR}{r}}{C'}$, or
  \item\label{en:lem:linpref4} $\exists \; \chanT{\_}{\_}{r}$ and  $\chanT{\_}{\_}{s} \in C$
  \end{enumerate}
\end{lemma}
\begin{proof}
  The proofs of \ref{en:lem:linpref1} and \ref{en:lem:linpref2}
  follow directly from Def.~\ref{def:linearity} and the definition
  of $\subtreeC{\_}$.
  The proof of \ref{en:lem:linpref3} follows by definition of $\subtreeC{\_}$
  and the fact that, by Def.~\ref{def:linrelations},
  there cannot be an output dependency from
  $\chanT{b}{s'}{r'}$ and $\chanT{a}{\GSTAR}{r}$ since
  \[
  \chanT{b}{s}{r} \not\linOO \chanT{a}{\GSTAR}{r} 
  \quad \text{and} \quad
  \chanT{b}{s}{r'} \not\linIO \chanT{a}{\GSTAR}{r}
  \quad
  \text{since } \ptp{s} \neq \GSTAR \neq \ptp{r}
  \]
  Therefore,  $\chanT{a}{\GSTAR}{r}$ must be the first prefix with label $\chanTA$ in $C$.
  The proof of~\ref{en:lem:linpref4} follows from the Def.~\ref{def:linearity}.
  In fact, since whenever the sender/receiver are different on two nodes with common
  channel there must be two dependency relation we have the following cases.
  In the following, we assume that there is no prefix on $\chan{a}$ in the ellipsis.
  If only the senders are different, i.e.\ the following appears on a path in $C$
  \[
  \chanT{a}{s'}{r} \ldots \chanT{a}{s}{r}
  \]
  then we have $\chanT{a}{s'}{r} \linII \chanT{a}{s}{r}$ and we must have at least
  a node  between the two such that, e.g.\ 
  $\chanT{a}{s'}{r} \linOO \chanT{b}{s'}{s} \linIO \chanT{a}{s}{r}$
  and we have the result. If only the receivers are different, we have
  \[
  \chanT{a}{s}{r'} \ldots \chanT{a}{s}{r}
  \]
  and we have $\chanT{a}{s}{r'} \linOO \chanT{a}{s}{r}$ and we must have at least
  one node between the two such that
  $\chanT{a}{s}{r'} \linIO \chanT{b}{r'}{r} \linII \chanT{a}{s}{r}$,
  and we have the result.
  If both sender and receiver are different, i.e.
  \[
  \chanT{a}{s'}{r'} \ldots \chanT{a}{s}{r}
  \]
  then we need two nodes $\nodeTN_1$ and $\nodeTN_2$ such that, 
  ($i$) $\lab{\nodeTN_1} = \chanT{b}{\_}{r}$, otherwise there would be a
  $\linII$ relation in the input dependency,
  ($ii$) $\lab{\nodeTN_2} = \chanT{c}{\_}{s}$ otherwise there would be a
  $\linIO$ relation in the output dependency (note that $\linOO$ is only defined
  if the channels are the same).
  In fact, there must an input dependency, e.g.
  \[
  \chanT{a}{s'}{r'} \linIO
  \chanT{b_1}{r'}{s_1} \linIO
  \chanT{b_2}{s_1}{s_2} \linIO
  \chanT{b_3}{s_2}{r} \linII
  \chanT{a}{s}{r}
  \]
  and an output dependency, e.g.\
  \[
  \chanT{a}{s'}{r'} \linIO
  \chanT{c}{r'}{s} \linIO
  \chanT{a}{s}{r}
  \]
  Observe that, in the first chain, we have  $\chanT{b_3}{s_2}{r}$,
  and $\chanT{c}{r'}{s}$ in the second.
  Actually, the shortest (input) chain when both pair of participants are different is
  \[
  \chanT{a}{s'}{r'} \linIO
  \chanT{c_1}{r'}{s} \linIO
  \chanT{c_2}{s}{r} \linII
  \chanT{a}{s}{r}
  \]
  where we also have
  $\chanT{c_1}{r'}{s} \linIO \chanT{a}{s}{r}$, for the output chain. Notice that in this
  case we have  $\chanT{c_1}{r'}{s}$ and  $\chanT{c_2}{s}{r}$ in $C$.
\end{proof}

\begin{corollary}\label{lem:linstar}
  If linearity holds on $C$ and $\chanT{a}{\GSTAR}{n} \in C$, then
  \[
  \subtreeC{\chanT{a}{\_}{\_}} = \onekidR{\chanT{a}{\GSTAR}{n}}{C'}
  \]
\end{corollary} 

\begin{lemma}\label{lem:linweak}
  
  If $\wfjudge{\addchan{C}{\chanT{a}{\GSTAR}{r}}}{\GA}$
  then  $\wfjudge{C}{\GA}$
  and
  if $\judgeC{\Gamma}{\addchan{C}{\chanT{a}{\GSTAR}{r}}}{\SYSV}{\GA}$
  then $\judgeC{\Gamma}{C}{\SYSV}{\GA}$.

\end{lemma}
\begin{proof}
  We have to show that if $\chanT{a}{\GSTAR}{r}$ is involved in a input or output
  dependency, then there is another dependency between the same two nodes
  without $\chanT{a}{\GSTAR}{r}$.
  Note first that by Lemma~\ref{lem:linpref}, we know that
  $\chanT{a}{\GSTAR}{r} \not\in C$, therefore, if there is a 
  need to have a chain of the form
  \[
  \chanT{a}{\GSTAR}{r} \linrel{\_} \ldots \linrel{\_} \chanT{a}{s'}{r'}
  \]
  this need disappears with $\chanT{a}{\GSTAR}{r}$ (dependencies are needed 
  only between nodes with a common channel). Thus, if
  $\chanT{a}{\GSTAR}{r}$ is the first node in a dependency chain, then
  the result holds trivially.
  Observe that $\chanT{a}{\GSTAR}{r} \not\linOO \chanT{b}{s}{r'}$ and
  $\chanT{b}{s}{r'} \not\linOO \chanT{a}{\GSTAR}{r}$ for any $\chan{a}$, $\chan{b}$ since
  $\GSTAR \neq \ptp{s}$, for the same reason, we have
  $\chanT{b}{s}{r} \not\linIO \chanT{a}{\GSTAR}{r}$.

  Finally, we have the following cases, where the left hand side describes
  the dependency involving $\chan{a}$ and the right hand side shows
  that the dependency between the two external nodes still exists without the
  node on $\chan{a}$.
  \[
  \chanT{b}{s}{r} \linII \chanT{a}{\GSTAR}{r} \linII  \chanT{c}{s'}{r}
  \quad \Rightarrow \quad
  \chanT{b}{s}{r} \linII  \chanT{c}{s'}{r}
  \]

  \[
  \chanT{b}{s}{r} \linII \chanT{a}{\GSTAR}{r} \linIO  \chanT{c}{r}{s'}
  \quad \Rightarrow \quad
  \chanT{b}{s}{r} \linIO  \chanT{c}{r}{s'}
  \]
\end{proof}
\subsection{Splitting systems}

\begin{lemma}\label{lem:sppsitheta}\label{lem:spresult}
\[
\{\ptp{n,m}\} \subseteq \Theta 
\Rightarrow
\exists N \neq M \in \Psi \st \ptp{n} \in N \land \ptp{m} \in M
\]
\end{lemma}
\begin{proof}
  Direct from rules \rulename{$\splepsilon$} and  \rulename{$\splax$}.
\end{proof}

\begin{lemma}\label{lem:spiff}
  If $\derive{\SYSV}{\GA}$ and
  $\spjudge{\Psi}{\Theta}{\SYSV}{\Omega}$ is derivable and coherent,
  then
  \[
  (\ptp{n,m}) \in \spR
  \iff
  \{\ptp{n,m}\} \subseteq N \text{ with } N \in \Psi
  \]
\end{lemma}
\begin{proof}
  ($\Leftarrow$)
  follows directly from the fact that the judgement is coherent.
  ($\Rightarrow$)
  The proof is by contradiction.
  Assume that there is $(\ptp{n,m}) \in \spR$ with $\ptp{n} \in N$ and $\ptp{m} \in M$,
  where $N \neq M \in \Psi$
  (assuming $\CHANS{\Omegan{n}} \cap \CHANS{\Omegan{m}} \neq \varnothing$,
  without loss of generality).
  
  Let us consider the following judgement:
  \[
  \spjudge{\Psi' \cdot N \cdot M}{\Theta}{
      \PBOX{n}{\PSEND{a}{\SORTV} \PSEP \PV_1} \PPAR
      \PBOX{m}{\PRECEIVE{a}{\SORTV} \PSEP \PV_2} \PPAR
      \SYSV'
      }
      {\Omega' \cdot \prtrack{n}{\PSEND{a}{\SORTV} \PSEP \spP}
        \cdot \prtrack{m}{\PRECEIVE{a}{\SORTV} \PSEP \spQ}
      } 
  \]
  Note that rule \rulename{$\epsilon$} is not applicable here since we do not have
  $\prtrack{n}{\epsilon}$ and $\prtrack{m}{\epsilon}$.
  Looking at the rules we see that only two applications of \rulename{$sync$}
  would introduce $\prtrack{n}{\PSEND{a}{\SORTV} \PSEP \spP}$ and
  $\prtrack{m}{\PRECEIVE{a}{\SORTV} \PSEP \spQ}$.

  For this rule to be applicable we must have
  $\SYSV' \equiv \PBOX{n'}{\PRECEIVE{a}{\SORTV} \PSEP \PV_3} \PPAR
  \PBOX{m'}{\PSEND{a}{\SORTV} \PSEP \PV_4} \PPAR
  \SYSV''$, and
  $\ptp{n'} \in N$, $\ptp{m'} \in M$.
  
  By definition of $\SsplitE$ and since rule $\gigws$ must be applied
  for  $\derive{\SYSV}{\GA}$ to hold,
  we should be able to derive
  \[
  \derive{\PBOX{n}{\PSEND{a}{\SORTV} \PSEP Q_1} \PPAR
    \PBOX{m}{\PRECEIVE{a}{\SORTV} \PSEP Q_2} \PPAR
    \PBOX{n'}{\PRECEIVE{a}{\SORTV} \PSEP Q_3} \PPAR
    \PBOX{m'}{\PSEND{a}{\SORTV} \PSEP Q_4} \PPAR
    \SYSV''}
  {\GA}
  \]
  However, this is not derivable due to the obvious race on channel $\chan{a}$!
\end{proof}

\begin{lemma}\label{lem:subredsplit1}\label{lem:splitrec}
If 
\[
\Ssplit{\PBOX{r}{\PV} \PPAR \QUEUE{a}{\rho} \PPAR T} = (\SYSV'_0, \SYSV'_1)
\quad \text{such that} \quad 
\restri{\SYSV'_0}{r} = \PV_0 
\quad \text{and} \quad
\restric{\SYSV'_0}{a} = \rho_0
\]
then
\[
\Ssplit{\PBOX{r}{\PRECEIVE{a}{\SORTV} \PSEP \PV} \PPAR \QUEUE{a}{\SORTV \cdot \rho} \PPAR T}
= (\SYSV_0, \SYSV_1)
\quad \text{such that} \quad 
\restri{\SYSV_0}{r} = \PRECEIVE{a}{\SORTV} \PSEP  \PV_0 
\quad \text{and} \quad
\restric{\SYSV_0}{a} = \SORTV \cdot \rho_o
\]
In addition, $\forall \ptp{n} \in \PARTS{T} \st \restri{\SYSV_i}{n} = \restri{\SYSV'_i}{n}$,
for $i \in \{0,1\}$.
\end{lemma}
\begin{proof}
  Straightforward (runtime rule)
\end{proof}

\begin{lemma}\label{lem:splitsend}
  If
  $\derive{\PBOX{s}{\PSEND{a}{\SORTV} \PSEP \PV \PINCH \PV'} \PPAR \QUEUE{a}{\rho} \PPAR T}{\GA}$
  and 
  \[
  \Ssplit{\PBOX{s}{\PSEND{a}{\SORTV} \PSEP \PV \PINCH \PV'} \PPAR \QUEUE{a}{\rho} \PPAR T}
  = (\SYSV_1, \SYSV_2)
  \]
  such that
  \[
  \restri{\SYSV_1}{s} = \PSEND{a}{\SORTV} \PSEP \PV_0 \PINCH \PV'_0
  \quad \text{and} \quad
  \restric{\SYSV_1}{a} = \rho
  \]
  then we have 
  \[
  \Ssplit{\PBOX{s}{\PV} \PPAR \QUEUE{a}{\rho \cdot \SORTV} \PPAR T}  = (\SYSV'_1, \SYSV'_2)
  \quad
  \text{such that}
  \quad
  \restri{\SYSV'_1}{s} = \PV_0
  \quad \text{and} \quad
  \restric{\SYSV'_1}{a} = \rho \cdot \SORTV
  \]
  and $\forall \ptp{n} \neq \ptp{s} \in \PARTS{\SYSV}.
  \restri{\SYSV_1}{n} = \restri{\SYSV'_1}{n}$ and 
  $\restri{\SYSV_2}{n} = \restri{\SYSV'_2}{n}$, ditto $\forall \chan{a} \in \CHANS{\SYSV}$.
\end{lemma}
\begin{proof}
  We must have a judgement of the form
  \[
  \spjudge{\Psi}{\Theta}{
    \PBOX{s}{\PSEND{a}{\SORTV} \PSEP \PV \PINCH \PV'} \PPAR \QUEUE{a}{\rho} \PPAR T
    }
    {
      \Omega \cdot \prtrack{s}{\PSEND{a}{\SORTV} \PSEP \PV_0 \PINCH \PV'_0}
      \cdot \prtrack{r}{\PRECEIVE{a}{\SORTV} \PSEP Q_0 \PINCH Q'_0}
      \cdot \prtrackc{a}{\rho} 
    }
  \]
  Since the system is derivable,
  we have to use \rulename{$\splqueue$} until the queue is empty (otherwise
  linearity would not be preserved).
  Then only use
  \rulename{$\splsync$} to remove the send on $\chan{a}$, therefore 
  both $\rho$ and the send should be on the same part of the split.
  After reduction, we have the following judgement
  \[
  \spjudge{\Psi}{\Theta}{
    \PBOX{s}{\PV} \PPAR \QUEUE{a}{\rho \cdot \SORTV} \PPAR T
  }
  {
    \Omega \cdot \prtrack{s}{\PV_0}
    \cdot \prtrack{r}{\PRECEIVE{a}{\SORTV} \PSEP Q_0 \PINCH Q'_0}
    \cdot \prtrackc{a}{\rho \cdot \SORTV} 
  }
  \]
  Since $\ptp{r}$ was able to receive $\SORTV$ from $\ptp{s}$ before, it
  must be able to receive it from the queue as well
  (note that there is not restriction on wrt $\Psi$ with action on queues).
\end{proof}

\begin{lemma}\label{lem:allinpsi}
  If
  $\derive{\SYSV}{\GA}$ then $\spjudge{\{\PARTS{\SYSV}\}}{\varnothing}{\SYSV}{\Omega}$
  and $\Omegan{n} \subs{\PEND}{\epsilon} \equiv \restri{\SYSV}{n}$.
  \end{lemma}
  \begin{proof}
    Straightforward induction on the derivation of 
    $\spjudge{\{\PARTS{\SYSV}\}}{\Theta}{\SYSV}{\Omega}$.
    Note that for each rule of the split, there is a rule in the inference system,
    whose premises are always \emph{weaker}.
    Also, \rulename{$\splepsilon$} is not applicable (since $\Theta$ is empty) and
    $\Omega$ keeps track of everything that happens (module the branches which are
    not taken).
  \end{proof}

\begin{lemma}\label{lem:splitunique}
  If $\derive{\SYSV}{\GA}$ and $\spjudge{\Psi}{\Theta}{\SYSV}{\Omega}$ is coherent,
  then $\forall \; \Psi' , \Theta' , \Omega'$ such that
  $\spjudge{\Psi'}{\Theta'}{\SYSV}{\Omega'}$ is coherent:
  $\Psi = \Psi'$, $\Theta = \Theta'$ and $\Omega = \Omega'$.
\end{lemma}

\begin{proof}
  First, recall that by Lemma~\ref{lem:spiff}:
   \[
  (\ptp{n,m}) \in \spR
  \iff
  \{\ptp{n,m}\} \subseteq N \text{ with } N \in \Psi
  \]
  We first show that $\Psi$ (such that the split is coherent) is unique.
  Note that because of condition~\eqref{eq:cohecond0} and the fact
  that $\derive{\SYSV}{\GA}$,
  it is not possible
  to have a coherent judgement with $\Psi'$ like $\Psi$ except for two
  sets in $\Psi$ being merged (subdivided) in $\Psi'$.
  Indeed, the number of interacting pairs of participants is fixed in $\SYSV$.
  The only changes one can do in $\Psi$ are as follows
  \begin{enumerate}
  \item\label{en:splitunique2} Add $\ptp{n}$ in $N \in \Psi$ (with $\ptp{n}$ not in $\Psi$).
  \item\label{en:splitunique4} Remove $\ptp{n}$ in $N \in \Psi$ (with $\ptp{n}$ in $\Psi$).
  \item\label{en:splitunique5} Permute $\ptp{n} \in N \in \Psi$ with $\ptp{m} \in M \in \Psi$.
  \end{enumerate}
  We now show that any of these changes makes the judgement not coherent.

  Case~\ref{en:splitunique2}.
  Assume $\Psi = N \cdot \Psi_0$ and  $\Psi' = \{\ptp{n}\} \cup N \cdot \Psi_0$.
  This means that $\Omeganp{n} \neq \PEND$ and  $\Omeganp{n} \neq \epsilon$
  (otherwise $\spR$ would not be total on $\{\ptp{n}\} \cup N$),
  and $\Omegan{n} = \PEND$, or $\Omegan{n} = \epsilon$, or $\ptp{n}$ only received
  from queues.
  This means that $\ptp{n}$ interact with participants in $\Omega'$ but not in $\Omega$,
  which means that at some point in the derivation of $\Omega$
  we have $\noonestep{\PBOX{n}{\_} \PPAR \SYSV_0}$ while we have
  $\onestep{\PBOX{n}{\_} \PPAR \SYSV_0}$ in the derivation for $\Omega'$.
  However, since $\GA$ is typable, there cannot be races in the systems and
  therefore, the interaction between pair of participants must be the same in 
  both derivations of $\Omega$ and $\Omega'$.

  Case~\ref{en:splitunique4}.
  Assume that  $\Psi = \{\ptp{n}\} \cup N \cdot \Psi_0$ and $\Psi' = N \cdot \Psi_0$,
  the case is similar to the previous one, we have that
  at some point in the derivation of $\Omega'$
  we have $\noonestep{\PBOX{n}{\_} \PPAR \SYSV_0}$ while we have
  $\onestep{\PBOX{n}{\_} \PPAR \SYSV_0}$ in the derivation for $\Omega$.

  Case~\ref{en:splitunique5}.
  Assume that $\Psi = \{\ptp{m}\} \cup N \cdot \{\ptp{n}\} \cup M \cdot \Psi_0$
  and  $\Psi' = \{\ptp{n}\} \cup N \cdot \{\ptp{m}\} \cup M \cdot \Psi_0$.
  We know that
  \[
  \forall \ptp{n'} \in N \, . \, 
  \CHANS{\Omegan{n'}} \cap \CHANS{\Omegan{m}} = \varnothing
  \quad \text{and} \quad
  \forall \ptp{m'} \in M \, . \, 
  \CHANS{\Omegan{m'}} \cap \CHANS{\Omegan{n}} = \varnothing
  \]
  since the 
  original judgement is coherent and the system typable. This implies
  that we cannot have $\spR$ total neither on $\{\ptp{n}\} \cup N$
  nor on $\{\ptp{m}\} \cup M$. Therefore the judgement is not coherent.

  We now consider the changes that one can make on $\Omega$. Note that changes on $\Omega$
  must be done on pairs of participants since they always interact by pair
  (except for those which interact only with queues but that has no effect on 
  $\Psi$ or $\Theta$).
  \begin{enumerate}
  \item There is $\ptp{n_1,n_2}$ such that
    $\Omegan{n_i}$ is a prefix of $\Omeganp{n_i}$.
    A pair of elements in $\Omega'$ can only be longer if
    one merges two in $\Psi$, which is possible (see above).
  \item  There is $\ptp{n_1,n_2}$ such that
    $\Omeganp{n_i}$ is a prefix of $\Omegan{n_i}$.
    A pair of elements in $\Omega'$ can only be shorter
    if two a set in $\Psi$ is subdivided into two sets which is not 
    allowed either.
  \end{enumerate}

  We consider changes that can be made on $\Theta$:
  \begin{enumerate}
  \item Add $\{\ptp{n,m}\}$ to $\Theta$,
    then, that pair will not allow the derivation to reach the axiom, unless
    sets in $\Psi$ are sub-divided, which is not possible.
  \item Remove  $\{\ptp{n,m}\}$ from $\Theta$,
    then, there will be a pair missing to reach the axiom (plus $\sprelT$ might
    not be total any more); unless two sets in $\Psi$ are merged, but this is not possible.
  \item Permutation in $\Theta$ would only possible if one could permute participants in sets
    of $\Psi$ which, by above, is not possible.
  \end{enumerate}
\end{proof}

\subsection{Others}

\begin{lemma}\label{lem:precon}
  If  $\GA \not\equiv \GA_0  \GPAR \GA_1 $ and $\wfjudge{\emptyTree}{\GA}$
  then
  $\conndef{\GA}$ is total on $\PARTS{\GA}$.
\end{lemma}
\begin{proof}
  We show this by induction on the structure of $\GA$.
  
  \proocase{$\GA \equiv \GSENDN \GSEP \GA'$}
  By definition $(\ptp{s,r}) \in \conndefG$ and by IH $\conndef{\GA'}$ is total
  on $\PARTS{\GA'}$.
  Since $\GA$ is well-formed we have
  \[
  \forall \; \chanT{\_}{n_1}{n_2} \in \GFIN{\GA'} \, . \,
  \{\ptp{n_1,n_2}\} \cap \{\ptp{s,r}\}
  \]
  Thus there is $(\ptp{s,n_i}) \in \conndefG$ or  $(\ptp{r,n_i}) \in \conndefG$
  and we have the required result by definition of $\conndefG$ and Def.~\ref{def:proj}.
  
  \medskip

  \proocase{$\GA \equiv \GA_0 \GCH \GA_1$}
  By IH, $\conndef{\GA_i}$ is total on $\PARTS{\GA_i}$ for $i \in \{0,1\}$.
  Since  $\GA$ is well-formed we have
  \[
  \forall \; \chanTN  \in \GFIN{\GA} .
  \forall \; \chanT{b}{s'}{r'} \in \GFIN{\GA'}
  \, . \, \ptp{s} = \ptp{s'}  \land \chan{a} \neq \chan{b}
  \]
  i.e.\ $\ptp{s} = \ptp{s'} \in \GA_i$, and we have the required result
  by definition of $\conndefG$ and Def.~\ref{def:proj}.

 \medskip
 
  \proocase{$\GA \equiv \GSENDN \GSEP (\GA_0 \GPAR \GA_1)$}
  By IH $\conndef{\GA_i}$ is total on $\PARTS{\GA_i}$.
  Since $\GA$ is well-formed we must have
  $\ptp{s} \in \PARTS{\GA_i}$ and  $\ptp{r} \in \PARTS{\GA_j}$ with $i \neq j \in \{0,1\}$.
  We have the required result since we have $(\ptp{s,r}) \in \conndefG$
  by definition of $\conndefG$ and Def.~\ref{def:proj}.

 \medskip

  \proocase{$\GA \equiv \GA_0 \GWS \GA_1$}
  Observe that by IH and by definition of $\GFINPE$, we have that
  $\forall N \in \GFINP{\GA_i} \, . \, \conndef{\GA_i}$ is total on $N$,
  with $i \in \{0,1\}$.

  Since the projection is defined as
  $\Proj{\GA}{n} = \Proj{\GA_0}{n}\subs{\Proj{\GA_1}{n}}{\PEND}$,
  we have that
  \[
  \forall \;  N_0 \times N_1 \subseteq \GFINP{\GA_0} \times \GFINP{\GA_1}
  \st
  \conndef{\GA} \text{ is total on }
  N_0 \cup N_1 \text{ if there is }\ptp{n} \in N_0 \cap N_1
  \]
  Since $\conndef{\GA}$ is a transitive relation, let us define
  a transitive relation on the intersection of sets 
  of participants from $\GA_0$ and $\GA_1$:
  \newcommand{\RPWF}{\mathtt{W}}
  \[
  \begin{array}{c}
    (N_0, N_1) \in \RPWF 
    \\
    \iff
    \\
    N_0 \cap N_1 \neq \varnothing
    \quad \text{or} \quad 
    \exists \, (M_0, M_1) \st (N_0,M_1) \in \RPWF \text{ and } (M_0,N_1) \in \RPWF 
  \end{array}
  \]
  It is easy to see that $\conndef{\GA}$ is total on any $N_0 \cup N_1$
  whenever $(N_0,N_1) \in \RPWF$, thus
  \[
  \conndef{\GA}
  \text{ is total on}
  \bigcup_{ (N_0, N_1) \in \RPWF } N_0 \cup N_1
  \]
  Let us show that
  \[
  (N_0,N_1) \in \GFINP{\GA_0} \times \GFINP{\GA_1} \Rightarrow (N_0, N_1) \in \RPWF
  \]
  Since  $\GA$ is well-formed we have
  \begin{equation}\label{eq:rpwf1}
    \forall \; \chanT{\_}{s}{r}  \in \GFIN{\GA_1} \, . \,
    \exists N_1 \neq N_2 \in \GFOUTS{\GA_0} \, . \,
    \ptp{s} \in N_1 \land \ptp{r} \in N_2
  \end{equation}
  and
  \begin{equation}\label{eq:rpwf2}
    \forall N \in \GFINP{\GA_0} \, . \,
    \exists N' \in \GFINP{\GA_1} \, . \,
    N \cap N' \neq \varnothing
  \end{equation}
  Therefore, by~\eqref{eq:rpwf1} and the fact
  that $ \chanT{\_}{s}{r}  \in \GFIN{\GA_1} \Rightarrow \{\ptp{s,r}\} \subseteq  N
  \in \GFINP{\GA_1}$, we have
  \begin{equation}\label{eq:rpwf3}
    \forall N_1 \in \GFINP{\GA_1} \, . \, 
    \exists N_0 \neq N'_0 \in  \GFINP{\GA_0} \st N_0 \cap N_1 \neq \varnothing
    \text{ and }  N'_0 \cap N_1 \neq \varnothing
  \end{equation}
  By~\eqref{eq:rpwf2}, we have
  \begin{equation}\label{eq:rpwf4}
    \forall N_0 \in \GFINP{\GA_0} \, . \, 
    \exists N_1 \in  \GFINP{\GA_1} \st N_0 \cap N_1 \neq \varnothing
  \end{equation}
  
  Now assume that there is $(N_0,N_1) \in \GFINP{\GA_0} \times \GFINP{\GA_1}$
  such that $(N_0,N_1) \not\in \RPWF$,
  this is contradiction with~\eqref{eq:rpwf3} and~\eqref{eq:rpwf4}.

 \medskip

  \proocase{$\GA \equiv \GRECN \GA$} By IH

 \medskip

  \proocase{Other cases}
  The cases where $\GA = \GEND$ or $\GA = \GRECV$ are trivial.
\end{proof}

\begin{lemma}\label{lem:chansetSG}
  If $\derive{\SYSV}{\GA}$ and $\wfjudge{\emptyTree}{\GA}$
  then the following holds
 \[
 \forall \ptp{n} \in \PARTS{\SYSV} \st
 \CHANS{\restri{\SYSV}{n}} \subseteq \CHANS{\Proj{\GA}{n}}
 \]
\end{lemma}
\begin{proof}
  Straightforward induction on the validation rules.
\end{proof}

\begin{lemma}\label{lem:RgimplieRs}
  If $\derive{\SYSV}{\GA}$ and
  $\conndefG $ is total on $\PARTS{\GA}$ then $\conndefS$ is total on $\PARTS{\SYSV}$.
\end{lemma}
\begin{proof}
  By Lemmas~\ref{lem:precon} and~\ref{lem:chansetSG} and the definition of $\conndefS$.
\end{proof}

\begin{lemma}\label{lem:conntotal}
  If $\wfjudge{\emptyTree}{\GA}$
  then $\forall N \in \GFOUTS{\GA} \; . \; \conndef{\GA}$ is total on $N$.
\end{lemma}
\begin{proof}
  By Lemma~\ref{lem:precon} and the definition of $\GFOUTE$.
\end{proof}

\begin{lemma}\label{lem:connGtoSpl}
  If $\derive{\SYSV}{\GA_0 \GWS \GA_1}$,
  $\wfjudge{C}{\GA_0}$,
  $\spjudge{\Psi}{\Theta}{\SYSV}{\Omega}$ is coherent,
  and
  $\Ssplit{\SYSV} \neq \nada$
  then
  \[
  \forall \ptp{n,m} \in \PARTS{\SYSV} \st
  (\ptp{n,m}) \in \conndef{\GA_0}
  \Rightarrow
  (\ptp{n,m}) \in \spR
  \]
\end{lemma}
\begin{proof}
  Straightforward by definitions of $\conndef{\GA_0}$ and  $\spR$.
  Note that external choice branches which do not appear in $\Omega$
  do not appear in $\GA_0$ either.
\end{proof}

\begin{lemma}\label{lem:recB}
  If $\boundn{\SYSV} \neq \varnothing$ then
  $\Ssplit{\SYSV} = \nada$.
\end{lemma}
\begin{proof}
  If $\boundn{\SYSV} \neq \varnothing$, we must have
  \[
  \SYSV \equiv \PBOX{n}{\PV} \PPAR \SYSV'
  \text{ where }
  \PRECN \PV' \text{ is a suffix of } \PV
  \]
  The result follows from the fact that there is no rule in Fig.~\ref{fig:split}
  which ``removes'' recursive definition. Therefore, it not possible to
  derive a split whenever there is a recursion definition in the system to be split.
\end{proof}

\begin{lemma}\label{lem:recF}
If $\derive{\SYSV}{\GRECN \GA}$ and $\GRECV \in \freen{\GA}$
then $\#(\GFOUTS{\GA}) = 1$.
\end{lemma}
\begin{proof}
  The proof follows from the fact that the context $\Gamma$ is emptied
  each time the rule \rulename{$\gipar$} is used in the derivation
  (this rule is the only one introducing concurrent branches).
  In addition, for the axiom \rulename{$\givar$} to be used in the derivation
  one must have $(\_,\_) :\GRECV \in \Gamma$.
  Therefore, the only way one could have $\#(\GFOUTS{\GA}) > 1$ (i.e.\
  at least two concurrent branches in $\GA$) is if $\GRECV$ does not appear
  in $\GA$.
\end{proof}

\begin{lemma}\label{lem:recBB}
  If $\derive{\SYSV}{\GA}$ and $\boundn{\SYSV} = \varnothing$
  then $\boundn{\GA} = \varnothing$
\end{lemma}
\begin{proof}
  By straightforward induction on the rules of Fig.~\ref{fig:gloinfer}.
\end{proof}

\begin{lemma}\label{lem:wfimplies}
  If $\wfjudge{C}{\GA}$ then  $\wfjudge{\emptyTree}{\GA}$ 
\end{lemma}
\begin{proof}
  This follows from the fact that
  $\addchan{\emptyTree}{C}$ is always defined.
\end{proof}

\begin{lemma}\label{lem:svsg}
If  $\judge{\Gamma}{C}{\SYSV}{\GA}$ and $\GA$ is well-formed
  \[
  \GA \equiv ((\GSEND{n}{m}{a}{\SORTV} \GSEP \GA_1 \GCH \GA_2) \GPAR \GA_3) \GWS \GA_4
  \iff
  \SYSV \equiv \PBOX{n}{\PSEND{a}{\SORTV} \PSEP \PV \PINCH \PV'}
  \PPAR
  \PBOX{m}{\PRECEIVE{a}{\SORTV} \PSEP Q \POUTCH Q'}
  \PPAR \SYSV'
  \]

\end{lemma}
\begin{proof}
\noindent
$(\Rightarrow)$
Assume that
  \[
  \judge{\Gamma}{C}{\SYSV}{((\GSEND{n}{m}{a}{\SORTV} \GSEP \GA_1 \GCH \GA_2) \GPAR \GA_3) \GWS \GA_4}
  \]
  is derivable.
  We show that either a rule introducing the corresponding operator
  is applicable or that an equivalent $\GA$ can be inferred.
  \begin{itemize}
  \item If $\judge{\Gamma}{C}{\SYSV}{\GA' \GWS \GA_4}$ is derivable
    then we have either $\Ssplit{\SYSV} = \nada$ 
    then $\GA_4 = \GEND$, thus 
    \[
    \judge{\Gamma}{C}{\SYSV}{\GA'}
    \]
    or $\Ssplit{\SYSV} \neq \nada$, in this case,
    we must have
    \[
    \Ssplit{\SYSV} = (\SYSV_1, \SYSV'_1)
    \]
    with
    \[
    \judge{\emptyctx}{C}{\SYSV_1}{\GA'}
    \quad \text{and} \quad
    \judge{\emptyctx}{\addchan{C}{\chanG{\GA'}}}{\SYSV'_1}{\GA_4}
    \]

  \item Assume $\GA' = \GA'' \GPAR \GA_3$,
    if $\judge{\Gamma}{C_1}{\SYSV_1}{\GA'}$ is derivable
    then we must have either 
    $\SYSV_1 \equiv \SYSV_2 \PPAR \SYSV_3$, $\chanset = \chanset_1 \cup \chanset_2$,
    and  $\chanset_1 \cap \chanset_2 = \varnothing$
    such that
    \[
    \judgeC{\chanset_1}{\emptyctx}{C}{\SYSV_2}{\GA''}
    \quad \text{and} \quad
    \judgeC{\chanset_2}{\emptyctx}{C}{\SYSV_3}{\GA_3}
    \]
    are derivable, or $\SYSV_3 = \PEND$ and $\GA_3 \equiv \GEND$.

    \item Assume $\GA'' = \GA''' \GCH \GA_2$, we must have either
      \begin{equation}\label{eq:svsg2}
        \SYSV_2 = \PBOX{n}{\PV_0 \PINCH \PV'_0} \PPAR \SYSV_4
      \end{equation}
      and
      \[
      \judgeC{\chanset_1}{\emptyctx}{C}{\PBOX{n}{\PV_0} \PPAR \SYSV_4}{\GA'''}
      \quad \text{and} \quad
      \judgeC{\chanset_1}{\emptyctx}{C}{\PBOX{n}{\PV'_0} \PPAR \SYSV_4}{\GA_2}
    \]
    are derivable, or
    $\PV'_0 \equiv \PEND$ and $\GA_2 \equiv \GEND$.

  \item Assume $\GA''' = \GSEND{n}{m}{a}{\SORTV} \GSEP \GA_1$, we must have
    \begin{equation}\label{eq:svsg3}
    \PV_0 \equiv \PSEND{a}{\SORTV} \PSEP \PV
    \quad \text{and} \quad
    \SYSV_4 \equiv \PBOX{m}{\PRECEIVE{a}{\SORTV} \PSEP Q \POUTCH Q'} \PPAR \SYSV'
  \end{equation}
  and
    \[
    \judgeC{\chanset_1}{\emptyctx}{C}{\PBOX{n}{\PV'} \PPAR
      \PBOX{m}{\PRECEIVE{a}{\SORTV} \PSEP Q \POUTCH Q'} \PPAR \SYSV' }{\GA_1}
    \]
    derivable.
\end{itemize}
Putting~\eqref{eq:svsg2} and~\eqref{eq:svsg3} together, we have that 
\[
\SYSV \equiv \PBOX{n}{ \PSEND{a}{\SORTV} \PSEP \PV \PINCH \PV'} \PPAR
\PBOX{m}{\PRECEIVE{a}{\SORTV} \PSEP Q \POUTCH Q'} \PPAR \SYSV'
\]

\noindent
$(\Leftarrow)$
Assume 
  \begin{equation}\label{eq:lem10:a}
    \judge{\Gamma}{C}{
      \PBOX{n}{\PSEND{a}{\SORTV} \PSEP \PV \PINCH \PV'}
      \PPAR
      \PBOX{m}{\PRECEIVE{a}{\SORTV} \PSEP Q \POUTCH Q'}
      \PPAR \SYSV'}
    {\GA}
  \end{equation}
  We show that either a rule introducing the corresponding operator
  is applicable or that an equivalent $\GA$ can be inferred.
  \begin{itemize}
  \item Either $\Ssplit{\SYSV} = \nada$ in which case $\GA \equiv \GA' \GWS \GEND$
    or 
    \[
    \Ssplit{\SYSV}
    = (\SYSV_1, \SYSV'_1) 
    \]
    and we must have $\GA_4 \not\equiv \GEND$ and
    \[
    \judge{\Gamma}{C}{
     \SYSV_1
    }
    {\GA}
    \]
    with
    \[
    \restri{\SYSV_1}{n} = \PSEND{a}{\SORTV} \PSEP \PV_0 \PINCH \PV'_0
    \qand
    \restri{\SYSV_1}{m} = \PRECEIVE{a}{\SORTV} \PSEP Q_0 \POUTCH Q'_0
    \qand
    \restri{\SYSV'_1}{n} = \PV_1
    \qand
     \restri{\SYSV'_1}{m} = Q_1
    \]
    such that 
    \[
    \PSEND{a}{\SORTV} \PSEP \PV \PINCH \PV'
    \equiv
    (\PSEND{a}{\SORTV} \PSEP \PV_0 \PINCH \PV'_0)\subs{\PV_1}{\PEND}
    \qand
    \PRECEIVE{a}{\SORTV} \PSEP Q \POUTCH Q'
    \equiv
    (\PRECEIVE{a}{\SORTV} \PSEP Q_0\POUTCH Q'_0 )\subs{Q_1}{\PEND}
    \]
  \item Either there is 
    \[ 
    \SYSV = \PBOX{n}{\PSEND{a}{\SORTV} \PSEP \PV_0 \PINCH \PV'_0}
    \PPAR
    \PBOX{m}{\PRECEIVE{a}{\SORTV} \PSEP Q_0 \POUTCH Q'_0}
    \PPAR \SYSV_1 \PPAR \SYSV_2
    \]
    $\GA' = \GA'' \GPAR \GA_3$
    and $\chanset = \chanset_1 \cup \chanset_2$,
    and  $\chanset_1 \cap \chanset_2 = \varnothing$
    \[
    \judgeC{\chanset_1}{\emptyctx}{C}{
      \PBOX{n}{\PSEND{a}{\SORTV} \PSEP \PV_0 \PINCH \PV'_0}
      \PPAR
      \PBOX{m}{\PRECEIVE{a}{\SORTV} \PSEP Q_0 \POUTCH Q'_0}
      \PPAR \SYSV_1
    }{\GA''}
    \quad \text{and} \quad
    \judgeC{\chanset_2}{\emptyctx}{C}{\SYSV_2}{\GA_3}
    \]
    where $\noonestep{\SYSV_1}$ (this is a sound assumption, since
    one could apply \rulename{$\gigws$} and \rulename{$\gipar$}
    as many times as necessary to obtain this),
    or $\GA_3 \equiv \GEND$.
    
  \item Either there is $\GA'' \equiv \GA''' \GCH \GA_2$ such that
    \[
    \judge{\emptyctx}{C_1}{
      \PBOX{n}{\PSEND{a}{\SORTV} \PSEP \PV_0}
      \PPAR
      \PBOX{m}{\PRECEIVE{a}{\SORTV} \PSEP Q_0 \POUTCH Q'_0}
      \PPAR \SYSV_1
    }{\GA'''}
    \]
    and
    \[
   \judge{\emptyctx}{C_1}{
      \PBOX{n}{\PV'}
      \PPAR
      \PBOX{m}{\PRECEIVE{a}{\SORTV} \PSEP Q_0 \POUTCH Q'_0}
      \PPAR \SYSV_1
    }{\GA_2}
    \]
    or $\GA_2 = \GEND$.

  \item For
    \[
    \judge{\emptyctx}{C_1}{
      \PBOX{n}{\PSEND{a}{\SORTV} \PSEP \PV}
      \PPAR
      \PBOX{m}{\PRECEIVE{a}{\SORTV} \PSEP Q \POUTCH Q'}
      \PPAR \SYSV_1
    }{\GA'''}
    \]
    to be derivable,
    we must have $\GA''' \equiv \GSEND{n}{m}{a}{\SORTV} \GA_1$.
  \end{itemize}
  Putting all the pieces together, we have the required result. 
\end{proof}

\begin{lemma}\label{lem:partis}
  If $\derive{\SYSV}{\GA}$ then
  \[
  \forall \ptp{n} \in \PARTS{\SYSV} . \, \restri{\SYSV}{n} \neq \PEND \st
  \ptp{n} \in \PARTS{\SYSV} \iff \ptp{n} \in \PARTS{\GA}
  \]
  and
  \[
  \forall \ptp{n} \in \PARTS{\SYSV} . \,  \restri{\SYSV}{n} = \PEND
  \iff
  \ptp{n} \not\in \PARTS{\GA}
  \]
\end{lemma}
\begin{proof}
 Straightforward.
\end{proof}

\begin{lemma}\label{lem:chans}
  If $\derive{\SYSV}{\GA}$ then $\CHANS{\GA} \subseteq \CHANS{\SYSV}$.
\end{lemma}
\begin{proof}
  Straightforward induction on the derivation.
\end{proof}

\begin{lemma}\label{lem:chansA}
  If $\derive{\SYSV}{\GA}$ then $\CHANS{\SYSV} \subseteq \chanset$.
\end{lemma}
\begin{proof}
  Straightforward induction on the derivation.
\end{proof}

\begin{lemma}\label{lem:gvsproj}
  If $\GA$ is well-formed and projectable then
  \begin{itemize}
  \item if $\Proj{\GA}{n} \equiv \PSEND{a}{\SORTV} \PSEP \PV \PINCH Q$
    then there is a branch of $\GA$ such that the
    first prefix on $\ptp{n}$ is 
    $\GSEND{n}{m}{a}{\SORTV}$.
  \item if $\Proj{\GA}{n} \equiv \PRECEIVE{a}{\SORTV} \PSEP \PV \POUTCH Q$
    then there is a branch of $\GA$ such that the
    first prefix on $\ptp{n}$ is 
    $\GSEND{m}{n}{a}{\SORTV}$.
  \end{itemize}
\end{lemma}
\begin{proof}
By definition of Projection.
\end{proof}

\begin{lemma}\label{lem:qstar}
  If $\derive{\SYSV \PPAR \QUEUE{a}{\SORTV \cdot \rho}}{\GA}$
  is derivable
  then for each branch in $\GA$ the first prefix on $\chan{a}$ is 
  $\GSEND{\GSTAR}{n}{a}{\SORTV}$.
\end{lemma}
\begin{proof}
  Follows from the fact that for a common channel,
  \rulename{$\giqueue$} must be used before
  \rulename{$\gipsep$} (see Lemma~\ref{lem:linstar}). 
\end{proof}

\begin{lemma}\label{lem:gvssys}
  If $\derive{\SYSV}{\GA}$ and there is a branch in $\GA$ such 
  that $\GSEND{n}{m}{a}{\SORTV}$ is the first prefix on $\ptp{n}$ (resp. $\ptp{m}$),
  then $\restri{\SYSV}{n} = \PSEND{a}{\SORTV} \PSEP \PV \PINCH \PV'$
  (resp.\  $\restri{\SYSV}{m} = \PRECEIVE{a}{\SORTV} \PSEP Q \POUTCH Q$).
\end{lemma}
\begin{proof}
  Straightforward.
\end{proof}

\begin{lemma}\label{lem:closeness}
  If $\freen{\SYSV} = \varnothing$, and  $\SYSV \ltsarrow{} \SYSV'$,
  then $\freen{\SYSV'} = \varnothing$, i.e.
  reduction of systems preserves closeness.
\end{lemma}
\begin{proof}
  Follows directly from the semantics of the calculus.
\end{proof}

\begin{lemma}\label{lem:qproj}
  If $\chan{a} \in \CHANS{\GA}$ then
  \[
  \derive{\QUEUE{a}{\rho} \PPAR \SYSV}{\GA} 
  \iff
  \text{ then }
  \Projc{\GA}{a} = \rho
  \]
 In addition, $\chan{a} \not\in \CHANS{\GA} \Rightarrow \rho = []$.
\end{lemma}
\begin{proof}
  Follows from rules \rulename{$\giqueue$} and \rulename{$\giend$}.
\end{proof}

\end{document}